\documentclass[twocolumn,english,aps,pra,longbibliography,superscriptaddress, floatfix]{revtex4-1}
\usepackage[T1]{fontenc}
\usepackage{inputenc}
\usepackage{color}
\usepackage[dvipsnames]{xcolor}
\usepackage{placeins}
\usepackage{tabularx}
\usepackage{tcolorbox}
\usepackage{amsmath,amsfonts,amsthm}

\usepackage[short]{optidef}

\usepackage{graphicx}
\usepackage{standalone}
\usepackage{braket}
\usepackage{amssymb}
\usepackage{todonotes}
\newcommand{\R}{\mathbb R}
\DeclareMathOperator{\tr}{tr}

\DeclareMathOperator{\poly}{poly}
 
\usepackage[normalem]{ulem}
\usepackage[colorlinks=true,urlcolor=blue,citecolor=blue,linkcolor=blue]{hyperref}
\usepackage[ruled,vlined]{algorithm2e}
\usepackage{setspace}
\usepackage{bbm}

\usepackage{tikz}
\usetikzlibrary{calc,patterns}

\DeclareMathOperator*{\argmin}{argmin} 

\usepackage{soul}

\newcommand{\C}[1]{\textbf{#1}}

\usepackage{tocloft}  
\usepackage{fancyhdr}
\makeatother

\usepackage{babel}

\begin{document}

\title{Challenges and Opportunities in Quantum Optimization}
\homepage[Summary of the Quantum Optimization Working Group.\\Authors are listed in alphabetical order.\\Contact: Stefan Woerner (\url{wor@zurich.ibm.com})]{}

\author{Amira Abbas}
\affiliation{QuSoft and University of Amsterdam}

\author{Andris Ambainis}
\affiliation{University of Latvia}

\author{Brandon Augustino}
\affiliation{Massachusetts Institute of Technology}

\author{Andreas B\"artschi}
\affiliation{Los Alamos National Laboratory}

\author{Harry Buhrman}
\affiliation{QuSoft and University of Amsterdam}

\author{Carleton Coffrin}
\affiliation{Los Alamos National Laboratory}

\author{Giorgio Cortiana}
\affiliation{E.ON Digital Technology GmbH}

\author{Vedran Dunjko}
\affiliation{Leiden University}

\author{Daniel J.~Egger}
\affiliation{IBM Quantum, IBM Research Europe -- Zurich}

\author{Bruce G.~Elmegreen}
\affiliation{IBM Research, IBM T.J.~Watson Research Center}

\author{Nicola Franco}
\affiliation{Fraunhofer IKS}

\author{Filippo Fratini}
\affiliation{Erste Group Bank}

\author{Bryce Fuller}
\affiliation{IBM Quantum, IBM T.J.~Watson Research Center }

\author{Julien Gacon}
\affiliation{IBM Quantum, IBM Research Europe -- Zurich}
\affiliation{\'Ecole Polytechnique F\'ed\'erale de Lausanne}

\author{Constantin Gonciulea}
\affiliation{Wells Fargo}

\author{Sander Gribling}
\affiliation{Tilburg University}

\author{Swati Gupta}
\affiliation{Massachusetts Institute of Technology}

\author{Stuart Hadfield}
\affiliation{Quantum Artificial Intelligence Lab, NASA Ames Research Center}
\affiliation{USRA Research Institute for Advanced Computer Science}

\author{Raoul Heese}
\affiliation{Fraunhofer ITWM}

\author{Gerhard Kircher}
\affiliation{Erste Group Bank}

\author{Thomas Kleinert}
\affiliation{Quantagonia}

\author{Thorsten Koch}
\affiliation{Zuse Institute Berlin}
\affiliation{Technische Universit\"at Berlin}

\author{Georgios Korpas}
\affiliation{HSBC Lab, Innovation and Ventures, HSBC, London}
\affiliation{Czech Technical University in Prague}

\author{Steve Lenk}
\affiliation{Fraunhofer IOSB-AST}

\author{Jakub Marecek}
\affiliation{Czech Technical University in Prague}

\author{Vanio Markov}
\affiliation{Wells Fargo}

\author{Guglielmo Mazzola}
\affiliation{University of Zurich}

\author{Stefano Mensa}
\affiliation{The Hartree Centre, STFC}

\author{Naeimeh Mohseni}
\affiliation{E.ON Digital Technology GmbH}

\author{Giacomo Nannicini}
\affiliation{University of Southern California}

\author{Corey O'Meara}
\affiliation{E.ON Digital Technology GmbH}

\author{Elena Pe\~{n}a Tapia}
\affiliation{IBM Quantum, IBM Research Europe -- Zurich}

\author{Sebastian Pokutta}
\affiliation{Zuse Institute Berlin}
\affiliation{Technische Universit\"at Berlin}

\author{Manuel Proissl}
\affiliation{IBM Quantum, IBM Research Europe -- Zurich}

\author{Patrick Rebentrost}
\affiliation{Centre for Quantum Technologies, National University of Singapore}

\author{Emre Sahin}
\affiliation{The Hartree Centre, STFC}

\author{Benjamin C.~B.~Symons}
\affiliation{The Hartree Centre, STFC}

\author{Sabine Tornow}
\affiliation{University of the Bundeswehr Munich}

\author{V\'{i}ctor Valls}
\affiliation{IBM Quantum, IBM Research Europe -- Dublin}

\author{Stefan Woerner}
\affiliation{IBM Quantum, IBM Research Europe -- Zurich}

\author{Mira L.~Wolf-Bauwens}
\affiliation{IBM Quantum, IBM Research Europe -- Zurich}

\author{Jon Yard}
\affiliation{Institute for Quantum Computing, Perimeter Institute for Theoretical Physics, University of Waterloo}

\author{Sheir Yarkoni}
\affiliation{Volkswagen AG}

\author{Dirk Zechiel}
\affiliation{Quantagonia}

\author{Sergiy Zhuk}
\affiliation{IBM Quantum, IBM Research Europe -- Dublin}

\author{Christa Zoufal}
\affiliation{IBM Quantum, IBM Research Europe -- Zurich}

\date{\today}

\begin{abstract}
Recent advances in quantum computers are demonstrating the ability to solve problems at a scale beyond brute force classical simulation. As such, a widespread interest in quantum algorithms has developed in many areas, with optimization being one of the most pronounced domains. Across computer science and physics, there are a number of different approaches for major classes of optimization problems, such as combinatorial optimization, convex optimization, non-convex optimization, and stochastic extensions. This work draws on multiple approaches to study quantum optimization. Provably exact versus heuristic settings are first explained using computational complexity theory  --  highlighting where quantum advantage is possible in each context. Then, the core building blocks for quantum optimization algorithms are outlined to subsequently define prominent problem classes and identify key open questions that, if answered, will advance the field. The effects of scaling relevant problems on noisy quantum devices are also outlined in detail, alongside meaningful benchmarking problems. We underscore the importance of benchmarking by proposing clear metrics to conduct appropriate comparisons with classical optimization techniques. Lastly, we highlight two domains -- finance and sustainability -- as rich sources of optimization problems that could be used to benchmark, and eventually validate, the potential real-world impact of quantum optimization.\\
\\
\end{abstract}

\maketitle

\tableofcontents

\section{Introduction\label{sec:introduction}}

Quantum computing could revolutionize numerous domains in business and science, with optimization frequently identified as a prime candidate to profit from such a revolution. Optimization problems arise almost everywhere, so any improvement over state-of-the-art classical algorithms with quantum computers could have a huge impact. Moreover, these improvements could occur across multiple dimensions, such as solution quality, solution diversity, time-to-solution, and cost-to-solution. But just how tangible are the benefits of \emph{quantum optimization}? Techniques like Grover's search~\cite{grover}, quantum annealing, and adiabatic quantum optimization~\cite{kadowaki1998quantum,farhi2000quantumAdiabatic} provided a promising start, but it has become clear that more tools are needed to realize a quantum advantage in optimization. In this work, we address the potential of quantum optimization from various angles, namely: complexity theory, problem classes and algorithmic design, execution on noisy hardware at scale, and fair benchmarking, while outlining illustrative examples derived from real-world applications. 
Each topic is self-contained and written for a diverse audience, from optimization researchers to industry professionals. Thus, readers may move directly to any section of interest.

For additional context, it is often stated that quantum computers can evaluate all possible combinations of an optimization problem simultaneously. This neglects that there remains an exponential list of possible solutions and that a quantum computer requires additional effort to find good ones. 
For some combinatorial optimization problems in a worst-case setting, the required effort of known classical approaches scales exponentially in the problem size~\cite{liskiewicz2014new}.
Here, quantum computers may offer a quadratic speedup, e.g., via Grover's Search, but a quadratic speedup over an exponential run time is still an exponential run time. 
The situation can, however, significantly change if we consider a concrete problem instance instead of the general worst-case scenario.
Classically, many algorithms and heuristics have been developed that -- even for some large problems -- can obtain an (almost) optimal solution in reasonable time. 
The most prominent example is the Traveling Salesperson Problem (TSP), where the goal is to find the shortest cycle to visit a given set of places.
It is often used to illustrate the difficulty of combinatorial optimization, but in practice, very large problem instances can be solved close to optimality classically~\cite{applegate1998solution, applegate1999finding, cook1998combinatorial,cook2011traveling,cook2012pursuit,cook2024startsp}. 
In contrast, there exist problems with less than $100$ variables that are difficult to tackle classically~\cite{puchinger2010multidimensional, Packebusch_2016}.
While quantum optimization algorithms will not necessarily improve the performance for \emph{all} problems, they provide additional tools with new properties that can have significant impact and improve performance for \emph{some} problem instances and thus, improve our overall capabilities in optimization. We probe these concepts more formally in Secs.~\ref{sec:complexity},~\ref{sec:paradigms} and~\ref{sec:problem_classes}, by laying the foundation for quantum optimization algorithms within a variety of well-studied problem classes.

Another important aspect to consider in quantum computing, is noise. The ultimate strategy to deal with noise is quantum error correction, which introduces a significant overhead in the number of qubits required~\cite{bravyi2023highthreshold}. 
Until fault-tolerant error-corrected quantum computers can be implemented at scale, other approaches to retrieve meaningful results from noisy devices have been considered. A particular strategy is quantum error mitigation~\cite{Temme2017, VandenBerg2023, Kim2023}, where one reduces noise by cleverly combining multiple noisy results. 
In Sec.~\ref{sec:scaling}, we investigate the possibility of using error-mitigated noisy quantum computing to bridge the gap between classical and fully fault-tolerant quantum computing for optimization problems.

In addition to noise, in practice, one often relies on optimization heuristics, where the performance of an algorithm on a given problem instance is not known upfront. This makes comparing quantum and classical algorithms rather tricky. To establish systematic, reproducible, and comparable benchmarks, Sec.~\ref{sec:benchmarks} introduces fair metrics and outlines relevant optimization problems for which different algorithms can be applied to.
This is crucial to identify what strategy works best and guide toward development of new quantum optimization algorithms. Sec.~\ref{sec:applications} then dives into two particular real-world problems to illustrate the nuance of optimization in practice. 
Combined with reasonable benchmarks, these problems highlight where we stand and what open questions must be answered before we can impact a certain domain.
Since quantum computing will not necessarily accelerate all problems, it is crucial to understand exactly where advantages might be found.
This needs to be probed theoretically, and empirically --- which is what we showcase in this work. Quantum optimization algorithms provide additional tools that complement existing ones, and synergies with classical algorithms can certainly lead to better performance.

\newcommand{\nc}{\newcommand}
\def\mathcc#1{{\mathbf{#1}}}
\renewcommand{\P}{\mathcc{P}}
\nc{\NP}{\mathcc{NP}}
\nc{\coNP}{\mathcc{coNP}}
\nc{\BQP}{\mathcc{BQP}}
\nc{\PH}{\mathcc{PH}}
\nc{\sharpP}{\mathcc{\#P}}
\nc{\FP}{\mathcc{FP}}
\nc{\PO}{\mathcc{PO}}
\nc{\FNP}{\mathcc{FNP}}
\nc{\FPTAS}{\mathcc{FPTAS}}
\nc{\PTAS}{\mathcc{PTAS}}
\nc{\NPO}{\mathcc{NPO}}
\nc{\APX}{\mathcc{APX}}
\nc{\PSPACE}{\mathcc{PSPACE}}
\nc{\BPP}{\mathcc{BPP}}
\nc{\FBPP}{\mathcc{FBPP}}
\nc{\VP}{\mathcc{VP}}
\nc{\VNP}{\mathcc{VNP}}
\nc{\HeurBPP}{\mathcc{HeurBPP}}
\nc{\HeurBQP}{\mathcc{HeurBQP}}

\section{Quantum Advantage \& Complexity Theory\label{sec:complexity}}

In theoretical computer science, complexity classes group computational problems together in terms of the amount of resources required to solve them. Naturally, the harder the computational problem, the more resources it will require, where the resources of most importance consist of time and memory. For this reason, complexity theory is often the guiding compass for attempting to find classes of problems that are hard for a classical computer to solve, but perhaps less resource-demanding on a quantum computer. This is a very reasonable and rigorous way to formulate a quantum advantage, but \textit{proving} that problems require strictly more resources classically than they would on a quantum computer is often a monumental task. In this section, we discuss notable strides made in complexity theory, but also highlight issues that arise when restricting our notion of quantum advantage to strict complexity-theoretic separations in the context of optimization. The main takeaway for this section can be summarized as follows: 
\begin{center}
    \textit{Understanding complexity theory is extremely useful for gauging possible quantum advantage in optimization, but rigorous complexity-theoretic separations are not necessary, nor sufficient when seeking a practical quantum advantage.}
\end{center}
A noteworthy example to drive home this point is the task of factoring a number into its prime constituents. This problem is interesting from a complexity point of view as it is not believed to be efficiently solvable by a classical computer --- although there is no formal proof that confirms this. A quantum computer, on the other hand, can solve this factoring task efficiently, thanks to Shor's algorithm~\cite{shor_1994_factoring}, and is expected to meaningfully demonstrate this efficiency once fault-tolerant quantum hardware is available~\cite{Gidney2021howtofactorbit}. Thus, our notion of quantum advantage in practice should cast a wider net than formal proofs, but within a reasonable range guided by complexity-theoretic arguments. We explain this further by delving into the basics of complexity theory, discussing pertinent results and extending the discussion to more practical scenarios for optimization. We refer the interested reader to Refs.~\cite{arora2009computational, watrous2008quantum} for a more comprehensive overview of quantum and classical complexity theory.

\subsection{Exact Solutions}
\subsubsection{Decision and relational problems}\label{sec:basiccomplexity}
Much of complexity theory deals with so-called \emph{decision problems}. As the name suggests, a decision problem involves determining (i.e.~deciding) whether there exists a particular solution to the problem or not. Without loss of generality, it is often convenient to formalize this idea using binary strings~\cite{de2019quantum}. A computational decision problem on binary strings corresponds to what is called a language in complexity theory. A language $L \subseteq \{0,1\}^*$ is a set of binary strings of arbitrary length where the
corresponding decision problem is to determine whether a given input string $x \in \{0,1\}^*$ is an element of $L$ or not. For example, the language could consist of the set of all prime numbers encoded in binary form and the problem would involve determining whether $x$ is prime ($x \in L$) or $x$ is composite ($x \not\in L$). For completeness, an alphabet $\Sigma$, can be thought of as the set of characters from which the language is formed. In this case, $\Sigma = \{0,1\}$. It is also fairly common to think of decision problems on binary strings as binary functions $f_n: \{0,1\}^n \to \{0,1\}$ where $f_n(x) = 1$ if and only if $x \in L$, which is referred to as a ``yes'' instance. The famous Traveling Salesperson Problem (TSP) can be formulated in such a way, where a list of cities, the distances between them, and a total distance $d$ is described by a bitstring $x\in \{0,1\}^n$. Then $f_n(x) = 1$ if and only if there exists a salesperson's tour through the cities whose  length  is  $\leq d$. This problem becomes increasingly difficult as the number of cities grows, since, even though the input size $|x|=n$ grows polynomially, the number of possible tours grows exponentially; and the best known classical dynamic programming solution still needs exponential time~\cite{bellman1962dynamic,held1962dynamic}. In fact, the notion of resource scaling in terms of the input size is how hardness of a problem is formulated in complexity theory. 

At this juncture, it is worth noting that decision problems only care about the existence of a particular solution, and not the actual solution itself. Clearly, in practical optimization settings, finding an actual solution to a problem is often necessary. Continuing with TSP as an example, merely knowing that there exists a path to travel to all cities and back is not enough to benefit in practice, and determining the actual path to take is needed. This leads to the notion of \textit{relational problems}, which generalize decision problems in this sense. More concretely, given a relation $R \subseteq \{0,1\}^* \times \{0,1\}^*$ and an input $x$, the goal of a relational problem is to output any $y$ such that $(x,y) \in R$. So one is concerned about finding a $y$ that satisfies the relation (sometimes called a ``witness'' to the original decision problem), not just determining the existence of such a $y$. For the discussions that follow, it is useful to keep in mind that the same problem can be formulated in different ways. The decision problem already encompasses the time complexity of the problem. Namely, having a procedure that determines whether a TSP tour of length $\leq d$ exists, can be used to efficiently, modulo the complexity of the decision procedure, find a tour of minimal length $\leq d$ by iterating over the decision problem after removing edges~\cite{bjoerklund2015fast}. Looking at the decision version of a problem may be relevant from a complexity point of view, but perhaps not so much from a practical one. Rather, one may explicitly interpret a relational problem as an optimization problem, making these closer to problems observed in practice.

\subsubsection{Deterministic, randomized and quantum computation}\label{sec:det_vs_rand}
Decision problems that are \textit{efficiently} solvable by a deterministic machine are grouped together into the complexity class \textbf{P}, for polynomial time.
Efficient implies the existence of an algorithm which solves the decision problem with a run time that grows no faster than a polynomial in the size of the input. Crucially, the degree and constants of the polynomial are important for practical purposes. For example, an algorithm with a run time scaling like $n^3$ is far more practical than one that runs in $n^{100}$ time, even though both approaches scale polynomially and are therefore deemed computationally efficient in complexity theory. The set of optimization problems whose decision versions are then in \textbf{P} is referred to as \textbf{PO}~\cite{ausiello2012complexity}. Switching to relational problems leads to the class \textbf{FP} (for functional), which consists of all relations $R$ for which a deterministic polynomial-time algorithm outputs a solution $y$, whenever one exists. In this regard, \textbf{FP} generalizes \textbf{P}, as it is the analog of \textbf{P} for functions with an $n$-bit output, and importantly, the set \textbf{FP} contains \textbf{PO}. 
While \textbf{FP} focuses on general relational problems that require finding a specific output, \textbf{PO} is specifically about optimization problems where the best solution under certain criteria needs to be found.

\begin{figure}
    \centering
    \includegraphics[scale=0.65]{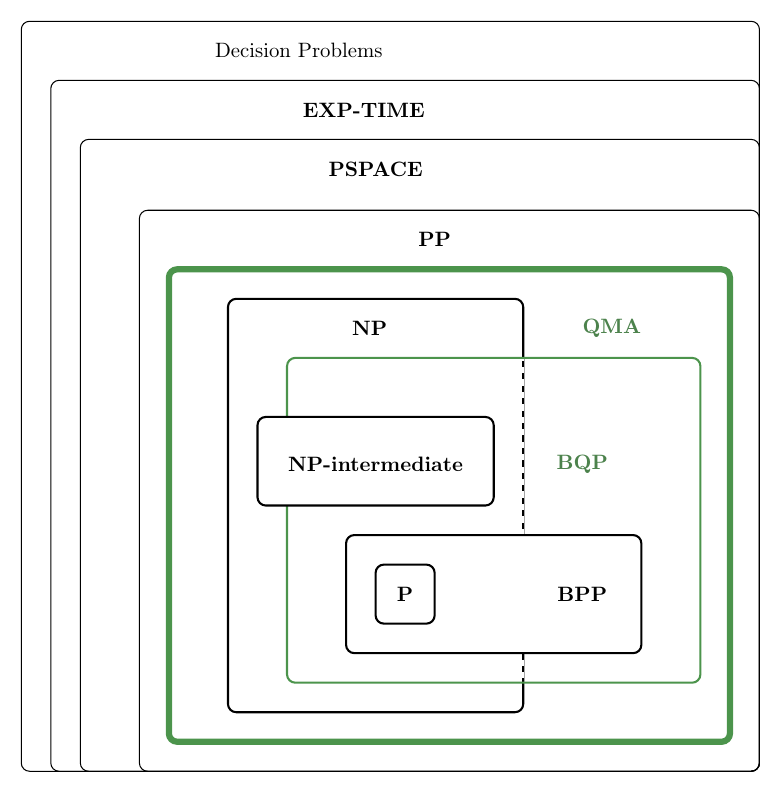}
    \caption{The inclusions of various complexity classes concerned with decision problems, as discussed in Sec.~\ref{sec:complexity}. While these inclusions are not known to be strict, this diagram captures the gist of prominent, canonical classes in complexity theory.}
    \label{fig:complexity_classes}
\end{figure}

Similar to \textbf{P}, \textbf{PSPACE} forms the class of decision problems  that can be solved with  a polynomial amount of space, but there is no restriction on the time used by the algorithm. Allowing for randomized computation brings us to the class \textbf{BPP}, which stands for bounded-error probabilistic polynomial time, where a classical probabilistic machine can solve all instances of a decision problem in polynomial time, with an error probability $\leq 1/3$. Generalizing further to a computational model that uses quantum mechanics, leads to \textbf{BQP}, the class of decision problems that quantum computers can solve in polynomial time with an error probability $\leq 1/3$ on every instance. Crucially, \textbf{BQP} contains problems that are not believed to be in \textbf{BPP} or \textbf{P}, like the decision version of factoring large integers for example, and it is widely  believed that  $\textbf{BQP}\neq \textbf{BPP}$~\cite{Bernstein1997}. Figure~\ref{fig:complexity_classes} illustrates the relationships between these classes. It is not known whether all inclusions are strict. However, there is still no proof that $\textbf{PSPACE} \not = \textbf{P}$. In the highly unlikely situation that $\textbf{PSPACE} = \textbf{P}$, it follows that $\textbf{BQP} = \textbf{P}$ and every efficient quantum algorithm can be simulated efficiently on a classical computer!

\begin{figure}[htb!]
\begin{tcolorbox}
\centering
\paragraph*{\textbf{Are quantum algorithms strictly more powerful than randomized algorithms?}}
{\color{blue}
The clear expectation is that the answer is affirmative, but in light of the difficulty of proving $\textbf{PSPACE} \not = \textbf{P}$, no proof to answer this question conclusively is known. A long line of research plausibly suggests that $\textbf{P} = \textbf{BPP}$. The reason for this is that good pseudo-random number generators are believed to exist and these in turn can be used to de-randomize \textbf{BPP} algorithms. 
There is also black-box evidence for this.
Bernstein and Vazirani~\cite{Bernstein1997} introduced the problem of Recursive Fourier Sampling (RFS) 
which is in \textbf{BQP} but not in \textbf{BPP} relative to an oracle. 
Interestingly, Yamakawa and Zhandry~\cite{yamakawa2022verifiable} show that, relative to a random oracle, there exist problems in \textbf{BQP} which are not in \textbf{BPP}. Arora \emph{et al.}~\cite{arora2023quantum} provide further results relative to a random oracle for shallow circuits and variational quantum algorithms, but it is not clear what practical implications, if any, these separations relative to random oracles mean. 
}
\end{tcolorbox}
\end{figure}

\subsubsection{Hard versus complete problems}
If \textit{useful} problems for optimization, outside of \textbf{P}, could be solved efficiently by quantum computers, such a finding would lead to huge breakthroughs in computer science, akin to that of Shor's algorithm. This highlights the importance of trying to design quantum algorithms for  classically hard optimization problems. To identify such problems, the class \textbf{NP} for nondeterministic polynomial-time, becomes relevant. A decision problem is in \textbf{NP} if, given an input $x$, we can efficiently check that $x$ is a ``yes'' instance of the problem. To do this, we are given a polynomial-size proof (sometimes called a witness) which certifies this fact. Intriguingly, there is also a relational problem extension of \textbf{NP}, namely \textbf{FNP}. As mentioned, problems in \textbf{NP} are concerned with the existence of a particular solution, such as a satisfying assignment. On the other hand, \textbf{FNP} versions of these problems are concerned with finding the actual value of a solution, if one exists, in keeping with practical requirements for optimization settings. In particular, the set of optimization problems within \textbf{NP} is termed \textbf{NPO}, and \textbf{NPO} $\subset$ \textbf{FNP}.

A problem is called \textbf{NP}-complete if all other problems in \textbf{NP} can be efficiently reduced to it, i.e., with a polynomial overhead in time. Hence, \textbf{NP}-complete problems are the most difficult problems in \textbf{NP}. Notable examples are TSP and Graph Colorability~\cite{Karp1972}. More broadly, a problem is called \textbf{NP}-hard if it is at least as hard as the hardest problems in \textbf{NP} --- but not necessarily in \textbf{NP} itself. Thus, \textbf{NP}-complete problems are merely a subset of \textbf{NP}-hard problems that belong to \textbf{NP}.

\begin{figure}[htb!]
\begin{tcolorbox}
\centering
\paragraph*{\textbf{Can quantum computers solve \textbf{NP}-hard problems?}}
{\color{blue}
It is generally believed that \textbf{NP}-hard problems are not in \textbf{BQP}. An intuition for this belief is given by the lower bound for search in the black-box~\cite{beals2001quantum,bennett1997strengths} setting. This proves that Grover's search algorithm~\cite{grover} is optimal in the black-box setting. The intuition is that this also carries over to the  Boolean satisfiability problem (SAT), which is the famous standard \textbf{NP}-complete problem. It is based on the feeling that a general instance of SAT  does not contain structure that can help finding a satisfying assignment just as in the black-box setting.  This would imply that  Grover's algorithm is also optimal for general instances of SAT.  Grover's algorithm can search over all $2^n$ possible assignments to an $n$-variable formula and obtains at most a quadratic speedup over classical unstructured search procedures. Since we do not know of either a quantum or a classical algorithm that can solve all SAT instances asymptotically faster than brute-force-search, it does not seem promising that quantum computers can efficiently solve \textbf{NP}-complete problems~\cite{bennett1997strengths, de2019quantum}. This does, however, advocate for quantum algorithms that exploit structure within certain problem instances that arise in practice, rather than attempting to try to develop fast quantum algorithms for all  instances~\cite{aaronson2022much}.
}
\end{tcolorbox}
\end{figure}

When thinking about problems in \textbf{NP}, it is tempting to imagine a quantum analog of this class, with complete problems of its own.  There are several possible ways to do this, cf.~\cite{gharibian20237} for a discussion.  We hone in on one in particular --- \textbf{QMA}, for quantum Merlin-Arthur. Intuitively, \textbf{QMA} encapsulates decision problems where ``yes'' instances can be verified efficiently and with high probability on a quantum computer, using a quantum state as a proof. This state should occupy no more than a polynomial number of qubits in the input size, however, finding such a state may of course, be computationally hard. For ``no'' instances all quantum proofs will be rejected (with high probability). Problems may then similarly be classified as \textbf{QMA}-hard or \textbf{QMA}-complete (see ~\footnote{There is a slight technical complication here, because we deal with a class that has a promise: either there exists a quantum proof that makes the quantum verifier accept with high probability or for every proof the verifier rejects with high probability. In the strict sense such a promise class is not known to contain complete problems. In order to still use the notion of completeness a framework of promise problems is introduced~\cite{goldreich2005promise} for additional technical details}).

Other classes relevant to quantum computing that are more exotic than \textbf{NP} and \textbf{QMA} exist, albeit not necessarily in the context of optimization. Due to a recent Gaussian Boson Sampling experiment~\cite{Madsen2022} the class \#\textbf{P} is worth mentioning. \#\textbf{P} is a counting class, defined as the set of all functions $f$ such that for an input $x \in \{ 0,1 \}^*$, $f(x)$ counts the number of accepting computation paths of a nondeterministic polynomial-time machine on input $x$. The complexity of problems in $\# \mathbf{P}$ relates to challenges like computing partition functions in statistical mechanics. The class \textbf{PP} may be thought of as the decision analogue of $\#\mathbf{P}$ and it is known that $\textbf{QMA} \subseteq \textbf{PP}$~\cite{Marriott2005}. 

\subsubsection{Polynomial Time Hierarchy}
A useful way to extend the complexity classes like $\C{P}$ and $\C{NP}$ is to give those classes access to an oracle/set $A$. This way one can define for example $\C{P}^A$, the class of languages decided by polynomial time Turing machines that have oracle access to $A$. This means that the machines have access to a subroutine that computes in a single step whether a query string $x \in A$. For a rigorous treatment of oracle Turing machines see for example~\cite{Arora2009}. One can also define oracle access to a whole class of problems: $\C{P}^{\C{NP}}$ and $\C{NP}^{\C{NP}}$. The $k^{th}$ level of the polynomial hierarchy $\bf{\Sigma^p_k}$  is $\NP$ with $k-1$ stacks of $\NP$ oracles. The third level for example is $\NP^{\NP^\NP}$. The polynomial hierarchy $\C{PH}$ is then defined as $\bf{\bigcup_k \Sigma_k^p}$. It is known that $\BPP \subset \NP^\NP$ but remains an open problem whether $\C{BQP}$ is in $\C{PH}$, and interestingly, there are oracles relative to whom this is not the case~\cite{raz2022oracle}. Another relevant computational model is presented by Chen \emph{et al.}~\cite{chen2023complexity}. The authors define the complexity class \textbf{NISQ} as the set of decision problems solvable by a \textbf{BPP} machine with has access to a noisy quantum device with $\text{poly}(n)$ qubits. They then prove relative separations with respect to a particular oracle $O$, such that $\mathbf{BPP}^O \subsetneq \mathbf{NISQ} \subsetneq \mathbf{BQP}^O$, which gives evidence that the power of \textbf{NISQ} lies strictly between \textbf{BPP} and \textbf{BQP}.

It is unlikely, that quantum computers provide exponential speedups for \textbf{NP}-hard (and thus, \textbf{NPO}-hard) problems. See~\cite{buhrman2021framework} for stronger hypotheses along these lines. But exponential speedups for an interesting subclass of \textbf{NP} problems, not complete but also not in $\C{P}$, called \textbf{NP}-intermediate, cannot be ruled out. For example, decision versions of factoring are believed to be of this type. As depicted in Figure~\ref{fig:complexity_classes}, these problems are outside of \textbf{P} and within \textbf{NP}, but are not \textbf{NP}-complete. Related  classes can similarly be studied for the optimization classes $\PO$ and $\NPO$, and relational classes $\FP$ and $\FNP$.  For instance, the relational version of TSP is $\FP^\NP$-complete, whereas the decision version is $\NP$-complete~\cite{cook2011traveling}. 
MAX-SAT is $\NPO$-complete, whereas the Vertex-Cover and Bin-Packing problems are not (even though their decision versions are $\NP$-complete)~\cite{10.5555/574848}.

\subsubsection{Worst-case complexity and fixed-parameter tractability}
It is important to emphasize the following point: 
\begin{center}
    \textit{Complexity theory is traditionally focused on worst-case problem instances. In practice, one often needs to solve a ``typical'' instance which may be much easier than the hardest, worst-case instance.}
\end{center}
This implies that, even when worst-case complexity-theoretic results rule out  quantum advantage for \textbf{NP}-hard problems,  
quantum computers may still exhibit exponential speedups on instances relevant for practical problems. More concretely, they may give efficient algorithms for instances that take classical computers exponential time. We discuss this in more detail in Sec.~\ref{sec:heuristics}.

On a similar note, while many problems of practical interest are \textbf{NP}-hard, if we are interested in measuring the computational complexity of problem which admits an algorithm $\mathcal{A}$, as a function of the input length $n$ in conjunction with some adjustable parameter $\epsilon$, then interestingly, parameterization could render otherwise \textbf{NP}-hard instances efficiently solvable --- although the dependence could be exponential with respect to $1/\epsilon$. Such a problem is termed fixed-parameter tractable (\textbf{FPT}) and the complexity is sometimes referred to as \emph{parameterized complexity}~\cite{Downey1999, cai1997fixed}. Therefore, $\textbf{FPT}$ amounts to the parameterized analogue of $\textbf{P}$. A common example is that of the Vertex-Cover problem, which is {\textbf{NP}}-complete when considering the input size in isolation. However, there exists an $\mathcal{O}(n)$
algorithm when the problem is restricted to constant-size covers~\cite{Chen2010}. Bremner \emph{et al.}~\cite{bremner2022quantum} introduced quantum parameterized complexity. In particular, the class \textbf{FPQT} is introduced as the class of parameterized decision problems tractable by a quantum computer, but intractable otherwise. It is well known that if a parameterized problem is in \textbf{FPT}, it admits kernelization, which could be seen as a pre-processing algorithm that reduces the size of an instance of the problem, often substantially~\cite[Theorem 1.4]{fomin2019kernelization}. For benchmarking purposes in optimization, cf.~Sec.~\ref{sec:benchmarks}, the notion of kernelization is of crucial importance. Formally, a kernelization (also known as kernelization algorithm) for a parameterized problem can be interpreted as follows: assume a language $L$ over an alphabet $\Sigma$ and parameters in $\mathbb{N}$, an algorithm given $(x, k)\in \Sigma^* \times \mathbb{N}$ outputs a string $x' \in \Sigma^*$ and an integer parameter $k' < k$, such that $(x',k') \in L$ --- if the algorithm runs in polynomial time and $k$ can be expressed as a polynomial of $k'$, the tuple $(x',k')$ is termed a polynomial kernel. Such polynomial kernels can be obtained for many combinatorial optimization problems considered in theoretical computer science~\cite{KernelizationMeta}. For \textbf{FPT} problems, it thus makes sense to benchmark on the kernel. 

\begin{figure}
    \centering
    \includegraphics[scale=0.7]{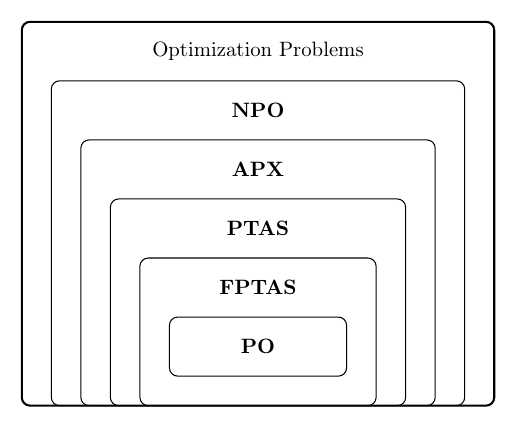}
    \caption{Complexity classes for optimization problems. Notice that, similar to $\sharpP$, the functional problem classes are outside of the decision problems, and hence, not displayed in this hierarchy.}
    \label{fig:optimization_classes}
\end{figure}

\subsection{Approximate Solutions}
Thankfully, the story does not end at obtaining \textit{exact} solutions efficiently, with high probability --- if it did, we would probably have a difficult time motivating classical computing, let alone quantum, as a useful paradigm in (combinatorial) optimization. In reality, finding a solution that is in some sense, close to the optimal is often sufficient. The shift from finding the optimal solution, to an approximate one, leads to the study of approximability --- in particular, how hard is it to approximate solutions to various problems? This leads to interesting complexity classes for optimization, induced by approximability.
Intuitively, it may seem like allowing for approximate solutions could reduce the complexity of a problem, similar to those in \textbf{FPT}. Unfortunately, it is not that straightforward when dealing with approximation and it seems that in many cases, approximating solutions to hard problems is just as hard as finding the optimal ones. We discuss this so-called hardness of approximation next, followed by classes of problems that can exhibit efficient approximation.

\subsubsection{Hardness of approximation}
In the context of approximate solutions for optimization problems, a longstanding result called the \textbf{PCP}-theorem informally states that, for many problems, computing an approximate solution with arbitrary precision is as hard as computing the exact solution. To discuss this a bit more formally, the approximation ratio, which --- as the name suggests --- is the ratio between the (expected) approximate solution and the optimal, quantifies the quality of a solution provided by an approximation algorithm. For many \textbf{NP}-hard problems, there are known bounds on the achievable approximation ratios, and these are referred to as ``inapproximability bounds''. Such bounds can be derived through many techniques, including information-theoretic arguments (e.g., any algorithm that achieves an approximation factor better than $c$ must make at least exponential queries to a function), or reductions to known \textbf{NP}-hard problems (e.g., if a problem can be approximated better than its inapproximability bound, then it is equivalent to solving another \textbf{NP}-hard problem). The former holds independent of whether \textbf{P} is equal to \textbf{NP}, but the latter inapproximability bounds are said to be conditional on $\mathbf{P}\neq \mathbf{NP}$.

For the MAX-E3-SAT problem, one of the earliest approximation algorithms showed that a random variable assignment satisfies each clause with probability $7/8$, and thus, achieves a $7/8$-approximation ratio ~\cite{johnson1974max3sat}. It was later shown that there is no classical polynomial-time algorithm that can achieve a ratio of $(7/8+\epsilon)$, for any $\epsilon>0$ unless \textbf{P}$=$\textbf{NP}~\cite{hastad2001inapproximability}. In such cases, where inapproximability bounds are achieved by classical algorithms, improving these ratios with quantum computers is highly unlikely, as this would imply that quantum computers can solve \textbf{NP}-hard problems.

Some inapproximability bounds require additional conditions. For example, in their seminal work in 1995, Goemans and Williamson introduced an approximation algorithm for the MAXCUT problem and proved that it achieves an approximation ratio of $c \approx 0.87856$~\cite{Goemans1995}. Later, it was shown by Khot \emph{et al.}~that there cannot exist a polynomial time algorithm that achieves $(c+\epsilon)$-approximation~\cite{khot2007inapproximability} under the assumption of the Unique Games Conjecture~\cite{khot2002uniquegames}. Again, in such cases, improving these ratios with quantum computers would imply that quantum computers can solve \textbf{NP}-hard problems \emph{or} that the Unique Games Conjecture is false. Similar results exist for other \textbf{NP}-hard problems, such as graph coloring problem and set covering~\cite{Lund1994}. 

There is also a quantum \textbf{PCP} conjecture proposed by Aharonov \emph{et al.}~\cite{aharonov2008detectability,aharonov2002quantum,aharonov2013guest}. At its core is the $k$-local Hamiltonian problem: given a Hamiltonian $H$ with $k$-local terms (i.e. operations acting on up to $k$ qubits), determine the smallest eigenvalue of $H$. This problem is \textbf{QMA}-complete~\cite{kitaev2002classical}, akin to SAT's \textbf{NP}-completeness. 
The local Hamiltonian problem comes with a promise denoted by $\gamma$. The ground state energy of $H$ is either less than some value $a$, or greater than $a+b$, where $\Gamma \coloneqq b$ is the absolute promise and the ratio $\gamma\coloneqq b/m$, where $m$ is the number of $k$-local terms, is the relative promise. The quantum \textbf{PCP} theorem then states that the $k$-local Hamiltonian problem is \textbf{QMA}-hard even with a constant relative promise gap $\gamma >0$, if the interaction graph is sparse and all $k$-qubit interactions have bounded norm. There is some supporting evidence for the quantum \textbf{PCP} theorem, perhaps most notably the recently proved No Low-Energy Trivial State (NLTS) theorem~\cite{nlts}. Under NLTS, there exist families of Hamiltonians for which all low energy states require circuits of super-constant depth to prepare. If NLTS were false, this would imply that all Hamiltonians have low energy states that are efficiently simulable --- effectively rendering the quantum \textbf{PCP} theorem invalid since approximating the ground state energy could be done efficiently using these low energy states. While the quantum \textbf{PCP} theorem implies NLTS, under the assumption that \textbf{QMA} is not equal to \textbf{NP}, the other direction does not hold, since there could be other types of classically simulable circuits that could prepare low energy states, other than low depth circuits. This remains a very open and interesting research area.

Indeed, the hardness of approximation seems to imply -- theoretically -- that quantum computers will not be very influential, even if approximate solutions are the goal. However, this is not necessarily true in general, and less so in practice. There is hope for quantum algorithms to outperform classical methods when there is a gap between known inapproximability bounds and provable approximation factors (e.g., in metric TSP, \cite{karpinski2015inapproximabilityTSP,karlin2021approximationTSP}). Further, quantum optimization methods might provide provable and computational speedups, even if they simply match the approximation factors of classical algorithms.

\subsubsection{Polynomial-time approximation schemes}\label{sec:approximationtheory}
If a problem is ``approximable'', it is said to belong to the class \textbf{APX} which contains the set of \textbf{NPO} problems for which there are polynomial-time approximation algorithms with approximation ratios bounded above by some constant $c$~\cite{lee2021classifying}. An interesting subclass of \textbf{NPO} problems that admit a so-called polynomial-time approximation scheme is \textbf{PTAS}~\cite{ausiello2012complexity} \cite[Chapter 8]{gonzalez2018handbook}. For any problem in \textbf{PTAS}, there is a polynomial-time algorithm that is guaranteed to find a solution whose value is within a $1+\epsilon$ factor of the optimum, where $\epsilon > 0$. For example, by restricting to the Euclidean plane, TSP is then in \textbf{PTAS}~\cite{arora1998polynomial}. However, as with similar issues in \textbf{FPT}, there are cases where the exponent of the polynomial might depend exponentially on $1/\epsilon$. To account for this, one may consider the class \textbf{FPTAS} for fully polynomial-time approximation scheme, which demands the run time to be polynomial in the input size, as well as in $1/\epsilon$. As such, $\mathbf{FPTAS} \subseteq \mathbf{PTAS}$, and the hierarchy of these classes are displayed in Figure~\ref{fig:optimization_classes}. When combined with quantum computing, this area makes for an exciting and rather unexplored research direction --- where quantum algorithms could serve as new approaches or improve over existing classical \textbf{PTAS} algorithms.

\subsection{Heuristics in Quantum Optimization}\label{sec:heuristics}
Thus far, the focus has been on worst-case instances. In practice, heuristic approaches can provide useful solutions to special, relevant and typical problem instances~\cite[e.g.~Chapter 28]{gonzalez2018handbook}. Heuristics have also proven useful in problems exhibiting structure and average-case instances --- which are oftentimes far more practical in real-world scenarios~\cite{spielman2009smoothed,gonzalez2018handbook}. The success of classical heuristic algorithms indeed motivates the study of quantum heuristic methods, where the aim of the latter is to design algorithms that leverage quantum computation to provide reasonable solutions to problems where classical approaches either cannot, or simply take too long to do so. Naturally, the word \emph{reasonable} depends very much on the context, making heuristic approaches difficult to analyze theoretically. More concretely, heuristic algorithms are understood as algorithms without provable performance guarantees, run time guarantees, or both (as in the case of so-called metaheuristic algorithms). By this negative definition, our understanding of heuristic algorithms is rather limited. Having said that, there are multiple attempts to provide more formal definitions. For instance Bogdanov \emph{et al.}~\cite{bogdanov2006average} define distributional problems, which are pairs $(L,D)$ of a language $L$ and a family of distributions $D = \{D_n \}$, one for each input size $n$. Then, a distributional problem $(L, D)$ is in the class $\HeurBPP$ (for heuristic \textbf{BPP}) if there exists a polynomial-time randomized classical algorithm $\mathcal{A}$ such that for all $n$ and $\delta>0$, $ \textup{Pr}_{x\sim D_n} \left[  \textup{Pr}\left[\mathcal{A}(x, 0^{1/\delta}) = L(x)\right] \geq 2/3  \right] \geq  1 - \delta $  where the inner probability is taken over the internal randomization of $\mathcal{A}$. In other words, algorithms that solve problems in $\HeurBPP$ are allowed to err on a small fraction of instances drawn from $D_n$. Analogously, we say that a distributional problem $(L, D)$ is in $\HeurBQP$ if there exists a polynomial-time quantum algorithm $\mathcal{A}$ that satisfies the same property.
This ``distributional'' view of heuristics has been rather popular in relation to quantum advantage. 

Notably, Pirnay \emph{et al.}~\cite{PirnayCO2022} prove a separation for a combinatorial optimization problem under very special distributions in the sense above, namely those which are Karp-reduction-images of factoring problems as pointed out by Szegedy~\cite{szegedy2022quantum}. The former proves that quantum computers can have a super-polynomial advantage over a certain class of NPO-problems and that this also holds in the approximation sense. It does so by making use of cryptographic tools and by means of notions of computational learning theory, in particular the Occam learning framework. The latter points out that similar separations hold for other NPO approximation problems, by resorting to the PCP theorem. The approach is via faithful reductions from well-established problems in \textbf{NP}, which have a large quantum advantage, namely factoring, to matching instances of $\NPO$ problems. Using techniques from complexity theory, it is possible to construct instances with a large approximation gap, which ensure that classical algorithms cannot approximate the correct solutions to these instances, unless factoring is classically easy. These findings, however, are unlikely to have much practical value as the proposed quantum solutions will only work on the contrived instances corresponding to factoring, and are not general optimization solvers which could solve other classes of instances efficiently as well. The results do, at least, contribute to understanding what kind of quantum advantages one can hope for in combinatorial optimization.

Independently, one could relate the distributional view of heuristics to numerous studies of so-called landscapes~\cite{ball2017average}.  
The term landscape here aims to generalize an intuitive mental picture where one associates the height of terrain to solution quality and traversing this terrain corresponds to moving through a space of solutions (which need not be two dimensional in general). 
For example, Refs.~\cite{zhang2022escaping,fontana2023adjoint,ragone2023unified} study the landscape associated with variational quantum algorithms for specific problems, 
with the aim of explaining when these heuristic methods converge.
An important feature in the study of landscapes are so-called \emph{barren plateaus}~\cite{Clean_2018_BarrenPlateaus, Cerezo_2021_costfunct, Cerezo_2021higher_order_bps, holmes2021AnsatzExpressBarrenPlateaus, Wiebe2020Barren, Wang_21_noiseinducedBPs, napp2022quantifying, Uvarov_2021BPs},
especially in conjunction with variational quantum algorithms,  
which correspond to loss landscapes that are exponentially flat in the number of system qubits. In fact, this phenomenon provides a central bottleneck in the scaling of variational quantum algorithms to practically relevant problem sizes.
There are multiple reasons for the appearance of barren plateaus. It was conjectured that they are induced by an ansatz that is (close to) a t-design~\cite{Clean_2018_BarrenPlateaus, holmes2021AnsatzExpressBarrenPlateaus} or exhibits too much entanglement paired with partial measurements~\cite{Wiebe2020Barren}. These findings, along with other causes of barren plateaus were united in theorems which either relate the occurrence to a mathematical object called the dynamical Lie algebra, which is induced by the structure of the generators and the form of the variational circuit~\cite{fontana2023adjoint, ragone2023unified}, or to the average behavior of the respectively induced light-cone \cite{letcher2023tight}.
Notably, these phenomena may be avoided if a good warm-starting strategy is known--in the sense of a parameterization that brings the system sufficiently close to the optimum. However, it is often unclear how such a warm-start should be chosen.

\begin{table*}[t!]
\caption{ An overview of the classical problem classes relevant to optimization problems. The models of analogue computation below the horizontal line are not considered in much detail in this section.}
\label{tab:optclassical}
\begin{tabularx}{\textwidth}{@{\extracolsep{\fill}}lllll}
\hline
Complexity class & Example problem & Key features & Year & Ref. \\ \hline
$\FPTAS$ & Knapsack, Subset-Sum & Efficiently approximable &      & \cite{Papadimitriou1991} \\
$\PTAS$ & Euclidean TSP & Approximable in poly-time &   & \cite{arora1998polynomial} \\
$\APX$-Complete & MAXCUT, MAX-3-SAT, Metric-TSP & Discrete decisions & 1991 & \cite{Papadimitriou1991,ausiello2012complexity} \\
$\NPO$-Complete & TSP, Binary IP & Discrete decisions &  1988& \cite{KRENTEL1988490}  \\
$\sharpP$     & Two-stage SP~\cite[Theorem 3.1]{dyer2006computational} & Stochastic objective function & 2006 & \cite{valiant1979complexity,dyer2006computational}\\
$\PSPACE$     & Multi-stage SP~\cite[Theorem 4.1]{dyer2006computational} & Stochastic objective function & 1985 & \cite{PAPADIMITRIOU1985288,dyer2006computational} \\
\hline 
Banach-Mazur computability   & Optimal Transport~\cite[Theorem 5]{lee2023computability} & Real-valued decisions & 2003 & \cite{lee2023computability}  \\  
$\NP$ over Reals   & Optimal Transport~\cite[Theorem 5]{lee2023computability} & Real-valued decisions &  1996 & \cite{blum1996complexity} \\ 
\hline
\end{tabularx}
\end{table*}

There have also been a number of works suggesting that quantum sampling could be a useful subroutine in a larger classical algorithm.
For example, there are quantum circuits for Metropolis-Hastings algorithms~\cite{Lemieux2020efficientquantum} and quantum enhancement of Markov-chain Monte-Carlo (MCMC) algorithms~\cite{layden2023quantum,intallura2023survey}.
These approaches should be possible to analyze, and thus, turn into randomized approximation schemes.

\subsection{Takeaways from Complexity Theory in Quantum Optimization}
Although it is widely believed that quantum computers will not offer exponential speedups for $\NP$-hard problems, the story does not end there. As we have seen, most complexity-theoretic statements deal with worst-case settings for exact solutions. In practice, typical optimization problem instances could differ substantially from the worst-case, and thus, exponential quantum speedups in these more realistic settings are still possible in principle. Importantly, the picture may drastically change when dealing with average-case hardness. Average-case settings could exclude particularly bad instances for which most complexity-theoretic statements hold, and thus, an advantage -- certainly one faster than simply the Grover approach -- becomes possible and it is an interesting open problem whether quantum algorithms can solve optimization problems that are average-case hard for classical algorithms. For example, there is evidence from~\cite{boulebnane2022solving} that quantum algorithms can solve random k-SAT faster than Grover and the state-of-the-art SAT solvers. Additionally, problems with a so-called Overlap Gap Property (OGP)~\cite{gamarnik2021overlap} have been proven to be average-case hard for certain families of classical algorithms (and local low-depth quantum algorithms), but it remains unclear whether these problems can be addressed by higher-depth and/or more sophisticated quantum algorithms~\cite{gamarnik2020hardness,huang2022tight,basso2022performance,chen2023local}. If so, this would be very convincing evidence for useful superpolynomial quantum advantage, since such problems naturally arise in combinatorial optimization, statistical physics and high-dimensional statistics. Beyond OGP in the average-case setting, there is an emerging literature on the ``computational-statistical gaps'' of many optimization and learning problems where we could also hope for a superpolynomial quantum advantage. Moving into the approximation realm also introduces interesting open questions for quantum computing. For example, can quantum algorithms saturate inapproximability bounds that have not yet been reached by classical methods? Moreover, if approximation ratios cannot be improved, there may still be room to speedup existing approaches with quantum computing. 

Another intriguing aspect of complexity theory, is that a restriction of some parameters of a problem, or some structure present, may drastically change the complexity. A perfect example of this is TSP, which -- in its most general decision version form -- is \textbf{NP}-hard and the optimization version is $\APX$-hard. If we restrict to Euclidean TSP, where all points are in $\mathbb{R}^d$ and edge weights equal their Euclidean distances, then, remarkably, there exists an explicit $\PTAS$ to solve such TSP instances~\cite{arora1998polynomial}. Shifting to heuristic algorithms for TSP, one can actually solve problems in practice with up to millions of variables. Lastly, if we were to consider the shortest path problem instead of TSP, where we just want the shortest path between two places, we would have a problem in $\P$. This illustrates the nuance of complexity theory that one should keep in mind when designing quantum optimization algorithms. Akin to TSP classically, if provable quantum speedups seem unattainable, quantum heuristic algorithms may still offer meaningful results. But developing quantum heuristic methods that exploit structure in a problem, or enhance classical approaches, is not straightforward either. This serves as the basis for the next section, where we lay out various building blocks to carefully design quantum optimization algorithms.

\section{Paradigms in the Design of Quantum Optimization Algorithms 
\label{sec:paradigms}}

There are multiple approaches that lead to exact quantum optimization algorithms, such as Grover (Adaptive) Search~\cite{durr1996quantum,Bulger2003,Baritompa2005} and  the Quantum Adiabatic Algorithm (QAA). There are lesser known approaches too, like Quantum Imaginary Time Evolution (QITE)~\cite{motta2020determining}, but much of the published work focuses on the more well-known approaches. In this section, we introduce the core concepts deployed as subroutines in various quantum optimization algorithms discussed in Sec.~\ref{sec:problem_classes}.

\subsection{Grover Search}
The seminal Grover algorithm for searching an unstructured database achieves a quadratic advantage over classical search in terms of the number of queries to the database (i.e. the query complexity).
In exact quantum optimization, this advantage translates to an advantage for function minimization~\cite{durr1996quantum} and can achieve a quadratic speedup over brute force search in many problems in combinatorial optimization \cite{Gilliam_2021_grover}, i.e., a run time of $\mathcal{O}(\sqrt{2^n})$ instead of $\mathcal{O}(2^n)$ for $n$-variable problems. In the most general case, without any further assumptions on the problem, for example in terms of oracle used in the query complexity, this is the best run time and advantage that we can hope for when using a quantum computer for discrete optimization~\cite{Bennett1997strenghts}. 
For a concrete problem class, however, brute force search is rarely the best available classical algorithm, and thus, the actual advantage achieved by Grover Search is usually sub-quadratic~\cite{cade2023quantifying}. Nevertheless, leveraging Grover Search as a sub-routine, together with other (quantum or classical) algorithms, can sometimes help to recover an overall quadratic speedup over the best available exact classical algorithm~\cite{shaydulin2023evidence}.

\subsection{Quantum Adiabatic Algorithm}\label{sec:algorithms_qaa}
The Quantum Adiabatic Algorithm (QAA)~\cite{kadowaki1998quantum,farhi2000quantumAdiabatic} is another exact optimization algorithm, reminiscent of homotopy methods of numerical mathematics \cite{sommese2005numerical}. It assumes a problem Hamiltonian $H$ whose ground state corresponds to the solution of a corresponding optimization problem, as well as a so-called mixing Hamiltonian $H_X$. The mixing Hamiltonian should have an easy-to-prepare ground state $\ket{\psi_0}$ that has a non-zero overlap with the ground state of $H$. A common choice is $H_X = \sum_{i=1}^n \sigma_X^{(i)}$, where $\sigma_X^{(i)}$ denotes the Pauli $X$ matrix on qubit $i$, with the ground state $\ket{+}^{\otimes n}$, i.e., the $n$-qubit equal superposition state that has non-zero overlap with every computational basis state. Starting in $\ket{\psi_0}$, QAA slowly drives the state according to $H(t) = \lambda(t/T) H + (1 - \lambda(t/T)) H_X$, and $d/dt \ket{\psi_t} = -i H(t) \ket{\psi_t}$, for some total annealing time $T > 0$ and an annealing schedule $\lambda: [0, 1] \to [0, 1]$ with $\lambda(0) = 0$ and $\lambda(1) = 1$. If the annealing is performed slowly enough, in other words, if $T$ is large enough, with a suitable annealing schedule, the Adiabatic Theorem \cite{born1928adiabatic,jansen2007bounds,lidar2009adiabatic,cheung2011improved} guarantees that the state $\ket{\psi_t}$ always remains in the ground state of $H(t)$. Thus, $\ket{\psi_T}$ represents the ground state of $H$, which corresponds to the solution of the encoded problem. It has been proven that the annealing time has to scale inversely with the minimal squared spectral gap $\Delta$ of $H(t)$ over all $t \in [0, T]$, i.e., $T = \mathcal{O}(1/\Delta^2)$.
During annealing, the system undergoes quantum phase transitions, and in the glassy phase, at low values of $t/T$, the energy gap closes exponentially with increasing system size \cite{FINNILA1994343,bapst2013quantum,knysh2016zero,isakov2016understanding, mohseni2023deep}. Consequently, this implies exponentially long run times if one aims to precisely follow the adiabatic path. Hence, we are limited to at most polynomial speedups.

The practical implementation of the quantum annealing protocol often involves repeated finite-time runs \cite{ronnow2014defining} or, potentially, diabatic annealing \cite{crosson2021prospects}. This variant relaxes the constraint that the system must always remain in the instantaneous ground state of $H(t)$.
Another variant is counter-diabatic annealing, where additional terms are included in the Hamiltonian to suppress the transition to excited states during the annealing~\cite{Demirplak2003, Campo2013}.
For example, the counter-diabatic terms can be built up from nested commutators~\cite{Claeys2019}, while the evolution can be engineered with optimal control~\cite{Chasseur2015}.
In its practical implementation, QAA often becomes a heuristic algorithm.
On a gate-based quantum computer, the time evolution to generate $\ket{\psi_t}$ 
can be implemented, e.g., through Trotterization \cite{trotter1959product,Suzuki1990trotter,suzuki1991general}. The typical long annealing times then translate to long quantum circuits.

\subsection{Quantum Phase Estimation}
Quantum Phase Estimation (QPE) is an algorithm to estimate the eigenphases of unitary operators \cite{nielsen2000}. It allows the estimation of the eigenvalues of certain matrices and can be used to find the ground state and corresponding energy of various Hamiltonians~\cite{Abrams1999,Aspuru2005,lee2208there}. 
Famously, QPE is also leveraged by Shor's algorithm.
Factoring is believed to be in \textbf{NP}-intermediate \cite{10.5555/574848} and Shor's algorithm \cite{shor} achieves an exponential speedup over classical algorithms, as discussed in Sec.~\ref{sec:complexity}.
Like many other problems, factoring can be cast into a quadratic unconstrained binary optimization problem (QUBO)~\cite{Jiang2018,phan2022quantum, Jun2023}.
Thus, this gives us an exponential speedup over a subset of QUBOs that happen to correspond to instances of factoring.
Similar reductions are possible from other classes of problems to factoring~\cite{pirnay2023inprinciple, szegedy2022quantum}.
In practice, however, we do not expect these instances to appear naturally, and it is an open question how to harness this relation in the context of optimization.

In general, QPE requires an initial state having a polynomial overlap with the ground state, i.e. an overlap greater than $1/\text{poly}(n)$. For a diagonal Hamiltonian, every computational basis state is an eigenstate, and so QPE does not increase the chance of sampling the ground state. Further, simply sampling from such an initial state provides a polynomial-time algorithm to find the ground state already. Thus, the true problem is finding such an initial state with a polynomial overlap. Nonetheless, some algorithms encode an optimization problem into a non-diagonal Hamiltonian, potentially making QPE useful in quantum optimization. The situation is similar for (Krylov) subspace expansions \cite{Suchsland_2021, Kirby_2023}, where an initial polynomial overlap is assumed to derive convergence properties towards the ground state of a Hamiltonian.

To make QPE and related techniques relevant for a large set of applications, it is important to have a quantum toolbox of matrix linear algebra operations  \cite{Lloyd1996universal,Childs2002qw,Berry2007sparse,Harrow2009linear,Berry2014trunc,childs2017quantum}, in analogy to the classical basic linear algebra subprograms (BLAS) \cite{Lawson1979BLAS}. 
Quantum signal processing~\cite{Low2017Optimal} enables the simulation of sparse matrices, i.e., the implementation of the time evolution operator $e^{-i H t}$ for a matrix $H$ and a time $t$, with optimal complexity. 
Phase estimation then allows one to extract eigenvalues and eigenvectors of $H$. 
Quantum singular value transformation \cite{gilyen2019quantum} generalizes this result and allows efficient simulation of matrix algebra such as $A+B$, $AB$, and singular value transformations $P(A)$, with $P$ being a low-degree polynomial applied to the singular values of $A$, where $A$ and $B$ are matrices. The block-encoding framework unifies input models for matrices given efficient descriptions, sparse access, or quantum memory.
These tools find applications in preparing Gibbs states, matrix problems, and continuous optimization.

\subsection{Gibbs Sampling}
Gibbs sampling corresponds to generating sample sequences from the joint probability distribution of multiple variables underlying a Gibbs distribution.
Geman and Geman~\cite{Geman1984-yy} introduced a Markov Chain Monte Carlo (MCMC) algorithm to facilitate Gibbs sampling. Its significance in optimization, particularly in classical MCMC algorithms, extends to solving problems like constraint satisfaction \cite{krzakala2007gibbs}.

Sampling from a classical many-body Hamiltonians' $H$ Gibbs distributions, and the related problem of approximating the partition function $\mathcal{Z}(\beta)=\operatorname{Tr}\left(e^{-\beta H}\right)$ or the (Gibbs) free energy of the system, $\log\mathcal{Z}$, falls within the complexity class $\textbf{BPP}^\textbf{NP}$ \cite{Stockmeyer1983,zhang2023dissipative}. Here, $\beta$ is the inverse temperature. Gibbs sampling, then, can be done with the Metropolis-Hastings algorithm  \cite{Metropolis1953, Hastings1970} which, despite its exponential scaling with particle number, often exhibits convergence.

In contrast, quantum Gibbs sampling \cite{poulin2009sampling} presents greater challenges and is believed to be at least \textbf{QMA}-hard \cite{aharonov2013quantum, bravyi2021complexity}. 
The quantum Gibbs state $\boldsymbol{\sigma}_\beta$ is (up to normalization) defined as
\begin{eqnarray}
\boldsymbol{\sigma}_\beta \propto {e}^{-\beta {H}}=\sum_{i=1}^{{\rm dim} H} \mathrm{e}^{-\beta E_i}\left|\psi_i\right\rangle\left\langle\psi_i\right|,
\end{eqnarray}
where $H$ is the Hamiltonian operator, $\ket{\psi_i}$ are the eigenstates, as before $\beta$ is the inverse temperature, and $E_i$ are the energy eigenvalues.  

The problem of thermalizing a quantum state has a long tradition and was studied earlier in \cite{terhal2000problem}. Quantumly, the difficulty does not only relate to the hardness of sampling but also to the difficulty to efficiently prepare Gibbs states. Quantum adaptations of the Metropolis-Hastings algorithm that involve quantum phase estimation (QPE) are challenging due to the algorithmic complexity\cite{Temme2011, Yung_2012}.
Furthermore, Gibbs sampling algorithms for Gibbs states of local Hamiltonians at all temperatures have been devised in~\cite{rouze2024efficient}. It
has also been seen that high-temperature Gibbs states are unentangled and efficiently preparable~\cite{bakshi2024hightemperature}.

A variational method for quantum simulators to generate finite temperature Gibbs states through the preparation of thermofield double states was introduced in \cite{PhysRevLett.123.220502}. A quantum algorithm to prepare the quantum Gibbs state, as well as the partition function within $\epsilon$-error, was proposed in \cite{poulin2009sampling}. A different VQA approach for the preparation of Gibbs states suitable for NISQ devices was presented in \cite{PhysRevApplied.16.054035} where the authors presented an algorithm in their paper by applying a truncated Taylor series to compute the free energy, and then selecting this truncated value as the loss function. Another VQA was proposed in \cite{consiglio2023variational} where the variational parameters are chosen by minimizing the free energy.
Warren \emph{et al.}~\cite{warren2022adaptive} introduced an adaptive VQA approach to prepare Gibbs states by introducing an objective function that is easier to measure than the free-energy and using dynamically generated, problem-tailored ans\"atze. 
Finally, a variational method for approximate Gibbs preparation via imaginary time simulation has been presented in~\cite{ZoufalVarQBM21}. The difficulty in these variational approaches either lies in the requirement to continuously compute the von Neumann entropy or to correctly propagate imaginary time dynamics via a variational ansatz.
If sampling from the Gibbs state is sufficient, approaches like ``Quantum Minimally Entangled Typical Thermal States''~\cite{motta2020determining} can be used, which are based on repeated imaginary time evolution of a weakly entangled initial state.
Wild \emph{et al.}~\cite{PhysRevLett.127.100504} introduced a set of quantum algorithms that provide unbiased samples by creating a state that encodes the quantum Gibbs distribution in its entirety and demonstrate that this method can outperform classical Markov chain algorithms in various scenarios, including the Ising model and sampling from weighted independent sets on two distinct graphs. Additionally, recent advancements in quantum Gibbs sampling, like the Dissipative Quantum Gibbs Sampling algorithm \cite{zhang2023dissipative}, offer a simpler and potentially more efficient alternative, as they utilize local update rules and are more feasible for near-term quantum hardware.

Quantum Gibbs sampling plays a crucial role in various optimization algorithms, particularly in convex optimization as discussed in Sec.~\ref{sec:convex_optimization}. Quantum speedups are possible because, under certain conditions, Gibbs states can be prepared more efficiently quantumly than classically. The block-encoding framework \cite{gilyen2019quantum} demonstrates this, where an $n \times n$ Gibbs state is constructed with complexity  ${\cal O}(\sqrt{n})$ in the dimension parameter, albeit other parameters also appear in the running time. This method and its applications in algorithms like the Matrix Multiplicative Weights Update for solving general semidefinite optimization problems are further elaborated in \cite{8104077,van2020quantum,brandao2019quantum,van2019improvements}. Notably, these algorithms leverage Gibbs states for computing trace inner products. 

Furthermore, the complexity analysis of quantum partition functions, as explored in \cite{bravyi2023quantum}, reinforces the importance of efficient Gibbs sampling methods in quantum computing. Encoding Gibbs distributions is at the heart of simulated annealing approaches for Gibbs partition functions \cite{montanaro2015MC,harrow2020adaptive,arunachalam2022simpler,cornelissen2023sublinear}, which can be used to approximately count certain combinatorial objects such as $k$-colorings or matchings of a graph, with genuine quantum speedups in theory. Gibbs sampling methods find application in specialized solutions to semidefinite relaxations of problems like MAXCUT \cite{brandao2022faster,augustino2023solving}, see Sec.~\ref{sec:convex_optimization}, and are integral to quantum algorithms for zero-sum games \cite{van2019zerosum,bouland2023quantum,gao2023logarithmicregret}. Finally, \cite{Li_2020_Gibbs_objective} introduced the notion of a Gibbs objective function which can be useful for quantum optimization problems. The objective can be evaluated by the means of Gibbs sampling, although it can be challenging, as discussed previously. We further mention this approach in Sec. \ref{sec:unconstrained_discrete_optimization}.

\subsection{Approximation Algorithms}
Grover Search, (Trotterized) QAA, QITE, (and QPE) are all expected to require Fault-Tolerant Quantum Computing (FTQC) due to the resulting circuit sizes. However, they also serve as theoretical motivation for approximation algorithms and heuristics, which may already lead to quantum advantages in optimization with noisy quantum computers, as discussed later in this section.

In addition to the exact algorithms discussed in Sec.~\ref{sec:complexity}, there exist also approximation algorithms that guarantee a certain approximation ratio.
Quantum Computing could be used to accelerate known classical approximation algorithms via Quantum subroutines. For instance, Semidefinite Programming (SDP) relaxations, as used in the famous Goemans-Williamson algorithm for MAXCUT \cite{goemans1995maxcut, Patti_2023} or more general variants for QUBO \cite{Brandao_2022}, may profit from Quantum SDP solvers, cf.~Sec.~\ref{sec:convex_optimization}. Eventually, for special cases, these may even achieve an exponential speedup for solving some SDPs compared to classical algorithms. However, unless the problem is given in a compact functional form and assuming we are not interested in reading out the full solution but only the optimal objective value, the end-to-end advantage will be reduced to polynomial due to input/output bottlenecks.
It also raises the question whether classical algorithms may be able to leverage this special setting to achieve similar improvements, as has been shown, e.g., in the context of recommendation systems \cite{Tang_2019_recommendation} or solving low-rank linear systems \cite{chia_2020_low_rank_linear}.
Further, there are many practical problems to achieve such an advantage for solving SDPs, as will be discussed in more depth in Sec.~\ref{sec:convex_optimization}, and very likely this will require FTQC.

\subsection{Variational Methods}\label{sec:paradigms_variational}
\begin{figure}
    \centering
    \includegraphics[scale=1]{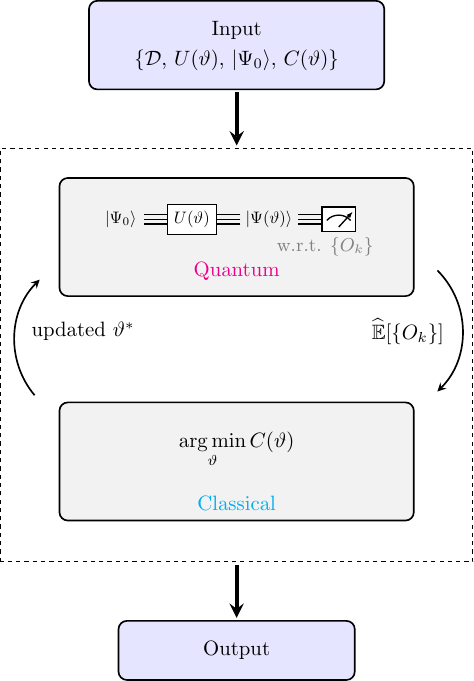}
    \caption{Variational methods iterate between the evaluation of empirical estimates of the expectations of a set of observables $\{ O_k \}$ associated with a parameterized quantum circuit, and classical computation of updated parameters for the parameterized quantum circuit.}
    \label{fig:vqas}
\end{figure}

Within the near-term intermediate-scale quantum (NISQ) devices, much work is focused on so-called Variational Quantum Algorithms (VQAs)~\cite{Peruzzo2014,Farhi2014,mcclean2016theory,cerezo2021variational,bharti2021noisy}, i.e., 
a family of randomized search algorithms that
solve certain optimization problems classically while evaluating involved expectations quantumly.
Specifically, a VQA is an algorithmic scheme where one provides some problem data $\mathcal{D}$, a parameterized ansatz $\ket{\Psi(\vartheta)}$ and a corresponding cost function $C(\vartheta)$ as input. Then, the parameterized quantum state is measured in some basis $\{O_k \}_{k\in \mathbb{N}} \in {\rm Herm}(\mathbb{C}^n)$ to evaluate the cost function.
The next step involves optimizing over the cost function in order to obtain updated parameters, which are then fed back onto the circuit. See Fig.~\ref{fig:vqas} for a schematic. 

Although finding the optimal parameters in a VQA is in general $\textbf{NP}$-hard \cite{Bittel_2021_vqa_np}, it can still result in a powerful heuristic to find good solutions, even potentially offering speedups \cite{golden2023numerical}. This situation is similar to the training of classical artificial neural networks, whose training to the optimum is also $\textbf{NP}$-hard \cite{Blum1992,khalife2023neural} (even with only $k=2$ layers, a ReLU neural network cannot be trained in time bounded by a polynomial in the dimension of the weights \cite{froese2023training}).
VQAs differ mainly in the choice of the ansatz, the cost function, and the optimizer, as discussed, for example, in Sec.~\ref{sec:discrete_optimization}.

Although there are analyses of 
asymptotic convergence relating variational methods to the QAA \cite{binkowski2023elementary},
bounds on iteration complexity of variational methods \cite{kungurtsev2022iteration,mastropietro2023flemingviot}, and in some cases,
bounds the objective function value at the limit point \cite{egger2020warmstarting,Tate2023WarmSDP}, VQAs are still mostly seen as heuristics.

\section{Problem Classes \& Algorithms\label{sec:problem_classes}}

There are many different classes of optimization problems that are defined through the types of variables involved, the objective function(s) used, presence of constraints, stochasticity, and so on. Furthermore, problems can be modeled in different ways, which can lead to representations of the same problem in different problem classes. 
Although these formulations might be formally equivalent, the chosen formulation can play a crucial role in how a problem is solved, as different formulations put more or less emphasis on certain properties of a problem. In this section, we introduce the most important classes of optimization problems and discuss existing algorithms to solve them. We then discuss key open questions in each domain that could help advance the field of optimization.

\subsection{Discrete Optimization}
\label{sec:discrete_optimization}

Discrete optimization is a branch of mathematical optimization that focuses on problems where, as the name suggests, variables take on discrete values. This often implies a combinatorial explosion of possible solutions with respect to size of the input parameters. We focus on discrete optimization problems with and without constraints. While problems of one class can be converted into problems of the other class, approaches to solve them can differ significantly and leveraging available structure might be beneficial in many cases. 
Moreover, the equivalence often only holds if a globally optimal solution is found.
The remainder of this section discusses exact algorithms, approximations, and heuristics for the considered problem classes within discrete optimization.

\subsubsection{Unconstrained Discrete Optimization}
\label{sec:unconstrained_discrete_optimization}

In the realm of unconstrained discrete optimization, Quadratic Unconstrained Binary Optimization (QUBO) serves as a foundational example and is given as follows:
\begin{mini}|s|[3]
{x\in \{0, 1\}^n}{ x^T Q x }
{}{},  \label{eq:qubo}
\end{mini}
where $Q \in \mathbb{R}^{n \times n}$ is the cost matrix and $n \in \mathbb{N}$ denotes the number of variables. 
Other types of unconstrained discrete optimization problems involve, for instance, high-order polynomial objective functions \cite{crama:hal-03795395}, also referred to as Polynomial Unconstrained Binary Optimization (PUBO), or black box objective functions \cite{Zoufal_2023_blackbox} as well as (bounded) integer variables.
While one could also define linear unconstrained discrete optimization, this class is of little interest as it can be solved by looking at every variable independently. Strictly speaking, QUBO also allows one to model linear objective functions when $Q$ is a diagonal matrix, because binary variables satisfy $x^2 = x$.

Some problems, such as MAXCUT, are naturally formulated as a QUBO, and most unconstrained discrete optimization problems can be mapped to QUBOs~\cite{Lucas2014,hadfield2021representation,glover2022,xavier2023qubo}, for example, by replacing integer variables by a suitable encoding~\cite{Karimi_2019_int2bin,sawaya2023encoding} or by converting higher-order polynomial terms to additional binary variables and quadratic terms \cite{biamonte2008nonperturbative,babbush2013resource}.
In addition, certain constraints can be modeled by translating them into penalty terms and adding them to the objective function \cite{Lucas2014,glover2022}.
However, these conversions usually assume that the problems are solved exactly, i.e., that the global optimum is found. In the case of approximate solutions, this has to be carefully analyzed. For instance, equality constraints might only be satisfied approximately when being represented via a penalty term, which, depending on the application, might be acceptable or not.
Furthermore, binary encodings of integers usually do not reflect the intrinsic order between values, i.e., they lack some important structure that a solver may leverage to find good solutions.

Within this section, we focus primarily on QUBO. QUBOs are a natural problem class to consider for quantum optimization as they can easily be converted into a ground state problem, often explored in quantum computing, usually in the context of quantum chemistry or quantum physics~\cite{Peruzzo2014}. 
The relationship between solving a QUBO and finding the ground state of a Hamiltonian is fundamental to many quantum optimization algorithms. 
The most common translation involves two steps: first, substituting binary variables $x \in \{0, 1\}$ with spin variables $z = 1 - 2x$, where $z \in \{-1, +1\}$, and secondly, replacing these spin variables with Pauli $Z$ matrices. The result is a diagonal Hamiltonian $H \in \mathbb{R}^{2^n \times 2^n}$ that encodes the objective values of Eq.~\eqref{eq:qubo} on its diagonal. The ground state of this Hamiltonian represents the QUBO's optimal solution~\cite{Lucas2014}.

QUBOs with spin variables $z \in \{-1, +1\}$ naturally arise in physics and are referred to as \emph{Ising spin glasses}~\cite{barahona1982computational}. Here, the present quadratic terms are usually given via a lattice or graph, and their weight corresponds to an interaction strength between neighboring sites. 
By allowing a polynomial resource overhead, many graph covering and coloring problems can be mapped to an Ising spin glass on a regular lattice~\cite{Lucas2014}.
However, to save qubit resources, many benchmarks use simpler Ising Hamiltonians on hardware-efficient lattices without a specific problem embedding \cite{ronnow2014defining}.

In general, QUBO problems are \textbf{NP}-hard~\cite{barahona1982computational}.
In principle, one could consider, for instance, Grover search or quantum annealing, as suggested in Sec.~\ref{sec:paradigms}, to solve QUBOs; however, these algorithms with provable performance guarantees are expected to require FTQC.
QUBO problems are also \textbf{APX}-hard, i.e. there cannot exist a PTAS for QUBO problems unless \textbf{P}=\textbf{NP} \cite{punnen2022quadratic}.
But, there exist PTASes for certain problems that can be formulated as QUBO problems, like the Euclidean Traveling Salesperson Problem \cite{Arora1998ETSP}.
Little is known about PTASes on quantum computers; there are many possibilities to try using quantum algorithms to accelerate known classical PTASes. An example might be to apply Grover Search or Quantum Dynamic Programming, cf.~Sec.~\ref{sec:dynamic_programming}, as sub-routines to achieve (sub-) quadratic speedups compared to purely classical PTASes. Like Grover Search, Quantum Dynamic Programming is also expected to require FTQC. Since exact algorithms and acceleration of PTASes most likely require FTQC, there is arguably more hope for a near-term quantum advantage in optimization sourced from quantum approximation algorithms and heuristics.

Originally proposed for quantum chemistry applications, the Variational Quantum Eigensolver (VQE) \cite{Peruzzo2014} denotes a class of heuristic VQAs to approximate ground states of Hamiltonians. 
This is achieved by choosing a parametrized quantum state $\ket{\Psi(\vartheta)} = U(\vartheta) \ket{\Psi_0}$, where $U(\vartheta)$ denotes a parametrized quantum circuit and $\ket{\Psi_0}$ denotes an initial state, and then using a quantum computer to evaluate $\braket{\Psi(\vartheta) | H | \Psi(\vartheta)}$ for given parameter values $\vartheta$, and employing a classical optimizer to minimize the expected value of this quantity. The variational principle guarantees that the expectation value is lower bounded by the ground state energy of $H$, with equality if and only if the ground state is reached. 

\begin{figure}[!htb]
    \centering
    \includegraphics[width=\linewidth]{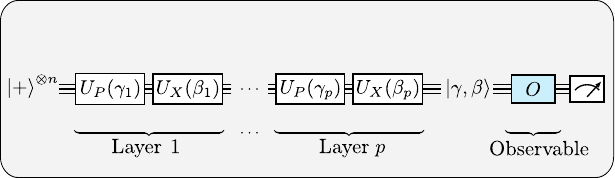}
    \caption{Structure of QAOA ansatz: every layer consists of a mixing and a problem unitary with corresponding parameters $\gamma_j$ and $\beta_j$ for layer $j = 1, \ldots, p$.}
    \label{fig:qaoa}
\end{figure}

The Quantum Approximate Optimization Algorithm (QAOA) \cite{Farhi2014} denotes a special case of VQE that can be applied to find good solutions to QUBOs. QAOA is motivated by QAA and has been shown to be a computationally universal algorithm \cite{lloyd2018quantum}.
More precisely, QAOA uses a particular problem-dependent ansatz given by pairs of unitaries in each layer. These unitaries have the form 
\begin{align} \label{eq:QAOAops}
    U_X(\beta) &= e^{-i \beta H_X},\\
    U_P(\gamma) &= e^{-i\gamma H},
\end{align}
and they are termed as mixing and problem unitaries, respectively. 
The unitaries are repeated alternatingly with new parameters for each layer and applied to the initial state $\ket{+}^{\otimes n}$, as shown in Fig.~\ref{fig:qaoa}.
Here, $H$ denotes the problem Hamiltonian, and $H_X$ denotes the mixing Hamiltonian, see Sec.~\ref{sec:algorithms_qaa} for more details.
Both unitaries are alternated and repeated $p$ times, with individual parameters $\beta_j, \gamma_j$ for $j=1,\ldots,p$.

QAOA is an approximation algorithm with worst-case performance bounds for certain problems and algorithmic settings.
For example, there exist worst-case performance guarantees for QAOA with $p=1,2,3$ layers for the MAXCUT problem on 3-regular graphs. Specifically, for $p=1$ Farhi \emph{et al.}~\cite{Farhi2014} derived a lower bound on the approximation ratio given by $0.692$, while Wurtz and Love \cite{PhysRevA.103.042612} found that a lower bound given by $0.7559$ for $p=2$ and (under certain assumptions) $0.7924$ for $p=3$.
Furthermore, the performance of QAOA for (large-girth) $d$-regular graphs has been analyzed numerically and it has been shown to outperform many known SDP-based relaxations from $p \ge 11$ \cite{farhi2022qaoaSK, basso2021quantum}. 
However, for general QUBO, QAOA is a heuristic without performance guarantees.

A crucial task for VQAs is to determine the optimal parameters, cf.~Sec.~\ref{sec:paradigms_variational}.
The problem-dependent approach QAOA to construct an ansatz implies a relatively small number of parameters, which is desirable for the corresponding classical optimization loop.
Considerable research has been published on training strategies for QAOA as well as on properties of the optimal parameters.
There are different strategies to solve this problem as we will discuss now.

The conventional approach to parameter optimization in a VQA is to leverage the quantum computer to evaluate the loss function and subsequently use a classical optimizer to improve the parameters.
Good parameter initialization is crucial for this task to succeed and there is increasing evidence that this requires some physics-inspired initialization, rather than generic random initialization \cite{Zhou2020,sack2021quantum,scriva2023challenges}.
Proposed strategies range from annealing-inspired parameters \cite{sack2021quantum}, to machine learning based approaches \cite{alam2020accelerating}. 
Similarly, one can try to evaluate the loss function for the optimization problem classically and only sample solutions from the quantum computer. 
The rationale here is that evaluating (or approximating) the (two-)local observables involved in QAOA for QUBO using say, tensor networks or Clifford perturbation theory, is computationally easier than sampling from the complete quantum state \cite{streif2019training, begusic023simulating}.
Alternatively, for some classes of QUBOs, QAOA has a property called \emph{concentration of parameters} or \emph{transfer of parameters} \cite{brandao2018fixed, streif2019training, Akshay_2021_concentration, galda2021transferability}.
For these problems, the optimal parameters become instance independent.
In other words, one can train the parameters for one instance, and reuse it for other instances. Depending on the setting, the instances can be of the same size, but also of larger size.
This may even allow for training of a smaller instance classically and reusing the parameters for larger problem instances where classical simulation is not possible anymore. 
Moreover, this could enable good solutions to a problem very quickly by just using pre-optimized parameters.
Sometimes, the opposite direction is also possible, i.e., optimal parameters for finite instances are derived from studying the limit of infinite size instances \cite{boulebnane2021predicting, farhi2022qaoaSK}.
Parameter optimization for QAOA in general, and transfer/concentration of parameters in particular, are crucial to progress towards a successful application of QAOA for practical settings.

Several variants of QAOA have been proposed over time.
Many of them propose different mixers and initial states to achieve certain goals. 
The Quantum Alternating Operator Ansatz, which shares the acronym QAOA, denotes a family of QAOA variants that encode constraints into the mixer such that they are preserved and the algorithm is restricted to feasible states. This will be discussed in more depth in Sec.~\ref{sec:constrained_discrete_optimization}.
Other variants adjust QAOA in order to warm-start it from solutions obtained by classical algorithms. The idea is to use a classical algorithm to compute an approximate solution, and then use it to warm-start QAOA to further improve upon the obtained solution \cite{egger2020warmstarting, Tate2023WarmSDP}. These approaches inherit some of the performance guarantees of the classical algorithms used for warm-starting and have been shown to significantly outperform standard QAOA on certain instances. With a modification of the ansatz of QAOA based on the warm-starts, one can still show the convergence to MAXCUT under the adiabatic limit (i.e., when the circuit depth $p \to \infty$). However, showing provable guarantees on warm-started QAOA for finite circuit depth with these custom mixers remains an open question. 
Other variants suggest to reuse qubits, leveraging mid-circuit measurements and dynamic circuits, to increase the number of variables for a limited number of qubits \cite{moses2023race}.
It seems also natural to restart the search, when one arrives in so-called barren plateau \cite{mastropietro2023flemingviot}, where the (sub)gradient is not informative enough. Mastropietro \emph{et al.}~\cite{mastropietro2023flemingviot} have shown that the more barren regions there are and the higher their proportion in the parameter space, the higher the speedup obtained by restarting.

Another alternative approach to QAOA is Recursive QAOA~\cite{Bravyi2019}, which uses the quantum computer to produce a sequence of reduced problems. 
Here, at each step QAOA is used to estimate correlations between variables, and the problem Hamiltonian is reduced by fixing the strongest one. This process is iterated until the reduced problem becomes small enough to for (provably exact or heuristic) classical solvers, which then yields an approximate solution to the input problem.  
Recent works~\cite{Bravyi2020b,Dupont_2023QuantumCombOpt,brady2023iterative,finvzgar2023quantum,patel2024reinforcement} have generalized recursive quantum approaches to problems beyond MAXCUT and to circuit ans\"atze beyond QAOA.

In the same paper that introduces Recursive QAOA~\cite{Bravyi2019}, the authors also show that, in the large problem limit, constant-depth QAOA cannot outperform the classical Goemans-Williamson approximation for certain instances of MAXCUT on $d$-regular graphs, i.e., the depth needs to grow with the problem size. However, these results are not applicable to alternative approaches, e.g., other QAOA algorithm variants or Recursive QAOA. 
There are further results about the limitations of QAOA and requirements on circuit depth with respect to problem size and structure. For instance, Farhi \emph{et al.}~\cite{farhi2020quantum} show that QAOA ``needs to see the whole graph'' for 
a particular QAOA approach to the Maximum Independent Set (MIS) problem. More precisely, the authors show that for $p < \log(n)/\log(d / \ln(2)) / 2$, where $n$ denotes the number of nodes in the graph and $d$ its degree, for large enough $d$, the approximation ratio of QAOA is upper bounded by $0.854$. However, for larger $p$, no such limitation has been found. 
Since the logarithm grows slowly in $n$, this is not too strong of a restriction. Similar obstructions to low-depth QAOA have also been shown to apply to $k$-local generalizations of MAXCUT~\cite{marwaha2022bounds}.

QAOA also inspired the development of classical algorithms, such as the the Mean Field Approximate Optimization Algorithm \cite{Misra_Spieldenner_2023}. 
Here, the QAOA circuit is approximated through mean-field approximations, which is demonstrated to perform well for certain problem instances. This may also be of interest as a classical alternative to train or initialize the QAOA parameters before sampling solutions from a quantum computer.

Since the ansatz for QAOA is derived from the problem, the structure of the problem directly affects the required qubit connectivity of the quantum circuit.
Given the often sparse-connectivity of quantum devices, this requires a transpilation step to map the problem/circuit to the hardware. The transpilation and execution of quantum circuits on noisy hardware will be discussed in depth in Sec.~\ref{sec:scaling}.
However, addressing this issue at the problem and modeling level might lead to significant improvements through hardware-optimized problem formulations.
A more detailed review of QAOA can be found in \cite{blekos2023review}.

Another family of near-term compatible approximation algorithms for MAXCUT is given by Quantum Random Access Optimization (QRAO, also called Quantum Relaxation Algorithms) \cite{fuller2021approximate, teramoto2023quantumrelaxation}. Here, the variable-to-qubit ratio is chosen to be larger than one and results in a non-diagonal Hamiltonian. This allows the encoding of larger optimization problems into a given number of qubits. However, depending on the chosen encoding and resulting ratio, there might be restrictions on the allowed quadratic terms in the QUBO, i.e., edges in a graph. Assuming a feasible graph, and that we can prepare a good enough approximation of the ground state of the corresponding non-diagonal Hamiltonian, which, in general, cannot be guaranteed, it has been shown that QRAO achieves approximation ratios for MAXCUT of at least $0.555$, $0.625$, and $0.722$, for variable-to-qubit ratios of $3$, $2$, and $1.5$ respectively.
Furthermore, \cite{teramoto2023quantumrelaxation} introduce another QRAO with ratio 2 that does not impose any constraints on the graph, but might not always achieve a non-trivial performance guarantee (i.e., $>1/2$). 
Since providing the sufficiently accurate approximation of the ground state of the Hamiltonian can in general not be guaranteed, QRAO is strictly speaking not an approximation, but a heuristic algorithm.
There also exist other encoding schemes that represent more variables than qubits and provide some variational heuristics for optimization algorithms \cite{patti2022variational}.
Note, that in principle, one could encode up to $2^n$ binary variables into $n$ qubits.
However, that would usually imply that writing down the problem (with a cost matrix in $\mathbb{R}^{2^n \times 2^n}$) would be similarly expensive as simulating the involved $n$-qubit circuits classically.  
Thus, only moderate, i.e., polynomial, encodings make sense unless the problem is given in a compact function form and access to the full solution is not needed, only the objective value. That is a similar situation, as for instance, in semidefinite programming, cf.~Sec.~\ref{sec:convex_optimization}.

While neither QAOA nor QRAO achieve the approximation ratios of classical approximations for MAXCUT, i.e., $0.878567$, they are nonetheless very interesting heuristics. As discussed in Sec.~\ref{sec:complexity}, there exist problems with approximation algorithms where the known lower bounds do not yet achieve the inapproximability bounds i.e., where there is room for improvement for classical and also quantum approximations. This represents a very interesting direction of research and potential quantum advantage over known classical approximations.

Another time evolution-based heuristic is quantum imaginary time evolution (QITE) \cite{mcardle2019variational,motta2020determining}. Like QAA, QITE assumes an initial state with a non-zero overlap with the ground state of $H$. Then, it applies the evolution 
\begin{eqnarray}
    \ket{\psi_{\tau}} &=& \frac{e^{-H \tau}}{Z_{\tau}} \ket{\psi_0},
\end{eqnarray}
where $\tau = i t$ and $Z_{\tau} = \sqrt{\bra{\psi_0} e^{-2H \tau} \ket{\psi_0}}$ denotes a normalization factor related to the partition function.
QITE exponentially suppresses amplitudes of eigenstates with larger eigenvalues and amplifies the amplitudes of the ground state. The time $T$ to exceed a fixed overlap with the ground state, assuming at least an exponentially small initial overlap with the ground state, scales inversely to the spectral gap $\Delta$ of $H$ and linear in the problem size, i.e., $T = \mathcal{O}(n/\Delta)$ \cite{zoufal2021generative}. Unlike QAA, the Hamiltonian is not time-dependent, and thus, the gap does not depend on the time and not necessarily on the problem size. In addition, a small gap might also imply that there are multiple very good solutions, which -- for shorter times -- would make it an interesting heuristic. 
However, QITE is a non-unitary evolution and can be challenging to implement on a quantum computer \cite{motta2020determining}.
If it were possible, it would allow \textbf{QMA}-hard problems such as generic ground state search or Gibbs state preparation to be solved \cite{aharonov2002quantum}.
Thus, only heuristic approximations of QITE exist to the best of our knowledge. These include projecting the non-unitary evolution to unitary operations~\cite{motta2020determining} or embedding it in a unitary of augmented dimension~\cite{lin_compressedtimeevo_2021}. 
Another family of algorithms are VQAs that map the time evolution to an ansatz by means of variational principles~\cite{mcardle2019variational, Yuan2019theoryofvariational} or directly projecting a Taylor-expansion of the time-evolution operator onto the ansatz~\cite{benedetti_projecttimeevo_2021, slattery_blocktimeevo_2022}. 
Variational approaches based on variational principles are also closely related to alternative optimizers, such as Quantum Natural Gradients \cite{Stokes2020quantumnatural} as well as stochastic approximations \cite{gacon2023stochastic}.
The advantage of these heuristics is that they do not require the ansatz to be derived from the problem, which makes it easier to implement them on real hardware and allows one to apply them also to unconstrained black box binary optimization \cite{Zoufal_2023_blackbox}. However, at the same time, that is a potential disadvantage since it is unclear what circuit structure might work and there is less theoretical foundation on why they may lead to a potential quantum advantage. The design of suitable circuits, possibly in an adaptive way, is a key open question for these type of algorithms.

In addition to variational approximations to QITE, there are also variational approximations to real-time evolution \cite{Yuan2019theoryofvariational, Barison2021efficientquantum}.
Those can be used to approximate QAA and form another family of heuristics.
However, like for QITE, the design of the ansatz is a crucial open question.
In addition, all of these variational time evolution-based algorithms introduce relatively high costs in terms of the number of circuits to be evaluated which may lead to a significant bottleneck to scale them \cite{OllitraultMolecularQuantDyn21} --- the cost is increased even further if bounds on the preparation accuracy are computed \cite{ZoufalErrorBounds23}.

While VQAs are often defined as minimizing an expectation value, in quantum optimization we might not care about an expectation value but rather about sampling good solutions, which gives us some flexibility in the choice of cost function.
Alternative cost functions can help to increase the robustness against noise or relax the requirements on an ansatz.
A frequently used example is the Conditional Value at Risk (CVaR) \cite{Barkoutsos2020, barron2023cvar_bounds}, where we do not average over the objective values corresponding to all samples obtained from measuring a quantum state. 
Instead, we sort them and only take the average over the best $\alpha$-fraction of samples, for a pre-defined $\alpha \in [0, 1]$, where $\alpha=1$ corresponds to the full expectation value and $\alpha=0$ corresponds to the single best observed sample. 
It has been shown by Barron \emph{et al.}~\cite{barron2023cvar_bounds} that choosing $\alpha$ based on the present noise when evaluating the corresponding circuit on a noisy quantum device, CVaR can provably bound noise-free expectation values.
Another proposal is the Gibbs objective function \cite{Li_2020_Gibbs_objective}. The rationale is the same as for the CVaR, it puts more emphasize on good samples than on bad samples, i.e., it biases the expectation value.

Within this section, multiple possible quantum optimization algorithms have been discussed.
All of them are limited by the number of available qubits on existing respectively near-term hardware.
However, if an empirical quantum advantage had been demonstrated for a smaller problem, it might be possible to extend it, e.g., using decomposition or multi-level schemes to leverage the quantum computer also for larger problems \cite{angone2023hybrid}.

Unconstrained discrete optimization is a rich research domain that has already led to plenty of results.
Nevertheless, there are still many open questions to be answered regarding the potential for quantum advantage for this problem class.
Important questions can be found at every level, from quantum-specific problem formulations, to better encodings of integer variables, and more efficient encodings of problems in general~\cite{sawaya2023encoding}. Furthermore, a key step in classical solvers is pre-processing which can simplify problems tremendously. However, little is known about the potential of quantum-specific pre-processing. In addition, classical algorithms, such as branch-and-bound, often come with a posteriori bounds on the optimality gap. This is crucial because in many cases this results in almost optimal solutions and also helps to determine how much compute resources to invest in finding good solutions to a given problem.
In addition to a priori or a posteriori performance bounds, it is crucial to understand how they carry over when the algorithm is executed on noisy quantum hardware and how problems can be optimally modeled and mapped to the hardware to minimize the impact of noise. Because these questions will often be difficult to answer, systematic benchmarking becomes even more important, cf.~Sec.~\ref{sec:benchmarks}.
To conclude, with the availability of quantum computers with more than one hundred qubits, it is crucial to develop and test quantum optimization heuristics on these devices for non-trivial problem sizes.
This is at a scale where exact simulation is not possible anymore, and even if approximate simulation might be possible, it does not help to learn how to deal with the noise, and thus, how algorithms are scaling in practice.
Developing this intuition is key to progress towards a practical quantum advantage in optimization.
While discrete unconstrained optimization already represents a very broad domain of optimization, many practically relevant problems naturally come with constraints, such as budget, capacity, or structural constraints. Thus, in the next section, we discuss different approaches to incorporate constraints into quantum algorithms for discrete optimization.

\subsubsection{Constrained Discrete Optimization}
\label{sec:constrained_discrete_optimization}

Constrained Discrete Optimization is defined by adding constraints of the type $g(x) = 0$ or $g(x) \leq 0$ to an Unconstrained Discrete Optimization problem. In the following, for simplicity, we consider QUBO and add linear inequality and equality constraints, although some of the algorithms could be applied to more complex types of constraints as well. The conversions for higher-order polynomial objective functions to quadratic functions and from integer to binary variables still apply as discussed in Sec.~\ref{sec:unconstrained_discrete_optimization}.
Thus, the considered problems are given by
\begin{mini}|s|[0]
{x\in \{0, 1\}^n}{ x^T Q x }
{}{}  \label{eq:quadratic_constrained_optimization}
\addConstraint{A x = b}
\addConstraint{C x \leq d}
\end{mini}
for matrices $A \in \mathbb{R}^{m_{e} \times n}$, $C \in \mathbb{R}^{m_{i} \times n}$, and vectors $b \in \mathbb{R}^{m_e}$, $d \in \mathbb{R}^{m_i}$, where $m_e$, $m_i$ denote the number of equality and inequality constraints, respectively.
Note, that unlike for QUBOs, linear objective functions (i.e., a diagonal $Q$) are interesting due to the presence of constraints.

Under mild assumptions on $A$ and $C$, this can always be converted to QUBO by adding slack variables to convert all inequality constraints to equality constraints and then converting equality constraints to penalty terms by squaring the right-hand-side and adding it to the objective function with a large weight \cite{glover2022}. Some constraints can also be encoded into penalty terms more efficiently (see, e.g., \cite{glover2022}). 
While this is always possible, it may come at a substantial cost in terms of additional binary variables and a dramatic increase in non-zero terms in the cost matrix in Eq.~\eqref{eq:quadratic_constrained_optimization}; i.e., it often maps an otherwise sparse matrix $Q$ to a dense one.
Further, adding penalty terms scaled by a large constant can tremendously amplify the range of values of an objective function, which can imply numerical instabilities or requirements to control certain parameters in the software or hardware stack up to an infeasible accuracy.
These are some of the reasons why explicitly keeping the structure can lead to more efficient algorithms. 
In the following, we discuss strategies to incorporate constraints directly.

Like for unconstrained discrete optimization, we can leverage Grover Search to achieve a quadratic speedup over brute force search \cite{Gilliam_2021_grover}. This requires not only an oracle for the objective function, but also one for each constraint such that we can mark all feasible solutions.
Since brute force search is rarely the most efficient classical algorithm the achieved quantum advantage is usually sub-quadratic or sometimes not present at all. 
However, Grover Search might still be useful as a subroutine in other algorithms.

Classically, constrained discrete optimization problems are often solved using branch-and-bound algorithms.
In \cite{montanaro2020quantum}, a quantum branch and bound algorithm is introduced that accelerates classical branch-and-bound algorithms. It is shown that it solves most instances of the Sherrington-Kirkpatrick model to optimality with a high probability in time $\mathcal{O}(2^{0.226n})$. The algorithm adapts Grover Search to be applicable to search branch-and-bound trees. While this has the potential to accelerate classical schemes, it has the disadvantage that no a posteriori bounds on the optimality gap are generated, which often allow an early termination of such search schemes. Further, it requires FTQC, which adds additional overhead.

There are multiple proposals to incorporate constraints in VQAs.
The most straight-forward approach is to design an ansatz that -- in the noise-free case -- is restricted to the feasible space.
For instance, suppose a constraint that fixes the Hamming weight of feasible bit strings, i.e., the number of ones. Then, we can prepare as initial state a uniform superposition of all these feasible bit strings, i.e., a Dicke state~\cite{baertschi2022shortdepth}. Or, we can use a parametric iSWAP or a Givens rotation gate, like in a particle preserving ansatz in quantum chemistry~\cite{google2020hartreefock}, to construct an ansatz that only generates feasible bit strings.
This is straight-forward for some type of constraints, but more difficult in general.
Further, it takes the constraints into account, but not yet the objective function.

The Quantum Alternating Operator Ansatz~(QAOA')~\cite{hadfield2017quantum,hadfield2019quantum} generalizes the Quantum Approximate Optimization Algorithm to much wider classes of problems and encodings, in particular to problems with hard constraints. 
Here, more general initial states are considered, such as ones from the subspace of feasible states for constrained problems. 
The QAOA operators of Eq.~\eqref{eq:QAOAops} are generalized to arbitrary families of phase of parameterized operators, again applied in $p$ alternating layers. In particular, mixers can be constructed as ordered products of non-commuting local operators such that problem hard constraints are guaranteed to remain satisfied. 
This has the advantage to keep the structure of the objective function and often leads to significantly cheaper implementations of cost operators than when adding constraints as penalty terms.
Alternatively, one can also directly replace the mixing Hamiltonian $H_X$ in Eq.~\eqref{eq:QAOAops}, e.g., for Hamming weight constraints by $XY$-model~\cite{wang2019xy,cook2020vertexcover} or Grover-based~\cite{baertschi2020grover,golden2023numerical} Hamiltonians.
However, all of this leads to more complex mixers, and implementing them can become challenging. 
Hardware-efficient ansatz variants~\cite{moll2018quantum,larose2022mixer,Dupont_2023QuantumCombOpt,maciejewski2023design} are one alternative approach toward alleviating this difficulty.

Another approach is to leverage quantum Zeno dynamics to repeatedly project the state back to the feasible subspace \cite{herman23}. This requires auxiliary qubits as well as projective mid-circuit measurements.
While this allows one to include multiple constraints of different types, the resulting circuits become deeper and are likely to require FTQC.

Instead of handling constraints explicitly in the quantum circuit, unconstrained black box binary optimization \cite{Zoufal_2023_blackbox} allows hiding constraints in the black box objective function.
Instead of requiring quadratic penalty terms for every constraint, a general metric for infeasibility can be defined, which allows one to add a more balanced penalty term to the objective without the disadvantages mentioned earlier in this section.
However, as mentioned in Sec.~\ref{sec:unconstrained_discrete_optimization}, it is unclear how to construct promising parameterized circuits.

Alternatively, Iterative or Recursive QAOA variants, cf.~Sec.~\ref{sec:unconstrained_discrete_optimization}, may be adapted to enforce constraints as suggested in \cite{brady2023iterative}. Here, as for the unconstrained case, problems are solved iteratively and usually one variable is removed at a time. Adjusting the selection rules for how variables are fixed can allow one to enforce constraints in certain cases. 

Adding constraints as penalty terms to the objective function, i.e., casting constrained binary optimization to QUBO, is often the main driver for the density of the cost matrix, and thus, the challenges to implement the algorithms on real hardware. If at least some constraints, e.g., cardinality, packing, or covering constraints, could be natively incorporated into an algorithm, this could lead to sparser problems, and thus, could lead to significant simplifications for implementations on real quantum hardware. This is sometimes called QUBO-Plus \cite{glover_2022_qubo_plus}, i.e., QUBO plus special types of constraints. For classical solvers, these constraints can even simplify the problem, since the knowledge about the optimal solution can be leveraged within the algorithm. This is also the idea behind most of the presented quantum algorithms within this section. However, often the complexity is just pushed to a different part of the algorithm, for instance, from the QAOA cost operator to its mixer.

The approaches for constrained discrete optimization mentioned in this section essentially leverage three strategies to handle constraints: they use Grover Search (and variants), they add penalty terms to the objective function, and they try to restrict or project states to the feasible space.
Since constraints are appearing almost everywhere in practice, further progress on how to efficiently handle them is crucial on the path towards quantum advantage in optimization.

\subsection{Continuous Optimization}
\label{sec:continuous_optimization}

\subsubsection{Convex Optimization}
\label{sec:convex_optimization}

Convex optimization problems can often be solved efficiently, both in theory and in practice, and find numerous applications in business, science and engineering. Quantum algorithms for convex optimization have been proposed for several relevant classes of problems. 

One of the most general formulations of a convex optimization problem is the task of minimizing a convex function $f: \R^n \to \R$ over all $x \in \R^n$ that have at most a certain size, say $\|x\|_2 \leq R$ for a given bound $R$, when one is provided black-box access to evaluate the function and possibly its derivatives. For example, given access to an oracle that on input $x \in \R^n$ outputs $(f(x), \nabla f(x))$, our task is to find an approximate minimizer of $f$ over the convex set $\{x \in \R^n: \| x\|_2 \leq R\}$. Gradient descent is an example of a well known algorithm that works in this framework. A natural question is whether quantum computers offer a speedup, be it in query complexity or run time. The answer to this question is not always positive; see, e.g., \cite{garg2021no,gargNoSmooth}, which consider black-box access to the function and its (sub)gradients. In this setting, classical and quantum algorithms have the same lower bound (no $n$-dependence, but a polynomial scaling in the inverse of the desired precision $\epsilon$) for the query complexity in general. 
The other regime, where we do allow a polynomial $n$-dependence, but only a polylogarithmic dependence on the desired precision, is equally well studied. Under mild assumptions, one can solve general convex optimization problems in polynomial time using e.g., the ellipsoid method~\cite{grotschel2012geometric}. In this very general setting one aims, for example, to solve a problem of the form $\min c^T x \text{ s.t. } x \in K$ for some bounded convex set $K$ to which we have various types of oracle access (membership, separation, optimization). Here mild quantum speedups (and no-go results) are known~\cite{van2020oracles,chakrabarti2020quantum}.

Often much more can be said when we consider convex optimization problems with a certain structure. The most notable such classes are arguably linear programming (LP) and semidefinite programming (SDP). Here the objective is to minimize a linear function subject to linear inequalities on non-negative vectors (LP) or on positive semidefinite matrices (SDP). An SDP can be expressed in the form 
\begin{maxi}|s|[0]
{X \succeq 0}{\tr(C X)\hspace{2.5cm}}
{}{} 
\addConstraint{\tr(A_i X) \leq b_i \, \forall i \in [m]}
\end{maxi}
where the problem is defined by symmetric matrices $C, A_1,\ldots,A_m$, each of size $n \times n$, and a vector $b \in \R^m$. Note that LPs correspond to a special case of SDP in which each of the input matrices is diagonal. Quantum algorithms for these problems usually trade off an improved dependence on the dimension $n$ and the number of inequalities $m$ with a worse dependence on the precision $\epsilon$ to which one solves the problem. We can categorize quantum algorithms for these problems based on the solution methodology. 

\textbf{First-order methods.} 
The first quantum algorithms for SDPs were first order methods, based on the Multiplicative Weights Update (MWU) framework \cite{arora2005fast}. In this framework one uses candidate solutions $X$ that are proportional to $\rho = \exp(\sum_{i=1}^m y_i A_i) / \tr(\exp(\sum_{i=1}^m y_i A_i))$ for some (sparse) vector $y \in \R^m$. That is, the candidate solutions are proportional to Gibbs states, which naturally correspond to trace-normalized positive semidefinite matrices and can be efficiently prepared on a quantum computer under some conditions. To implement the MWU framework, one additionally needs to compute trace inner products $\text{tr}(A \rho)$ between such Gibbs states $\rho$ and a matrix $A$ (which is either one of the $A_i$'s or $C$), and essentially solve a search problem based on these values. All these steps can be performed on a quantum computer, with various levels of efficiency; the initial run time bounds derived in the pioneering papers \cite{8104077,van2020quantum} were subsequently improved \cite{brandao2019quantum,van2019improvements,van2020phd}, and further to \cite{van2019improvements}. 

The run time of quantum MWU methods depends on several natural parameters; the dimension $n$, the number of constraints $m$, the desired (additive) error $\epsilon$ in the objective function; and on two instance-specific parameters, typically denoted by $R$ and $r$, that are related to the diameter of the feasible region for the primal ($\tr(X) \leq R$) and dual problems ($\|y\|_1 \leq r$), and bound the so-called ``width'' of the oracle. The initial works had a run time scaling as $\widetilde{\mathcal{O}}\left( \sqrt{mn} s^2 \poly\left(\frac{R r}{\epsilon}\right) \right)$, where $s$ is the row sparsity of the input matrices~\cite{8104077,van2020quantum} and $\widetilde{\mathcal{O}}$ ignores logarithmic factors, later works improved this to $\widetilde{\mathcal{O}}\left( \left(\sqrt{m} + \sqrt{n} \right) s^2 \poly \left(\frac{R r}{\epsilon}\right) \right)$.
Throughout, the polynomial was of a high degree; the state of the art in a certain input model is $\widetilde{\mathcal{O}}\left(\left(\sqrt{m} + \sqrt{n}\frac{R r}{\epsilon}\right) s \left(\frac{R r}{\epsilon}\right)^4 \right)$, see \cite{van2019improvements}. In all these bounds, the parameters $r, R$, and $\varepsilon$ appear together as one ``scale-invariant'' parameter $\gamma = Rr/\epsilon$~\cite{van2020quantum}. Often, this parameter $\gamma$ scales poorly in terms of $n$ and $m$~\cite{van2020quantum}, 
especially when parameters $r, R$ scale linearly or superlinearly with $n$. (For example, in the SDP relaxation of MAXCUT \cite{goemans1995maxcut}, cf. \eqref{eq:maxcutsdp}, $\tr(X)$ scales linearly with $n$.) 
Notable exceptions are applications to shadow tomography, quantum state discrimination, and E-optimal design (i.e., optimal design in which one maximizes the smallest eigenvalue of the information matrix)~\cite{brandao2019quantum,van2019improvements}. As an aside, we mention a second famous appearance of the MWU framework in the quantum (complexity) literature: it is a key tool in the proof of the equality \textbf{QIP}~=~\textbf{PSPACE}~\cite{qippspace2011}.

For comparison, the classical complexity of a general-purpose SDP-solver based on the MWU framework is $\widetilde{\mathcal{O}}\left(mns \left(\frac{R r}{\epsilon}\right)^4 + ns \left(\frac{R r}{\epsilon}\right)^7\right)$, see \cite{van2020quantum}, but the input models of the classical and quantum algorithms are not necessarily equivalent so these results are difficult to compare. Namely, quantum algorithms often rely on the use of Quantum Random Access Memory (QRAM), which is a RAM addressed by qubits. QRAM accessed by a unitary, swaps the state of the addressed memory cell with the state of a target qubit. Although concrete proposals exist \cite{GLM08a, GLM08}, a scalable fault-tolerant implementation is yet to be found \cite{Aar15,DGM20,CDS+22,JR23}, making this at best a long-term prospect. The QRAM model of computation is common in the quantum optimization literature, and it is also studied in some of the papers mentioned in other sections: we mention it here because the frameworks discussed in this section are particularly reliant on it. Also, in the classical literature the MWU framework was specialized (and significantly accelerated) for the SDP relaxation of structured combinatorial problems such as a form of QUBO, balanced separator, and sparsest cut, see \cite{arora2005fast,arora2009computational}. 

So far, we have discussed the MWU framework in the context of SDPs. One can equally well apply this machinery to the simpler class of linear programs, often obtaining much better run times, see, e.g., the works~\cite{van2019zerosum,bouland2023quantum,gao2023logarithmicregret} that are based on a zero-sum game approach \cite{grigoriadis1991approximate, grigoriadis1995sublinear}. In particular, \cite{bouland2023quantum} successfully implemented a dynamic data structure in the quantum setting. It is an interesting open question to extend such a result to SDPs. 
One could also study further first-order methods \cite[e.g.]{marecek2021cuttingplane}, beyond MWU, to SDPs. While these have been prototyped \cite{marecek2021cuttingplane},
and there may be some speed-up in per-iteration complexity, their iteration complexity remains the same as in the classical case. 

\textbf{Second order methods.}    
Interior Point Methods (IPMs) are popular second order methods for solving structured convex optimization problems both from a theoretical and practical perspective. At their core, IPMs start from a point lying in the interior of a convex feasible set $\mathcal{X} \subset \R^n$, and use Newton's method to solve a sequence of barrier problems of the form 
\begin{equation}\label{e:barrierFcn}
    \min_{x \in \R^{n}}~\eta \cdot \langle c, x \rangle  + f(x),
\end{equation}
where $\langle \cdot, \cdot \rangle$ is an inner product on $\R^n$, $\eta \geq 0$ and $f: \text{int}\left( \mathcal{X}\right) \mapsto \R$ is a barrier function encoding the constraints defining $\mathcal{X}$. That is, $f(x) \to \infty$ as $x$ approaches the boundary of $\mathcal{X}$, and thus the only ``constraint'' present in Eq.~\eqref{e:barrierFcn} is that $f$ is only defined on the interior $\text{int}\left( \mathcal{X}\right)$ of $\mathcal{X}$. In each iteration, the value of $\eta$ is increased, and the set of minimizers $\left\{ x(\eta) : \eta \geq 0 \right\}$ of Eq.~\eqref{e:barrierFcn} define the so-called \textit{central path}, an analytic curve extending throughout the interior of the feasible region to the set of optimal solutions that is uniquely determined by the starting point. By iteratively applying Newton's method, IPMs approximately follow the central path to the optimal solution via a sequence of local linearizations (i.e., Newton steps). Computing the Newton step requires one to solve a linear system of equations, and constitutes the dominant operation at each iterate of an IPM. It is therefore natural to try to use quantum linear system solvers~\cite{childs2017quantum,chakraborty2019power} in an effort to accelerate the computation of the Newton step~\cite{rebentrost2019quantum}. Work on this line of research was initiated by \cite{kerenidis2018quantum}, with several follow-up works trying to address the shortcomings in the initial proposal \cite{augustino2023quantum,mohammadisiahroudi2023efficient,Wu2023}.
Because the direction computed in the Newton step is necessary to process the subsequent iteration, the quantum linear system algorithm is followed by the application of a quantum state tomography algorithm.     
Specifically, two issues have emerged: the use of state tomography introduces inexactness in the search direction and a $1/\epsilon$ dependence in the run time, where $\epsilon$ is the precision to which tomography is performed \cite{van2023quantum}; and the use of the quantum linear systems algorithm introduces a run time dependence on the condition number of the Newton linear system, which, unfortunately, goes to $\infty$ as we approach optimality. The first issue can be dealt with by developing an IPM framework that allows for inexact search directions \cite{augustino2023quantum}, and iterative refinement techniques can alleviate (but not necessarily eliminate) the remaining issues \cite{mohammadisiahroudi2023efficient,Wu2023}. Overall, assuming access to QRAM, quantum IPMs achieve favorable dependence on the size of the problem ($m$ and $n$) for linear and semidefinite optimization compared to classical algorithms, but the remaining dependence on some numerical parameters (e.g., a condition number bound for the Newton systems) makes it unclear if an overall speedup is achieved. 

Very recently, a quantum speedup for interior point methods was achieved without introducing a dependence on a condition number~\cite{apers2023quantum}. Under some mild assumptions and access to QRAM, this IPM achieves a quantum speedup for linear programs in which the number of linear inequalities $m$ is much larger than the number of variables $n$, i.e.~LPs of the form $\max c^T x$ s.t.~$Ax \geq b$ where $A \in \R^{m \times n}$, $b \in \R^m$ and $c \in \R^n$ with $m \gg n$.  In particular, the best classical solver achieves a scaling of $\widetilde{\mathcal{O}}(mn + n^{2.5})$~\cite{van2021minimum}, whilst the quantum analogue scales as $\sqrt{m} \cdot \poly (n, \log(1/\epsilon))$. The key new subroutine here is a faster quantum algorithm to compute spectral approximations of matrices of the form $B^T B$, provided query access to a tall matrix $B \in \R^{m \times n}$. A spectral approximation of such matrices suffices to speedup the costly Newton step. This subroutine can be seen as a generalization of a quantum graph sparsifier (by letting $B$ be the edge-vertex adjacency matrix)~\cite{apers2022quantum}, which in turn has been used in optimization algorithms in both the continuous~\cite{qscaling2STACS} and discrete settings (e.g.~cut approximations)~\cite{apers2022quantum}. 

\textbf{Quantum SDP-solvers for approximating MAXCUT.}  
Given the challenges in evidencing end-to-end speedup for general SDPs posed by the scale invariant error parameter $Rr/\epsilon$, the MWU framework has been specialized to the semidefinite approximation of MAXCUT:
\begin{maxi}|s|[0]
    {X \succeq 0}{\tr(C X)\hspace{1.75cm}}{}{}\label{eq:maxcutsdp}
    \addConstraint{}{X_{ii} = 1, \, \forall i \in [n].}
\end{maxi}
Brandao \emph{et al.}~\cite{Brandao_2022} demonstrated that, after normalizing the diagonal constraints by $n$ and relaxing slightly the constraints, feasibility reduces the task of solving the MAXCUT SDP to a possibly simpler task: preparing a Gibbs state that is (i) approximately indistinguishable from the maximally mixed state when measured in the computational basis, and (ii) whose trace inner product with $C$ is close to an estimate of the optimal objective value, determined via binary search. This enables  the design of specialized oracles for testing feasibility within the MWU scheme, and leads to an algorithm that, provided access to QRAM, outperforms previous algorithms in the problem dimension and sparsity of $C$. However, the algorithm exhibits a poor $\mathcal{O} (\epsilon^{-28})$ dependence on precision, prohibiting an overall speedup.

The impractical error scaling was addressed by Augustino \emph{et al.}~\cite{augustino2023solving}, where iterative refinement techniques are used to exponentially improve the dependence on the inverse precision, for both the quantum and classical algorithms proposed by Brandao \emph{et al.}~\cite{Brandao_2022}. Moreover, it is shown that when $C$ is stored in QRAM, one can remove the dependence on the sparsity parameter by using Gibbs sampling techniques based on Quantum Singular Value Transformation \cite{gilyen2019quantum}. This yields a quantum algorithm that runs in time $\widetilde{\mathcal{O}}\left(n^{1.5}\right)$, plus the time it takes to read $C$, obtaining a polynomial speedup over the best known classical algorithm. Due to the lack of a classical lower bound for the problem, however, it is possible that better (and faster) classical algorithms for this problem could be developed. Note that the speedup in Augustino \emph{et al.}~\cite{augustino2023solving} is reliant on QRAM: the authors analyze their algorithm in the standard gate model and show that the resulting algorithm is outperformed by classical approaches.

\textbf{Quantum algorithms for matrix scaling and matrix balancing.}
Aside from the SDP relaxation of MAXCUT, another convex optimization problem that has enjoyed a provable quantum speedup for first- and second order methods is the matrix scaling problem. One version of the matrix scaling problem can be stated as follows: given an entrywise non-negative matrix, find positive weights for the rows and columns such that the reweighted matrix becomes doubly stochastic. This problem has many applications, both in theory (e.g.,~approximating the permanent~\cite{lsw00}) and in practice (used by default as a preconditioner in LAPACK~\cite{lapack}). Recent works~\cite{qscalingICALP,qscaling2STACS} have shown how to speedup both first- and second-order methods for a natural convex formulation of the problem, using a variety of techniques (basic amplitude amplification, but also graph sparsification). Importantly, these quantum algorithms do not introduce a dependence on a condition number. 

\textbf{Quantum approaches with no direct classical equivalent.} 
Some classical continuous optimization algorithms can be phrased in terms of the solution of a certain dynamical system, i.e., ODEs. These ODEs can then be cast as the Schr\"odinger equation, and solved by means of a quantum algorithm, thereby solving the optimization problem. The first work in this line of research is Quantum Hamiltonian Descent (QHD) \cite{leng2023quantum}, which can be viewed as a quantization of the Bregman-Lagrangian framework for continuous-time accelerated gradient descent introduced by Wibisono \emph{et al.}~\cite{wibisono2016variational}. By properly designing a family of Hamiltonians that, as time progresses, lead to the optimal solution of the dynamical system, quantum Hamiltonian descent can be shown to converge to the optimal solution even on certain non-convex problems (although it may not do so efficiently, as these problems are provably hard). With regard to output, QHD returns a classical description of the solution without the need for state tomography. The final state is a probability distribution concentrated near the global minimizer of the objective function, and thus a bitstring corresponding to a solution is obtained upon sampling from the QHD final state.

Along this line, Augustino \emph{et al.}~\cite{augustino2023centralPath} describe a fully quantum algorithm for solving linear optimization problems by quantum-mechanical simulation of the central path, which the authors call the Quantum Central Path Method (QCPM). While IPMs and QIPMs approximately track the central path using successive linearizations of the perturbed KKT conditions (i.e., Newton steps), the QCPM consists of a single simulation that works directly with the nonlinear complementarity equations. This is achieved by designing a Hamiltonian which encodes the central path in its ground state. Like QHD, no intermediate measurement is required, and a classical description of the solution is obtained upon sampling from the final state. This approach is faster than any classical or quantum IPM in a certain sparsity/condition number regime, while also avoiding the use of QRAM, block-encodings, quantum linear system algorithms, and state tomography.

\subsubsection{Non-Convex Optimization}\label{sec:non-convex}

In non-convex optimization, one minimizes a function:
\begin{mini}
    {x \in \mathbb{R}^n}{ f(x) }{}{}\label{eq:non-convex}
\end{mini}
in dimension $n \in \mathbb{N}$.
Many variants are undecidable \cite{liberti2019undecidability}, as long as one allows for periodic functions $f$, or sufficiently non-smooth functions. (As suggested in Tab.~\ref{tab:optclassical}, some are inapproximable in even in the Banach-Mazur model~\cite{lee2023computability}.) One can hence hope to study $k$-th order critical points of non-convex functions, in general. In smooth non-convex optimization problems, this means points where necessary conditions of optimality involving first $k$ partial derivatives are satisfied. In non-smooth non-convex optimization problems, this often involves necessary conditions of optimality involving the Clarke subdifferential \cite{Clarke1990,Li2020Clarke}. 

Echoing the situation in convex optimization \cite{garg2021no}, 
there are clear limits as to the speedup in non-convex optimization \cite{zhang2023quantum}.
In particular, when first $k$ partial derivatives are available via an oracle or when one has access to stochastic gradients (SG), and no further assumptions are made, classical lower bounds apply to quantum algorithms too. One hence wishes to find suitable special cases, such as by considering the numbers of local minima, local behavior of the functions around those, and behavior of the trajectories connecting nearby local minima (``the barriers''). 

Consider the problem of finding an $\epsilon$-stationary point $x\in \mathbb R^n$ of a non-convex function $f$, which means that
\begin{equation}
\Vert \nabla f(x) \Vert \leq \epsilon.
\end{equation}
For $x\in \mathbb R^n$, consider the set of $k$-order derivatives $\nabla^{(0,\cdots,k)}f(x) := \{ f(x), \nabla f(x), \cdots \nabla^k f(x)\}$ with universal Lipschitz  constants, and consider also the stochastic gradient $g(x, \xi)$, with $\xi$ a random seed and satisfying $\mathbb E_\xi [g(x,\xi)] = \nabla f(x)$ and $\mathbb E_\xi \Vert g(x,\xi) - \nabla f(x)\Vert \leq 1$.
Assuming quantum access $O_f^{(k)}\ket {x} \ket {y} = \ket {x} \ket { y \oplus \nabla^{(0,\cdots,k)}f(x)} $, which outputs a binary representation of $\nabla^{(0,\cdots,k)}f(x)$,  Zhang and Li~\cite{zhang2023quantum} prove the lower bound $\Omega(\epsilon^{-(k+1)/k})$.
Assuming quantum access $O_g \ket x \ket \xi \ket y = \ket x \ket \xi \ket{y \oplus g(x, \xi)}$, produces a lower bound of $\Omega (\epsilon^{-4})$.
Both query lower bounds are the same as the classical case.
On the other hand, Sidford and Zhang~\cite{sidford2023quantum} show query upper bounds for finding critical points with quantum algorithms for stochastic gradient descent (SGD). With access to a probability-weighted superposition over the bounded-variance stochastic gradient, the algorithm requires
$O(\sqrt {n} \epsilon^{-3})$ queries, which is only better than the classical algorithm  $O(\epsilon^{-4})$ when the dimension is $<\epsilon^{-2}$. This can be improved with access to mean-squared smooth SGs, from $O(\epsilon^{-3})$ to $O(\sqrt n \epsilon^{-5/2})$. 
The run time for all of these algorithms both classically and quantumly will in most cases be polynomial in $n$.

In high-dimensional situations, the availability of the quantum linear systems algorithm and of quantum linear-algebra methods motivates the direct quantum implementation of gradient descent over $O(\text{polylog}(n))$ qubits 
\cite{rebentrost2019quantum, Kerenidis2020gradient} to provide descriptions of stationary points of non-convex functions. With a non-convex differentiable function $f(x) :\mathbb R^n \to \mathbb R$ and an initial point $x^0 \in \mathbb R^n$, consider the gradient descent update $x \to x - \eta \nabla f(x)$ with $\eta>0$. For $T\in \mathbb N$, the task is to output a quantum state proportional to $x^T$, where the update is applied $T$ times starting from $x^0$.
To mimic a classical gradient descent step for fixed $t \in [T]$, define the $\log(n)$-qubit states $\ket{x^t} := \tfrac{1}{\Vert x^t\Vert_2} \sum_{j=1}^n x^t_j \ket j$ and $ \ket{\nabla^t} := \tfrac{1}{\Vert \nabla f(x^t)\Vert_2} \sum_{j=1}^n \nabla f(x^t)_j \ket j$. 
Let us be given a circuit to prepare $ \alpha^t\ket 0_1 \ket{x^t}  + \ket {\phi^t}$, where $\ket {\phi^t}$ is some non-normalized orthogonal state and $\vert \alpha^t\vert \leq 1$, and a circuit $U_\nabla$ such that $U_\nabla \ket{x^t} =  \ket{\nabla^t}$. 
Using an auxiliary qubit and Hadamard gates, the state 
$\alpha^{t+1} \ket {00}_{12} \ket{x^{t+1}} + \ket {\phi^{t+1}}$ can be prepared, where 
\begin{align}
\alpha^{t+1} := \alpha^t \tfrac{\Vert x^t - \eta \nabla f(x^t)\Vert_2}{\sqrt 2 \sqrt{\Vert x^t \Vert_2^2 + \Vert \nabla f(x^t)\Vert_2^2} }
\end{align}
and $\ket {\phi^{t+1}}$ is an orthogonal non-normalized state. 
Hence, the $\ket {00}_{12}$ part of the state proportionally encodes the desired gradient step and we may continue with the next step. 
In this setting, Rebentrost \emph{et al.}~\cite{rebentrost2019quantum} consider certain classes of multivariate polynomials, and includes Newton steps.
While the algorithm may admit a $\text{polylog} (n)$ complexity, 
amplifying the $\ket{\bar 0}$ state at the end of the gradient procedure can cost $2^T$, since 
at every step $\vert \alpha^{t+1} \vert \leq \vert \alpha^{t} \vert/\sqrt{2}$.
An efficient circuit $U_\nabla$ may be constructed only for special cases \cite{rebentrost2019quantum,Li2021polynomial},
or could be approximated via pre-training a parameterized quantum circuit. 
For affine gradients corresponding to quadratic optimization problems, Kerenidis and Prakash~\cite{Kerenidis2020gradient} show an efficient implementation using phase estimation.

The idea of simulated annealing, finding global minima assisted by thermal fluctuations, applies to continuous non-convex optimization as well. 
Quantum computers obtain polynomial speedups in executing the classical simulated annealing process via quantum walks \cite{somma2008quantum}.
Quantum fluctuations provide an additional resource allowing quantum tunneling through barriers between different minima \cite{morita2008mathematical}.  Recently, Liu \emph{et al.}~\cite{liu2023quantum} investigated quantum tunneling for SGD, where the SGD process is approximated by a continuous-time stochastic differential equation. Mastropietro \emph{et al.}~\cite{mastropietro2023flemingviot} investigated restarting the search, when one arrives in so-called barren plateau, which allows for speed-up when there are many or large barren regions.
These quantum walks can provide a speedup over classical SGD when the barriers between different local minima are high but thin and the minima are flat.

There are several promising directions 
and open questions.
What are lower bounds for various settings of quantum polynomial optimization?
Beyond lower bounds, more in-depth focus on possible run time advantages of quantum algorithms is required. 
One direction is about the quantum speedup could one obtain by utilizing moment/sum-of-squares approaches to polynomial optimization and quantum algorithms for SDPs, in analogy to mixed-integer programming discussed below. 
Extending to larger classes of polynomials and beyond, what classes of non-convex functions would allow for amplitude-encoded embeddings, efficient quantum circuits for gradient computations, and converging gradient procedures.

\subsection{Mixed-Integer Programming}
\label{sec:milp-miqp}

Combining discrete decision variables and continuous decision variables,
Mixed-Integer Programs (MIP) are non-convex and at least as hard as either
of the discrete optimization (cf.~Sec.~\ref{sec:discrete_optimization} above)
and continuous optimization (cf.~Sec.~ \ref{sec:continuous_optimization} above) 
problems they subsume.  
We differentiate between Mixed Integer Linear Programs (MILP), usually used synonymously to MIP, and its generalization Mixed Integer Quadratic Programs (MIQP).
In practice, some instances with unbounded variables are notoriously hard \cite{Bomze2019}.
Having said that, in the case of mixed-integer linear programming, 
especially when restricted to binary variables, 
classical solvers based on branch-and-bound-and-cut have made an astonishing progress over the past three decades, often solving 
instances of up to one million binary variables to proven optimality, cf. Fig. \ref{fig:gurobi_scaling}, 
and sometimes solving structured problems with tens of millions of binary variables \cite{van2018large} 
to proven optimality.  

Following Eq.~\eqref{eq:qubo}, let us consider the following illustrative MIQP:
\begin{mini}|s|[0]
    {x\in \{0, 1\}^{n_b}, y \in \mathbb{R}^{n_c}}{ x^T Q x + y^T R y \hspace{0.25cm}}{}{}\label{eq:milcqp}
    \addConstraint{}{A x = a}
    \addConstraint{}{B y = b}
    \addConstraint{}{C x \leq c}
    \addConstraint{}{D y \leq d}
\end{mini}
where $n_b,n_c \in \mathbb{N}$ denote the numbers of binary and continuous variables, respectively,
$Q \in \mathbb{R}^{n_b \times n_b}$ and $R \in \mathbb{R}^{n_c \times n_c}$ are cost matrices,
and matrices 
$A \in \mathbb{R}^{m_{e} \times n_b}$, 
$B \in \mathbb{R}^{m_{e} \times n_c}$, 
$C \in \mathbb{R}^{m_{i} \times n_b}$, 
$D \in \mathbb{R}^{m_{i} \times n_c}$, 
and vectors 
$a \in \mathbb{R}^{m_e}$, 
$b \in \mathbb{R}^{m_e}$, 
$c \in \mathbb{R}^{m_i}$, 
$d \in \mathbb{R}^{m_i}$, 
define $2 m_e$ equality constraints and
$2 m_i$ inequality constraints.

There are several potential algorithms to approach such a problem with a quantum computer, whose short overview is provided in Tab.~\ref{tab:mio}.
In more detail:

\textbf{Branch-and-Bound-and-Cut:} the main workhorse in classical optimization. One possibility for a hybrid implementation is that discrete decisions are handled by branching and generation of cuts implemented classically, while the continuous, possibly convex relaxations could be solved on the quantum computer. Considering the progress in quantum algorithms for convex optimization (cf.~Sec.~\ref{sec:convex_optimization}), this seems plausible, although distant. Perhaps even more ambitiously, one could determine the branching decisions on the quantum computer as well, which would yield an additional quadratic speedup \cite{montanaro2015quantum,ambainis2017quantum,montanaro2020quantum,chakrabarti2022universal} over the speedup of quantum algorithms for convex optimization, while increasing the requirements on the numbers of qubits and fault tolerance substantially. We note that the earlier papers \cite{montanaro2015quantum,montanaro2020quantum} consider the speedup in the case of locating \emph{all} optima (measured against the classical cost of finding all optima), whereas Refs.~\cite{ambainis2017quantum,chakrabarti2022universal} consider the speedup of obtaining one optimum (out of possibly multiple ones), measured against the classical cost of finding one optimum. Recent research by Dalzell \emph{et al.}~\cite{dalzell2023mind} suggests that super-quadratic speedups may be possible in some cases. 

\textbf{Decomposition/Splitting/ADMM:}
this is a large class of algorithms that is based on decomposing the problem. Refs.~\cite{gambella2020mixed,gambella2020multiblock} suggested solving the QUBO sub-problem using a quantum computer, while solving the continuous sub-problem classically. This approach is based on the hope that a practical quantum speedup for some type of QUBO can be proven. Under some restrictive conditions \cite{gambella2020multiblock}, such an iterative approach can be shown to be globally convergent. A similar approach has been tested by Chang \emph{et al.}~\cite{chang2020hybrid}, in the context of Benders decomposition.

\textbf{Reformulation to an unconstrained optimization problem:} 
Braine \emph{et al.}~\cite{braine2021quantum} experimented with a decomposition minimizing values of slack variables used to convert inequality constraints to equality constraints, and a Lagrangian relaxation of the resulting equality-constrained continuous-valued problem. 
Another, approach that optimizes over the Lagrangian function classically over the circuit parameters and dual variables has been introduced by Le and Kekatos~\cite{le2023variational}.
Such Lagrangian methods are well understood \cite{bertsekas2014constrained,birgin2014practical}, but global-convergence results are limited by  very restrictive conditions, in general, or require some form of backtracking. One could also solve the problem approximately by utilizing a test done by Aspman \emph{et al.}~\cite{aspman2023hybrid} for the active set remaining fixed subsequently, and only then utilize the Lagrangian relaxation. This could allow for easier global convergence proofs without backtracking. It should also be noted that there are multiple options for working with both the slack variable, penalties for its non-negativity \cite{kovcvara2003pennon}, as well as multiple options for augmenting the Lagrangian with quadratic or higher-order terms in order to improve numerical performance. These are to be explored, yet. 

\textbf{Reformulation to high-dimensional LPs:} 
using techniques that go back to Dantzig and Wolfe \cite{dantzig1960decomposition}, one can reformulate mixed-integer programs as in Eq.~\eqref{eq:milcqp} as a very large linear program (LP). While quantum algorithms for LPs have been somewhat overshadowed by quantum algorithms for SDPs so far (cf. Sec.~\ref{sec:convex_optimization}), a continuous development of quantum algorithms for LPs may render this a viable avenue. 

\textbf{Reformulation to high-dimensional SDPs:} 
using moment/sum of squares (SOS) techniques \cite{parrilo2000structured,lasserre2001global}, one can reformulate mixed-integer programs as in Eq.~\eqref{eq:milcqp} to an SDP in a much higher dimension than $n + d$. 
Indeed, there are instances where this dimension needs to be exponential in $n + d$~\cite{schoenebeck2008linear}. 
If quantum algorithms for SDPs (cf. Sec.~\ref{sec:convex_optimization}) made it possible to solve such instances
independent of the dimension, it may offer an interesting avenue for quantum algorithms for mixed-integer programming.

\textbf{Reformulation to a completely positive problem:} a mathematically elegant reformulation of a mixed-integer programming problem results in a linear program over the dual of the cone of copositive matrices~\cite{burer2009copositive}.
Recall that copositive matrices are real symmetric matrices $Q \in \mathbb{R}^{n \times n}$, corresponding to quadratic forms $x^T Q x$ non-negative on the positive orthant, i.e., for $x$ element-wise non-negative.
For a fixed $n$, these matrices form a convex cone~\cite{dickinson2011geometry}, but even the test of membership in this cone is \textbf{NP}-hard~\cite{dickinson2014computational}. These cones have recently attracted attention in mathematical optimization \cite{bomze2012copositive}. A possible way to exploit this reformulation in the context of a hybrid solver is to use a quantum computer to solve the difficult cut generation problem, whereas the classical computer solves a convex problem. This approach has been tried in \cite{brown2023copositive}, using a QUBO solver to generate cuts. We note that to prove optimality one eventually needs to show that no cut exists, and this in turn requires an exact solver for the QUBO subproblem. Solving the cut generation problem exactly can be difficult even for classical solvers, see \cite{guo2021copositive} for a discussion of such an approach in the context of power systems.

\begin{table*}[t!]
\caption{ A short history of proposed quantum algorithms for mixed-integer optimization. For approaches without available guarantees, we write \emph{N/A}, although this should not be construed as ruling out guarantees. }\label{tab:mio} 
\centering
\begin{tabularx}{\textwidth}{@{\extracolsep{\fill}}llll}
\hline
Algorithm  & Guarantees & Year & Ref. \\
\hline
ADMM & Asymptotic convergence under restrictive conditions. Speedup of QUBO? & 2020 & \cite{gambella2020mixed,gambella2020multiblock} \\ 
Benders decomposition & N/A & 2020 & \cite{chang2020hybrid} \\
Branch-and-bound & Quadratic speedup + speedup of convex opt. & 2020 & \cite{ambainis2017quantum,montanaro2020quantum}\\ 
Lagrangian methods & N/A & 2021 & \cite{braine2021quantum,le2023variational}   \\
Branch-and-bound & Quadratic speedup + speedup of convex opt. & 2022 & \cite{chakrabarti2022universal} \\
``Dantzig-Wolfe'' & Finite convergence, speedup of convex opt. & 2023 & Here  \\ 
Moment/SOS & Asymptotic convergence, speedup by SDPs & 2023 & Here \\ 
Subadditive duals & Asymptotic convergence & 2023 & Here\\  
Completely-positive &  N/A & 2023 & Here \\ 
\hline
\end{tabularx}
\end{table*}

Next, we review multiple proposals for quantum algorithms for MIP and propose a list of new directions. There are still many open questions to be answered, however, in order to achieve a quantum advantage, which we express in the remainder of this section. 

First, can we provide general statements on provable quantum advantage for MIP? Is there a general meta-theorem for combined speedup of solving a convex relaxation and branch-and-bound?   
What would be the best possible speedup for branch-and-bound? It seems plausible that it could allow for super-quadratic speedup for \textbf{NP}-hard problems quite generically, a tantalizing prospect. Quartic speedup has been shown, so far, only for learning majorities \cite{belovs2015quantum} and the quantum counterfeit coin problem \cite{iwama2012quantum}.
Further, how could the Lagrangian approaches benefit from an approximate solution and the test from Aspman \emph{et al.}~\cite{aspman2023hybrid} for the active set remaining fixed subsequently? How does one make design choices within Lagrangian approaches (penalties for non-negativity of slack variables, augmenting the Lagrangian with quadratic or higher-order terms) optimally?  And out of the reformulation approaches sketched out above (Dantzig-Wolfe, Moment/SOS, Co(mpletely)-positive), which ones would allow for the best speedup in the continuous optimization (cf.~Sec.~\ref{sec:continuous_optimization} above)? This entails bounding the ratio of the inscribed to outscribed balls of the reformulations within continuous optimization. 

Given the practical importance of MIP, it is crucial to advance our understanding of how to approach this problem class with quantum computers. Most of the proposals in the literature consider a quantum sub-routine inside a known classical algorithm. However, it is unclear what advantage the sub-routine needs to provide to imply an overall advantage of the optimization algorithm for MIP. This could be a combination of speed, quality, but also diversity of solutions in order to accelerate branching. Further, there might opportunities to have more quantum-native algorithms. In any case, it is important to further investigate this problem class and systematically benchmark the performance of available proposals.

\subsection{Dynamic Programming}
\label{sec:dynamic_programming}

Dynamic Programming (DP) is a very generic mathematical optimization method that breaks down large optimization problems into smaller ones and combines results for the overall problem in a recursive fashion.

\begin{table*}[t!]
\caption{Most important provable quantum speedups for dynamic programming, where the notation $\mathcal{O}^*(f(n))$ hides a polynomial factor in $n$ \cite{ABIKPV19}.}
\label{table:dp}
\begin{tabularx}{\textwidth}{@{\extracolsep{\fill}}lllll}
\hline
Problem  & Quantum Algorithm & Best Classical Algorithm & Year & Ref. \\ \hline
Generic Vertex Ordering Problem & $O(1.817...^n)$ & $O^*(2^n)$ & 2018 & \cite{ABIKPV19} \\ 
Traveling Salesperson Problem & $O(1.728...^n)$ & $O^*(2^n)$ & 2018 & \cite{ABIKPV19} \\ 
Minimum Vertex Cover & $O(\text{poly}(m, n) 1.728...^n)$ & $O(nm2^n)$ & 2018 & \cite{ABIKPV19} \\ 
Graph Coloring & $O(1.7956...^n)$ & $O(2^n)$ & 2023 & \cite{gaspers2023quantum}  \\ 
Minimum Steiner tree & $O(1.812...^k \text{poly}(n))$ & $O(2^k \text{poly}(n))$ & 2020 & \cite{miyamoto2020quantum} \\
Treewidth & $O(1.538...^n)$ & $O(1.755...^n)$ & 2022 & \cite{kpv22} \\
\hline
\end{tabularx}
\end{table*}

A dynamic programming algorithm consists of repeatedly optimizing the 
Bellman equation
\begin{equation}
\label{eq:bellman}
V(x) = \min_{a} T(x, a) + V(x_a) 
\end{equation}
where $a$ are the actions that are available in the state $x$ and $x_a=F(x, a)$ is the state to which we get by applying $a$ in the state $x$. 
The entries $V(x)$ can be arranged into a dynamic programming table. A dynamic programming algorithm then computes all the entries $V(x)$ in this table. The order in which the entries are computed is chosen so that, whenever the algorithm has to compute $V(x)$, all $V(x_a)$ have already been computed and $V(x)$ can be computed from Eq.~\eqref{eq:bellman}. For example, in the case of TSP \cite{bellman1962dynamic}, $x$ are subsets of the set of city and the entry $V(x)$ is the length of the shortest path through $x$. Thus, the DP algorithm for TSP consists of computing the shortest paths through every subset of cities.

The complexity of a DP algorithm is $O(Sm)$ where $S$ is the number of possible states $x$ for which we have to optimize the Bellman equation and $m$ is the maximum number of actions in one state. In the case of TSP, this is $\widetilde{O}(2^n)$.
DP can be used to address many $\NP$-complete problems, some of which are listed in Tab.~\ref{table:dp}. For example, the best algorithm with provable running time for TSP is based on DP \cite{bellman1962dynamic}.
It is also often used as a sub-routine in approximation algorithms (e.g., some PTAS), and
it can be used to solve Markov Decision Problems (MDPs), as discussed in more detail in Sec.~\ref{sec:optimal_control}.

Quantum speedups are known for a substantial number of classical DP algorithms. The most important examples are shown in Tab.~\ref{table:dp}. The quantum algorithms are a combination of classical dynamic programming and Grover's search. For example, the quantum algorithm for TSP by Ambainis \emph{et al.}~\cite{ABIKPV19} first uses DP to compute the shortest paths through sets of cities that contain up to 24\% of all cities and then uses Grover's search to find the shortest combination of those paths that visits all the cities and returns to the starting point. Similar ideas can be applied to other problems in which the classical algorithm uses an exponentially large DP table indexed by subsets of an $n$-element set. This includes graph coloring \cite{gaspers2023quantum}, minimum Steiner tree \cite{sm20,miyamoto2020quantum}, tree-width \cite{kpv22} and other problems. The problems amenable to this approach are typically vertex ordering problems (in which the task is to find the optimal ordering of vertices in a graph, as in TSP) or set partitioning problems (in which the task is to find the optimal partition of a set, as in Set Cover or Graph Coloring). Such problems lead to DP tables in the form of Boolean hypercube ($2^n$ entries indexed by $x\in\{0, 1\}^n$) which can often be handled by computing part of the table and then using Grover's search to find the best solution.

One can formulate a generic DP problem with the DP table in the form of hypercube and show that, in the generic case, the quantum speedup can be at most quadratic \cite{ABIKPV19}. It is an open problem whether quantum speedups can be obtained for algorithms with a DP table of a different structure. For example, the DP algorithm for the Edit Distance problem \cite{wagner1974string} (in which one is given two strings and has to find the smallest number of symbol insertions/deletions/replacements to transform the first string into the the second string) uses a DP table that is a two-dimensional array. If the size of the array is $n\times n$, the classical DP algorithm takes $O(n^2)$ steps. Quantumly, no quantum speedup is known but an $\widetilde{\Omega}(n^{1.5})$ quantum lower bound is known for the generic case \cite{AmbainisBIKKPSSV23,buhrman2021framework}.

The broad applicability of DP renders it a very interesting approach. However, many open questions remain in terms of a potential quantum advantage. What structures of state space (besides hypercube) allow general quantum DP algorithms? 
Can we add stochasticity to quantum DP algorithms or extend them to MDPs (cf.~Sec.~\ref{sec:optimal_control})?
What about the explosion of state space?
Can we prove better lower bounds for quantum speedups of generic DP algorithms? 
In addition to provable algorithms, can we find good quantum approximations or heuristics, such as reinforcement learning?
Answering any of these questions will help to further understand the potential and practical applicability of quantum DP algorithms and some of them will also be discussed in the following Sec.~\ref{sec:optimal_control}.

\subsection{Optimal Control}
\label{sec:optimal_control}

Optimization with differential equations as constraints is a subject of optimal control theory. A standard finite horizon optimal control problem takes the following form: find a control function $u$ with values in a given set $\mathcal{U}$ solving the following optimization problem 
\begin{mini}|s|[0]
    {u(t)\in \mathcal{U}}{ \int_0^T \ell(x(t), u(t)) d t+h(x(T)) }{}{}
    \addConstraint{\dot{x}=f(x, u), x(t)\in\mathbb{R}^n, \quad t \in[0, T)}{}
\end{mini}
for some \emph{nice} functions $\ell, f, h$, e.g., see Ref.~\cite{bardi}. A globally optimal solution of this problem, the so-called feedback optimal control policy $u^\star(x,t)$ can be found by introducing a value function 
\begin{align}
    V(z,t)=\min_{u(t)\in \mathcal{U}} \biggl\{\int_t^T &\ell(x(s), u(s)) ds+h(x(T)), \nonumber \\
    &\dot{x}=f(x, u), \, x(t)=z \biggr\} \, .
\end{align}

Generalizing classical dynamic programming ideas one can demonstrate that $V$ solves the following Hamilton-Jacobi-Bellman (HJB) equation backwards in time: 
\begin{equation}\label{eq:HJB}
\frac{\partial V}{\partial t}+\min_{u \in \mathcal{U}}\left\{\ell(x, u)+f^\top (x,u)\nabla_x V(x, t)\right\}=0\, ,
\end{equation}
subject to the \emph{terminal condition} $V(x, T)=h(x)$, for all $x \in \mathbb{R}^n$. Given the solution of Eq.~\eqref{eq:HJB} one computes the optimal control policy in feedback form as follows: 
\begin{align}
u^{*}(t,x)=\underset{u\in\mathcal U}{\operatorname{argmin}}\left\{\ell(x, u)+f^\top(x,u)\nabla_x V(x, t)\right\} \, .
\end{align}
It should be noted that in many practical situations, $\nabla_x V$ does not exist in the classical sense. For this reason the solution of HJB Eq.~\eqref{eq:HJB} is understood in the viscosity sense: a viscous term represented by $\varepsilon\Delta V$, $\varepsilon>0$ is introduced in the right-hand-side of Eq.~\eqref{eq:HJB} so that the resulting solution $V_\varepsilon$ becomes differentiable, and then taking the limit of $V_\varepsilon$ as $\varepsilon\downarrow0$ one gets the viscosity solution of Eq.~\eqref{eq:HJB}. This construction relates $V_\varepsilon$ to the stochastic optimal control~\cite{bardi} as the Laplacian $\varepsilon\Delta V$ can be seen as adding "white noise" to the dynamics of $x$ of variance $\sqrt{\varepsilon}$. 

In addition to optimal feedback control design HJB has many applications: for example, it allows one to solve the nonlinear filtering / state estimation problem by constructing reachability sets of nonlinear dynamical systems~\cite{kurzh}.
The most relevant applications of HJB equations in the context of this paper are in discrete optimization: HJB equation for discrete-time dynamics can be seen as the famous Bellman equation of dynamic programming: 
\begin{equation}\label{eq:hjb-discrete}
    V(t,x) = \min_{u\in\mathcal{U}} \{\ell(x,u) +  V(t+1, x+f(x,u))\}
\end{equation}
with final time condition $V(T,x)=h(x)$. Here time $t$ is discrete, and position $x(t)$ is assumed to take only finitely many values (a so-called piece-wise constant approximation), so $x(t)$ can be seen as an index taking values in the set of integers $\mathcal X$~\cite{bardi} and hence $V(t,x)$ can be interpreted as a stack of \emph{dynamic programming tableaus} indexed by $(t,x)$ (cf. Sec.~\ref{sec:dynamic_programming}).  

Yet another important application is in optimization for discrete stochastic processes: as was mentioned above, continuous time viscosity solution $V$ of Eq.~\eqref{eq:HJB} is connected to stochastic optimal control for Markov diffusion processes via regularization. Hence it is not surprising that the discrete-time HJB in Eq.~\eqref{eq:hjb-discrete} could be connected to discrete stochastic processes too: namely, if instead of deterministic transitions $x(t+1)=x(t) + f(x(t),u(t))$, which are singular in terms of probability distributions, one introduces Markov Decision Process (MDPs)~\cite{puterman2014markov}, which jumps from the state $x(t)=x$ to state $x'=x(t+1)$ with known probability $P(x'|x,u)$ provided the action $u$ was applied, then Eq.~\eqref{eq:hjb-discrete} transforms into the famous backwards induction for episodic MDPs: 
\begin{align}
V(t,x) = \min_{u\in\mathcal{U}} \left\{\ell(x, u)+\sum_{x'\in \mathcal{X}} P(x'|x,u) V(t+1, x'))\right\}, 
\end{align}
with a final time condition $V(T,x)=h(x)$,~\cite{puterman2014markov}. MDPs are central object of the Reinforcement learning (RL), a family of algorithms in which an agent aims to learn optimal decisions in unknown environments through the experience of taking actions and observing the rewards gained. In some cases, the environment is not influenced by the actions of the RL agent, in which case $P$ is independent of action $u$ then MDP turns into the uncontrolled MDP~\cite{Calafiore2007,epperlein2022reinforcement}, and if the transition matrix $P(x'|x)$ is of rank 1 such an MDP becomes the so-called contextual multi-armed bandit and in this case finding $V$ amounts to solving the optimization problem $\min_{u\in\mathcal{U}} \ell(x, u)$ for each $x$.  

The key problem associated with HJB equation, continuous or discrete, is the \emph{curse of dimensionality}. Solving HJB in Eq.~\eqref{eq:HJB} by conventional methods of numerical analysis quickly becomes intractable even in rather modest dimensions, e.g., if $x(t)\in \R^n$ for $n\ge 10$. In other words, direct approach to solving HJB equation has slim chances of succeeding. In what follows we discuss potential quantum optimization approaches to find global solutions to the three aforementioned problems: 
\begin{enumerate}
    \item Contextual multi-armed bandit problems.
    \item Markov decision processes (MDPs).
    \item Optimization problems constrained by differential equations.
\end{enumerate}

Finding ways of solving the above problems will in turn suggest ways of solving the corresponding HJB equations, continuous or discrete. In fact, solving the discrete HJB in Eq.~\eqref{eq:hjb-discrete}  sheds some light back on the solution of the continuous one and vice-versa, suggesting that algorithms developed in the discrete setting can help construct solutions of continuous optimization problems. 
Potential algorithms include:
\begin{enumerate}
    \item Value iteration and policy iteration boil down to optimization problems, wherein the convex optimization of Sec.~\ref{sec:convex_optimization} could
    be applicable, perhaps combined with the quadratic speedup of approaches of Sec.~\ref{sec:dynamic_programming}. 
    \item Reinforcement learning is a broad family of approaches, rather than a single algorithm, which approximate value or policy iteration, 
    for instance using Least-Squares Temporal-Difference (LSTD). Again, the convex optimization of Sec.~\ref{sec:convex_optimization} could
    be applicable, perhaps together with the tabular method of Sec.~\ref{sec:dynamic_programming} providing additional speedup. 
    \item Quantum reinforcement learning (heuristics). 
    In the unitary oracle setting, from the perspective of exact methods, that is, algorithms that are guaranteed to identify optimal policies in arbitrary environment,  tight quantum upper and lower bounds have been identified. 
    In Refs.~\cite{Wang2021a, Wang2021b}, lower bounds on the more special task of Multi-armed bandit problems were found, along with essentially matching quantum upper bounds, for a number of RL-related tasks, including finding the optimal q-function. These works achieve a quadratic speedup over best possible classical methods, assuming appropriate oracle access to the task environment.
\end{enumerate}

We can break the above into two main classes of approaches, differing in whether the learning agent has a standard classical access to the environment, or if the interaction itself is quantum. In the latter line, numerous types of quantum oracular access to the task environment have been explored, e.g., where coherent access to a unitary encoding the transition function is possible \cite{Cornelissen18}, or where the environment closely mimics classical RL settings, and allows only actions as inputs and has memory \cite{Dunjko16}. 

We note that the algorithms listed above have significant limitations, including the fact that they require fault tolerance, and that quadratic speedups may not translate to wall-clock speedups in early fault tolerant quantum computers. The main conceptual limitation is the requirement to a coherent access to the task environment, which could only be possible in very restricted or constructed cases, and the fact they deal with exact solutions, and in real-world problems, the search spaces are simply too large, such as the state space of all 1 megapixel images.
The up-side of the above results is that the speedups concern the number of interactions with the environment, and are not contingent on any assumptions in complexity theory.

In practice, one often foregoes optimal policies for optima within a parameterized policy family, as is found by policy iteration, and policy gradient algorithms. Although the method cannot find an optimal policy in general, one can still be interested in finding optima with respect to the function approximation family available. In Jerbi \emph{et al.}~\cite{Jerbi2023} it was shown that quadratic speedups in policy optimization were possible as long as the policies satisfy certain regularity conditions, which are, e.g., satisfied for policies defined by parameterized quantum circuits. 
The lower bounds for this problem have been established for a more general setting by Cornelissen \emph{et al.}~\cite{Cornelissen2022} and they show that these quadratic improvements cannot be improved further. However, if the environments are not arbitrary, but special, more significant improvements are possible, including provable exponential speedups \cite{Dunjko2017}. The contrived nature of these environments makes these results unlikely to have implications in practice. 

However, the second research line investigates  quantum enhancements in reinforcement learning when the access to the task environment is restricted to be classical. Here too speedups are provably possible, at least in special cases, but now subject to standard complexity assumptions. For example in Refs.~\cite{Jerbi2021, Skolik2022} it was proven that there exist task environments with a super-polynomial advantage for quantum reinforcement learning algorithms (with respect to finding optimal policies and optimal q-functions), unless the discrete logarithm problem has an efficient classical solution. 
Again, in reality, we are most often not interested in finding provably optimal policies, but rather more effective heuristic solutions which perform well empirically, as is done in the famous examples of AlphaGo \cite{silver_2016_alphago} and AlphaStar \cite{Arulkumaran_2019_alphastar}. 
In this domain, there has been substantial effort in defining various types of parameterized-circuit based QRL algorithms \cite{Lockwood2020,Chen2020,Jerbi2021}.

Closely related to MDPs are optimal stopping problems. These problems are concerned with timing an action for maximizing a reward and are often solved with dynamic programming. Optimal stopping appears, for example, in sequential parameter estimation \cite{muravlev2020bayesian}, sequential hypothesis testing \cite{daskalakis2017optimal}, and, most famously, in financial derivative pricing \cite{longstaff2001valuing}. The pricing of American options involves stochastic modeling of asset prices and performing a combination of Monte-Carlo estimation, least-squares regression, and dynamic programming for the optimal stopping time \cite{longstaff2001valuing}. Given quantum access to a sampler for the stochastic process, a quantum algorithm with a polynomial speedup using Quantum Amplitude Estimation (QAE) \cite{brassard2002quantum,montanaro2015MC} 
was shown by Doriguello \emph{et al.}~\cite{doriguello2022stopping}. 
Key questions are whether we can extend such stopping time algorithms to other problem classes and to problems with more decisions, and whether we can make the algorithms more near-term quantum friendly. 

In the following, we outline open questions and promising directions to further advance our understanding on how quantum computers may provide advantages for optimal control and related problems.
In the era of noisy quantum computers, convex optimization algorithms may be hard to use directly. Nevertheless, one may considering sketching the value iteration \cite{pan2017effective} or (LSTD) \cite{li2018finite}, wherein both the random projections and the convex optimization could be implemented quantumly.
What MDPs could have a bound on the radius of the ball outscribed to the (primal, dual) feasible set? Such MDPs could benefit from speedup of quantum algorithms based on MWU, cf.~Sec.~\ref{sec:convex_optimization}.
Approximating the value function, discrete or continuous, using the concept of inductive bias recently introduced in the context of function approximation by means of quantum kernels~\cite{ind-bias}. Basic idea is to embed a prior knowledge of certain properties of the value function into the design of the approximator: for example, a rather efficient deep learning approximation of value function implemented in AlphaGo, or an assumed linear approximation of the value function implemented in LinUCB~\cite{chu2011contextual}. In this direction one would benefit from maintaining connections between discrete and continuous HJB equations for the optimization problem at hand in order to extract inductive bias, e.g. piece-vise smoothness structure of continuous value function might help in designing a kernel-based function approximation.

\subsection{Robust Optimization}
\label{sec:robust_optimization}

Robust optimization addresses the issue of data inaccuracy in  optimization and allows one to model uncertainty in the constraints. The convex variant was introduced by Ben-Tal and Nemirovski~\cite{ben1998robust} and lead to many follow-up works, for example Refs.~~\cite{ben2002robust, ben2008selected, bertsimas2011theory}. The main feature is a set of parameterized constraints and the requirement that these constraints hold for every choice of parameters chosen from some uncertainty set. The problem is formally defined as
\begin{mini}|s|[0]
    {x\in \mathcal{D}}{f_0(x)\hspace{3.5cm}}{}{}
    \addConstraint{}{f_i(x, u_i)\leq 0,\, \forall u_i \in \mathcal{U}, \, i \in [m],}
\end{mini}
where $f_0,f_1,\dots,f_m$ are convex functions in the parameter $x$ and $f_1,\dots,f_m$ are concave in the noise parameter $u$. Moreover, $\mathcal{D}\subseteq\mathbb{R}^n$ and the uncertainty set $\mathcal{U}\subseteq\mathbb{R}^d$ are both convex. Despite many advances, robust optimization for large-scale problems comes with a significant computational overhead. Ben-Tal \emph{et al.}~\cite{Ben-Tal2015a} developed meta-algorithms to approximately solve the robust version of a given optimization problem using an oracle for the solving of the original optimization problem. A quantum version was developed by Lim \emph{et al.}~\cite{lim2023quantum}, using quantum access to the functions, extending the original algorithm to include a stochastic gradient, and using quantum sampling subroutines with polynomial advantage.

\subsection{Multi-Objective Optimization}
\label{sec:multi_objective_optimization}

The previous classes were all considering a single objective function, possibly with constraints. However, in practice, decision makers often have to find trade-offs between different objectives. 
This is addressed by Multi-Objective Optimization (MOO) \cite{gunantara_20018_moo}.
MOO is defined as
\begin{mini}|s|[0]
{x\in X}{ (f_1(x), \ldots, f_m(x)) }
{}{},  \label{eq:moo}
\end{mini}
where $X$ denotes the feasible set and $f_i: X \rightarrow \mathbb{R}$, $i=1, \ldots, m$, denote the multiple objective functions, for instance. 
In this setting, we first need to define what we consider an optimal solution.
This is done using the concept of \emph{Pareto optimality}. A solution $x$ is called Pareto optimal for a problem with multiple objective functions, if there exists one objective function $f_i$, that cannot be decreased any further without increasing at least one other objective function $f_j$. The set of all Pareto optimal solutions is called the Pareto frontier or \emph{efficient frontier}.

It is well known that the cardinality of the Pareto set can be substantial \cite{papadimitriou2000approximability}, and thus, one would like to obtain
a PTAS. For certain bi-objective problems, such as the shortest paths problem, Diakonikolas and Yannakakis~\cite{Yannakakis2010} have shown that computing a set of solutions that approximates the Pareto curve within $\epsilon$ is $\NP$-hard, if the set is to be smaller than twice the optimum number of solutions,
while it is possible to obtain the 2-approximation in polynomial time. 

Depending on the application, there are different goals in MOO.
Either, the goal is to determine or approximate the complete Pareto front to be able to analyze possible trade-offs. Alternatively, a single Pareto-optimal solution is sought. For instance, objectives might be prioritized or have fixed weights, which allows one to convert MOO into a (series of) single objective optimization problems. If the full Pareto frontier is required, the number of objective functions plays a crucial role. For a small number of objective functions, there exist efficient strategies, assuming the underlying single objective optimization problems can be solved reasonably well.
This is usually achieved by constructing a new objective function as a weighted sum of the individual objective functions and adjusting the weights based on the previous solutions. Alternatively, a single objective can be optimized while the values of the other objectives are controlled via additional constraints. 
While the former might sometimes be easier to implement, it only finds the convex hull of the efficient frontier. In contrast, the latter can find also Pareto efficient solutions that do not lie on the convex hull.
Neither of these strategies is efficient for growing numbers of objective functions.
If the number of objective functions increases, MOO problems quickly become very hard. 

In contrast to single-objective optimization, there is no approximation ratio that can be used to conveniently quantify the performance of a given solution candidate. However, there are multiple other metrics that allow one to compare different approximations to the efficient frontier.
The most commonly used ones are the Hyper Volume (HV) and the Generational Distance (GD) \cite{riquelme_2015_moo_metrics}. 
The HV measures the volume spanned by the found solutions, which is measured as the total volume covered by all boxes defined by a reference point upper bounding (or lower bounding in case of maximization) all objective functions, and the solution candidates.
The GD, in contrast, measures the distance between a set of candidates and the optimal Pareto front or a reference solution.

There is only very little literature considering quantum optimization algorithms for MOO. Most of the proposals apply classical strategies and use quantum optimization sub-routines for the resulting single objective optimization problems \cite{chiew2023multiobjective} or sample from quantum states to approximate the efficient frontier \cite{ekstrom2024variational}. 
Thus, quantum-native algorithms for MOO present a very interesting domain for future research.

Within this section, we introduced the most important classes of optimization problems, reviewed existing quantum algorithms to approach them and key open questions to be answered to further advance our understanding. In the following, we discuss how to execute them on noisy quantum devices.

\section{Execution on Noisy Digital Hardware at Scale\label{sec:scaling}}

We now discuss best practices to execute quantum optimization on noisy quantum hardware.
There are many different platforms to consider for optimization tasks~\cite{Barends2016, dlaska2022rydberg, Harrigan2021, EbadiEtal2022, Pagano2020, Lotshaw_2023_qaoa,pelofske2023highround}, here, we focus on superconducting qubits~\cite{Krantz2019}.
More precisely, we focus on universal gate-based quantum computing platforms accessible via the cloud, where quantum algorithms researchers can directly test their algorithms.
While there are additional algorithmic challenges to scale to large problems, cf.~Sec.~\ref{sec:problem_classes}, the goal of this section is to discuss how to achieve best possible results from available quantum devices.

Superconducting qubits come in many forms, such as frequency-tunable qubits~\cite{Barends2014} and fixed-frequency qubits with static~\cite{Rigetti2010} or tunable coupling elements~\cite{McKay2016, Ganzhorn2020}.
Over the past years, superconducting qubits have greatly improved.
The number of qubits has increased with system sizes featuring up to 433 qubits.
Coherence times have breached the $100~\mu{\rm s}$ barrier with recent devices exhibiting median $T_1$ and $T_2$ times approaching $300~\mu{\rm s}$ and $150~\mu{\rm s}$, respectively. It should be noted that these figures are based on IBM Quantum Eagle devices with 127 qubits~\cite{ibm_quantum_platform}.
Importantly, the qubits are arranged in planar lattices where each qubit is only connected to its immediate neighbors.

We review each stage of the execution pipeline with scalability and best practices in mind.
In Sec.~\ref{sec:hardware_stack}, we describe how a quantum algorithm is translated down the quantum stack and executed on hardware.
We discuss the impact of modeling, quantum algorithm choice and qubit encoding in Sec.~\ref{sec:hardware_model}.
Next, Sec.~\ref{sec:hardware_transpilation} discusses transpilation and Sec.~\ref{sec:hardware_em} reviews error mitigation methods in the context of optimization.
Sec.~\ref{sec:hardware_pulse} touches on the pulse-level and the opportunities it presents.
In Sec.~\ref{sec:hardware_benchmarks} we discuss hardware benchmarks and then briefly discuss an execution example in Sec.~\ref{sec:hardware_example} before summarizing in Sec.~\ref{sec:hardware_summary}.

\subsection{Quantum Stack\label{sec:hardware_stack}}
\begin{figure}[htbp!]
    \centering
    \includegraphics[width=\columnwidth]{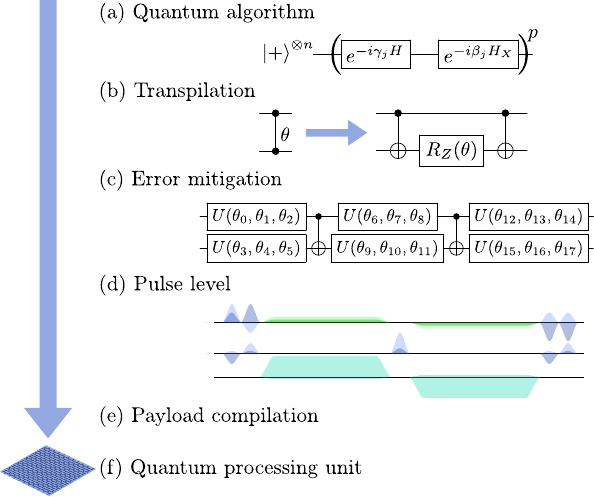}
    \caption{High-level summary of the quantum stack.
    (a) A quantum algorithm, including an encoding of decision variables in qubits and any hardware-guided problem simplifications, results in a high-level quantum circuit.
    (b) The instructions in this circuit are then transpiled to hardware native instructions.
    (c) Error mitigation methods, such as PEC, are often encoded as circuit instructions.
    (d) At the pulse-level, circuit instructions are represented by the physical pulses that are played on the qubits.
    (e) Finally, the circuit is compiled into machine-executable waveforms, (f) that are run on the quantum processor.}
    \label{fig:scaling}
\end{figure}

Hardware constraints and noise limit the scalability and performance of quantum optimization algorithms.
An analysis of the quantum stack, the stages shown in Fig.~\ref{fig:scaling} through which an algorithm evolves from its abstract design to its tangible execution on a quantum processor, offers insight into where scalability and performance bottlenecks arise. 

At the top of the stack is an abstract expression of the quantum algorithm.
It is usually free from any hardware constraints and expressed as mathematical operations and logical constructs that embody the problem-solving approach. 
The transpilation process then maps the abstract operations to the machine-executable instructions of the chosen target hardware resulting in a hardware-native circuit.
Here, circuit instructions may be inserted or modified as part of an error mitigation scheme.
The pulse-level compiler converts the circuit instructions to finely-tuned electromagnetic pulses. 
These pulses manipulate the quantum states to enact the chosen algorithm.
Therefore, as the algorithm is lowered through the layers of the quantum stack its representation changes.
Each level offers optimization opportunities to extract the most out of the noisy hardware.

\subsubsection{Modeling, quantum algorithms, and encoding\label{sec:hardware_model}}
At the top of the stack, an algorithm is described by a high-level circuit which may not be explicit in the number of qubits used and only provides an abstract representation of the quantum gates to run.
For example, Fig.~\ref{fig:scaling}(a) shows a high-level representation of the QAOA.
To arrive at such a circuit one must first appropriately model the optimization problem to solve and correspondingly choose the quantum algorithm.
For instance, linear equality constraints can be included in a QUBO by adding them as quadratic penalty terms which then requires appropriately sizing the corresponding multiplicative factor, see  Sec.~\ref{sec:constrained_discrete_optimization}.
This modeling step also presents opportunities to consider the impact of choices on the lower levels of the quantum stack.
For example, due to hardware noise a problem model that results in more qubits with less connectivity could lower the impact of noise compared to a model with fewer densely-connected variables.
Alternatively, Dupont \emph{et al.}~\cite{Dupont_2023QuantumCombOpt} project a fully-connected QUBO onto the sparse connectivity of a quantum computer.
Looking forward, additional research on quantum tailored modeling of optimization problems is needed to find the right balance between model approximations and noise susceptibility.

Since the hardware is noisy, algorithms that efficiently use the quantum resources are required.
For example, a warm-start reduces the number of QAOA layers~\cite{egger2020warmstarting, Tate2023WarmSDP}. 
In addition, the QAOA mixer can be changed to include constraints of the optimization problem~\cite{hadfield2019quantum,Hadfield2022Framework}, e.g., for equality constraints~\cite{wang2019xy,cook2020vertexcover}.
Altering the mixer improves the approximation ratio compared to penalty term constraints (soft constraints) for constrained optimization problems \cite{Niroula2022}. 
Furthermore, counter diabatic terms may also improve the approximation ratio~\cite{Chandarana2022}.
While these mixers may keep the structure of the objective function and lead to cheaper implementations of the cost operators, they may increase mixer circuit depth thereby limiting their near-term utility~\cite{Baker2022,Niroula2022}.
Crucially, from a hardware perspective, the best practice is to choose the methods that limit low-level circuit depth and preserve solution quality.
For example, adaptive bias-fields~\cite{Yu2022} and warm-starts~\cite{egger2020warmstarting} only require additional single-qubit rotations in the QAOA mixer which are implemented on hardware with a negligible overhead.
Furthermore, it may be useful to research methods that modify the original optimization problem to facilitate its mapping to the hardware.
For example, in QAOA one may investigate techniques that modify the problem Hamiltonian $H$ to produce an $H'$ approximating $H$ which is however easier to implement on hardware,
i.e., $e^{-i\gamma H'}$ has a lower circuit depth than~$e^{-i\gamma H}$.

While both scale and quality are improving over time, quality is typically the limiting factor rather than scale for optimization~\cite{weidenfeller2022scaling, Sack2023, pellow2023qaoa}.
Nevertheless, for many practically relevant optimization problems the scale of current Quantum Processing Units (QPUs) is likely insufficient to reach a classically hard regime.
It may therefore be necessary to consider denser encoding schemes that encode multiple decision variables in a qubit.
The Holevo bound states that $n$ qubits are required to faithfully transmit $n$ bits of information \cite{Holevo1979}.
However, encoding schemes such as Quantum Random Access Codes (QRAC) can encode $m$ bits of information in $n$ qubits where $m > n$.
The cost is that each bit of information can only be retrieved using Projective Operator Valued Measures (POVMs) with some probability $p > 1/2$ \cite{Ambainis1999,Ambainis2002,Imamichi2018, Fischer2022}.
There are many QRACs that, rather than encoding the two states of a single bit into the $\ket{0}$ and $\ket{1}$ states of a qubit, encode more information by utilizing more of the possible states on the Bloch sphere. 
For example, the $(2,1,0.85)$ QRAC encodes $n=2$ bits into a single qubit ($m=1$) with the 4 possible states of 2 bits encoded in states that form the corners of a square on the Bloch sphere.
Here, each bit is recovered with probability $p \geq 0.85$.
For all QRACs, there is a trade-off between the ratio of classical to quantum bits and the retrieval probability. 
The more bits encoded into a given number of qubits, the lower the probability $p$ of being able to retrieve a given bit and the more complicated the code becomes.
As discussed in Sec.~\ref{sec:unconstrained_discrete_optimization}, this allows the hardware to tackle problems with more decision variables than it has qubits \cite{fuller2021approximate, teramoto2023quantumrelaxation, Patti_2023, dupont2024extending}.
However, denser encodings tend to have a lower approximation ratio and introduce additional overhead to readout solution candidates.

Research for quantum optimization algorithms running on noisy hardware must carefully consider the hardware constraints.
Crucially, many of these methods are heuristic and can thus only be properly researched at scale on actual quantum hardware.

\subsubsection{Transpilation\label{sec:hardware_transpilation}}
A transpiler transforms a quantum circuit into another quantum circuit, typically built from instructions native to the target quantum hardware.
Such low-level circuits can be expressed in domain specific languages which include OpenQASM~\cite{cross2022openqasm}, Quil~\cite{smith2016practical}, or Blackbird for continuous-variable quantum computers~\cite{killoran2019strawberry}. 
Since each quantum hardware bears unique characteristics, tailoring the transpiler to individual architectures is fundamental to maximize operational efficacy.
Furthermore, the design and optimization of hardware-executable quantum circuits must include all constraints of the target hardware such as the native gate-set and the qubit connectivity.
On noisy hardware the transpiler must also account for properties such as gate error-rates, cross-talk, and qubit properties, e.g., $T_1$ and $T_2$ times.
The transpiler has a large impact on the gate count and, thus, the performance of the algorithm when run on the hardware.
In general, the transpiler is responsible for (i) selecting the qubits, (ii) decomposing the high-level circuit into hardware native instructions and (iii) routing the qubits to overcome any limited qubit connectivity.

Step (i) depends on the properties of the hardware and the problem being solved.
For example, if we do not need all the qubits on the processor then we may select the best ones according to a metric~\cite{Nation2023}, such as gate fidelity~\cite{Sack2023}, cross-talk~\cite{hua2022synergistic}, or layer fidelity \cite{mckay2023benchmarking}.
This metric should be based on the properties of the expected circuit.
E.g., if the expected circuit is predominantly composed of two-qubit gates then selecting qubits based on their gate fidelity is reasonable~\cite{Sack2023}.
However, if the algorithm requires many mid-circuit measurements then measurement quality should also impact qubit selection~\cite{hua2023exploiting}.
Furthermore, in approximate optimization a decision variable is often encoded in a qubit.
The choice of which decision variable is mapped to which qubit significantly impacts circuit depth~\cite{Matsuo2023sat}.

In step (ii), the possibly high-level gate descriptions are synthesized into hardware native instructions~\cite{zhang_bgate_2004, drury_qsd_2008, maslov_ion_2017}.
To decompose arbitrary unitaries, a quantum computer must support a universal set of basis gates~\cite{DiVincenzo2000}, which impacts the optimal synthesis technique.
These compilation methods are either deterministic or approximate.
For example, any exponential of Pauli operators can be implemented using a local 2-qubit gates on the involved qubits with a Pauli-$Z$ rotation and single-qubit Cliffords~\cite{Li2022}, as exemplified in Fig.~\ref{fig:scaling}(b).
However, due to hardware noise it might be beneficial to reduce the circuit depth by allowing for an approximate compilation~\cite{madden_aqc_2022, smith_leap_2023}.

Finally, the routing step (iii) must overcome any limited qubit connectivity of the target QPU.
Here, general-purpose heuristic algorithms exist to route arbitrary circuits~\cite{Li2019, Zulehner2019mapping, Cowtan2019Routing}.
However, certain algorithms may have a predetermined structure.
For example, QAOA results in circuits with blocks of commuting two-qubit gates.
Exploiting this structure results in shallower quantum circuits~\cite{Lao2022, jin2022structured, weidenfeller2022scaling, Matsuo2023sat, Alam2020}.
More precisely, if the topology of the problem graph does not perfectly match the coupling map of the QPU, SWAP gates are necessary to enable two qubit gates between unconnected qubits.
The denser the graph, the more SWAP gates are necessary which results in a deeper circuit~\cite{weidenfeller2022scaling}.
Nevertheless, limited qubit connectivity remains an issue and the problems that scale best on noisy superconducting devices are therefore likely to be those that are not overly dense~\cite{Harrigan2021, Sack2023}.
A key challenge is thus to identify problem instances that are not dense but are classically hard.

Modern hardware also exposes new circuit instructions such as dynamic circuits~\cite{Corcoles2021}.
In a dynamic circuit some of the qubits are measured and the outcome of these measurements changes the following circuit instructions within the coherence time of the qubits.
For instance, this offers additional possibilities to address limited qubit connectivity~\cite{baeumer2023efficient}.
As an example, the SWAP based implementation of the QAOA problem graph may be replaced on grid-like QPU coupling maps by constant depth circuits with a quadratic auxiliary qubit overhead~\cite{messinger2023constant}.
Additional research is needed to see how to leverage dynamic circuits to reduce the circuit depth and width.

Crucially, not all types of hardware face the same challenges.
For example, trapped ion devices typically have all-to-all connectivity which permits denser graphs to be studied without increasing circuit depth~\cite{Niroula2022}. 
However, at the time of writing trapped ions have not reached the same scale as superconducting devices meaning problems sizes are limited.

\subsubsection{Error suppression and mitigation\label{sec:hardware_em}}

After transpilation, the quantum circuit may be modified to include additional instruction for error suppression and mitigation, as shown in Fig.~\ref{fig:scaling}(c).
Error suppression methods reduce the noise in quantum circuits without having to gather more shots.
Most prominently, Dynamical Decoupling (DD) suppresses non-Markovian errors by adding pulses in idle times of a quantum circuit~\cite{viola1998dynamical, vitali1999using, Pokharel2018}.
This can improve circuit execution quality without introducing additional overhead.
In some cases, such as QAOA transpiled with SWAP networks~\cite{Sack2023}, the resulting quantum circuits may be dense which leaves little space for DD pulse sequences.
By contrast, DD has a large impact when the circuit has many idle times, e.g., in hardware native circuits on a heavy-hex lattice~\cite{Pelofske23QAOA}.

Error mitigation techniques use pre- and post-processing of quantum circuits and measured results to reduce the impact of hardware noise. 
While fully eliminating the error comes at an exponential classical overhead, these methods extend the reach of current devices even at a limited classical cost~\cite{Temme2017, Kandala2018, Piveteau_2022_qpd}.
This is in contrast with error correction, which aims to correct errors by encoding logical qubits in vast arrays of physical qubits~\cite{gottesman2009introduction}.
Typically, most error mitigation methods, such as Probabilistic Error Cancellation (PEC)~\cite{VandenBerg2023} and Zero-Noise Extrapolation (ZNE) \cite{Li2017,Kandala2018, Kim2023}, produce noise-mitigated expectation values.

PEC learns the noise channel $\Lambda$ of layers of Pauli-twirled gates~\cite{VandenBerg2023}.
The inverse of the noise channel $\Lambda^{-1}$ is usually not physical, but can be implemented by quasi-probability decomposition.
PEC thus produces an unbiased estimate of a noiseless observable $O$.
The cost is an increase by a factor of $\gamma^2$ in the measured variance.
Here, $\gamma\geq1$ depends on the strength of the terms in the learned noise model and is a measure of the fidelity the quantum circuit can be executed with.
The noise strength $\gamma$ is also closely related to the Layer Fidelity (LF) and the Error Per Layered Gate (EPLG), two related measures of noise in large-scale quantum devices~\cite{mckay2023benchmarking}.
If $\bar\gamma$ is the average $\gamma$ per qubit then a circuit with $n$ qubits and $d$ layers will have an error mitigation cost that scales as $\bar\gamma^{2nd}$.

In ZNE multiple logically equivalent copies of a quantum circuit are executed at different noise levels to compute an expectation value.
From the noisy expectation values one extrapolates to the zero-noise limit to obtain a (usually biased) error mitigated expectation value.
The first proposals implemented ZNE by stretching pulses~\cite{Kandala2018}, requiring a large calibration overhead, or by folding gates, resulting in large stretch factors.
Methods such as partial-gate folding~\cite{LaRose2022} or stretching of cross-resonance pulses~\cite{Vazquez2023} alleviates some of these issues.
Crucially, ZNE does not require more circuits as the size of the problem is increased.
However, the circuit depth limits the range of stretch factors which may render the extrapolation unusable.
PEC is unfeasible in large scale-experiments since they have a very large $\gamma$.
For instance, the $\gamma^2$ of 60 layers of CNOTs on 127 qubits was $10^{128}$ in Kim \emph{et al.}~\cite{Kim2023}.
This prompted the development of PEA~\cite{Kim2023}, a probabilistic ZNE, in which the noise is learned like in PEC but then amplified (instead of canceled) to then extrapolate to the zero-noise limit.
This may also reduce the bias in the extrapolation.
Importantly, circuits designed with regular networks of SWAP gates alternate a small number (e.g., two) of identical layers of CNOT gates~\cite{weidenfeller2022scaling}.
This minimizes the noise-learning overhead of PEC and PEA.

Error mitigation methods that act on expectation values help train variational parameters~\cite{Kandala2018, Sack2023}.
Crucially, to find good solutions, quantum (approximate) optimization requires sampling from the optimized circuit.
Indeed, the task is to find a good sample $x$ that, e.g., minimizes a cost-function $f(x)$.
Therefore, developing new error mitigation methods to produce error-mitigated samples is crucial for the success of quantum optimization.

\subsubsection{Pulse-level and compiler\label{sec:hardware_pulse}}
Ultimately quantum instructions are implemented by physical pulses, scheduled in time, to manipulate the quantum states~\cite{Alexander2020}, as exemplified in Fig.~\ref{fig:scaling}(d).
The pulse-level makes the differences between quantum architectures more apparent.
For instance, tunable couplers~\cite{Ganzhorn2020} can natively implement exchange-type operations which, on cross-resonance hardware~\cite{Rigetti2010}, are synthesized from $R_{ZX}$ rotations.
Exchange-type gates may help reduce schedule duration for applications like QAOA that need SWAP networks with phases to overcome limited device connectivity. 

Pulse control enables many different types of optimizations to reduce schedule durations.
This may be as simple as a pulse-efficient transpilation in which variational parameters are encoded in durations of cross-resonance pulses~\cite{Earnest2021}.
The resulting reduction in schedule duration improves the sampling performance of QAOA~\cite{Sack2023}.
Furthermore, the extra energy levels of the transmon can help to engineer gates that would otherwise require many CNOTs~\cite{galda2021implementing, Fischer2023}.
Finally, advanced methods of quantum optimal control can help synthesize gates~\cite{Egger2014, Gokhale2019}, measurements~\cite{Egger2014, McClure2016}, and adiabaticity~\cite{Chasseur2015} which may be useful in the context of approximate quantum optimization.
However, such methods are not always compatible with quantum error mitigation as discussed by Egger \emph{et al.}~\cite{Egger2023}.

The pulse-level is also responsible for quantum program compilation which entails loading the pulse-schedule into waveform memory.
Care must be taken at this step and inefficiencies can be costly to the total run time.

\subsection{Hardware Benchmarks}\label{sec:hardware_benchmarks}
In general, the performance of quantum hardware is measured by scale, quality, and speed~\cite{wack2021quality}.
Scale is simply the number of qubits.
Quality is tracked by low-level performance metrics, e.g., qubit coherence and gate fidelity, and holistic benchmarks such as Layer Fidelity and Error per Layered Gate~\cite{mckay2023benchmarking}, Quantum Volume (QV)~\cite{Cross2019} and the $\gamma$ and $\bar\gamma$ of PEC~\cite{VandenBerg2023}.
A QPU with a $2^n$ QV can reliably execute a $n$ qubit circuit with $n$ layers of random ${\rm SU}(4)$ gates~\cite{Cross2019}. 
The QV accounts for gate error rates, error suppression and mitigation, and qubit connectivity.

Neither QV, Layer Fidelity nor EPLG, however, account for the speed of the computation. 
Speed is crucial for optimization tasks if the quantum hardware is supposed to deliver a solution with a certain quality \emph{faster} than classical hardware.
Furthermore, speed enables the hardware to achieve more accurate estimations of observables within a certain time or to pay the cost of error mitigation which often requires more shots.
Therefore, Circuit Layer Operations Per Second (CLOPS) complement quality metrics.
The CLOPS metric is inspired from the floating point operations per second metric used to benchmark classical computers.
In essence, CLOPS quantifies the throughput of a device by measuring the number of gate layers that can be executed within a certain time. This measure includes the classical overhead of updating parameters in the circuit and is therefore a crucial indicator for the run time of variational workloads.

QV, EPLG, and CLOPS are by design only proxies for the utility of a quantum device.
For example, EPLG and CLOPS on \emph{ibm\_sherbrooke} are 1.7\% and 2700, respectively.
These metrics are general enough to draw comparisons between different hardware architectures and track improvements of a platform.
However, these general metrics could fail to predict which quantum device will obtain the best performance, be it in solution quality, execution speed, or any other criterion, for a specific application.
Therefore, the best possible measure of a quantum computer's usefulness for optimization is how well it performs on tasks similar to the problem of interest, according to the performance criterion of choice. 
For example, the ability to create spin-entangled states, i.e., squeezing, relates directly to a QAOA solving a fully connected MAXCUT problem~\cite{Santra2023}.
We foresee that more application tailored benchmarks are needed, especially since many algorithms are heuristic in nature.

\subsection{Execution Example\label{sec:hardware_example}}
We now exemplify some of the considerations discussed in the previous sections by examining the work by Sack and Egger~\cite{Sack2023}.
Here, the authors executed depth-two QAOA on hardware with Random-Three-Regular (R3R) graphs with up to forty nodes on superconducting qubit processors with 127 qubits.
First, since only a third of the physical qubits are needed, the QAOA is run on the best qubits as measured by the two-qubit gate fidelity; the most used and error-prone circuit instruction.
Second, the R3R connectivity is engineered with a network of predetermined SWAP gates~\cite{Harrigan2021, weidenfeller2022scaling}.
This results in dense and shallow quantum circuits that transpile quickly~\cite{weidenfeller2022scaling}.
Third, Sack and Egger~\cite{Sack2023} also choose a decision-variable-to-physical-qubit mapping that minimizes the number of SWAP gates following the approach of Matsuo \emph{et al.}~\cite{Matsuo2023sat}.
These best practices for QAOA are found open source on GitHub~\cite{BestPractices}.

Finally, a machine-learning-based error mitigation of expectation values allowed the optimization of the QAOA variational parameters resulting in $\boldsymbol{\gamma}^\star$ and $\boldsymbol{\beta}^\star$.
A noiseless simulation of the expected mean of the samples drawn with $\boldsymbol{\gamma}^\star$ and $\boldsymbol{\beta}^\star$ showed a large gap with respect to samples drawn from the hardware.
This highlights the need for an error mitigation of samples for quantum approximate optimization.

\subsection{Summary\label{sec:hardware_summary}}
Sec.~\ref{sec:scaling} provides an overview of the crucial parts of the software stack affecting the execution of quantum algorithms at scale on noisy hardware.
When considering this stack, it becomes evident that users possess varying degrees of control over the scalability of quantum algorithms on noisy hardware.
At the higher levels, such as problem recasting, encoding, and high-level circuit design, users enjoy a favorable degree of control (also depending on the problem being considered), allowing them to shape the algorithm's scalability to an extent.
Best practices for noisy hardware are to choose algorithms and their variants that lower the circuit depth without compromising on algorithmic speed and quality.

Moving lower into the quantum stack, transpilation and pulse-level control also offer the opportunity for optimization.
SWAP networks, SAT mappings, pulse-efficient transpilation, and problem-tailored gate-sets all help reduce circuit depth and duration.
These lower-level aspects demand a profound understanding of quantum mechanics and hardware intricacies.
In some cases, these layers may even conceal these complexities behind a primitive such as a sampler or an estimator.
It is nevertheless crucial to have transparency on the methods employed.
This also enables the design of hardware-optimized algorithms and problem formulations, which may further improve the performance of executing quantum optimization algorithms on noisy hardware.

\section{Benchmarks\label{sec:benchmarks}}

Benchmarking can provide insights into realistic algorithmic performance when purely analytical methods fail, and it can provide new perspectives that motivate deeper analytical studies. 
In particular, theoretical complexity analysis typically pertains to general problem classes and considers mostly asymptotic scaling behavior or worst case scenarios.
In fact, theoretical worst case bounds, as they are discussed in Sec.~\ref{sec:complexity}, often fail to describe the difficulties of solving real-world applications that focus on a specific problem instance of practical use, nor do they faithfully describe how well practically relevant problems can be solved for fixed size and data.
In this case, we need to rely on meaningful and systematic benchmarking of optimization algorithms with well-defined assumptions considering implementation, data, and metrics --- a challenging task. Clear benchmarks that study the utility and robustness of optimization algorithms can also serve to provide trustworthy and verifiable performance indicators for a broader audiences, including policy makers and industry leaders.
For a general introduction to computational benchmarking, we refer the interested reader to Refs.~\cite{dunning2018works, JainComputerPerformance91, KochBertholdPedersenetal2022,bartzbeielstein2020benchmarking,Muller2011Spec,dongarra1979linpack}. 

The aim of this section is to describe best practices for benchmarking and provide a set of interesting benchmarking problems that provide a thorough, robust, fair, and reproducible setting for comprehensive comparisons of different classical and quantum approaches.
The remainder of this section is structured as follows. First, an overview of related work is given in Sec.~\ref{sec:benchmarks_relatedwork}. Next, Sec.~\ref{sec:benchmarking_goals} discusses various reasons for running a benchmark. Notably, the goal behind running a benchmark is crucial for the definition of the setting. Benchmark design choices such as the metrics to be evaluated are presented in Sec.~\ref{sec:settings}. Then, we elaborate on a variety of optimization problems which are difficult for state-of-the-art solvers and may provide interesting benchmarking problems for quantum methods in Sec.~\ref{sec:benchmarking_problems}. Finally, Sec.~\ref{sec:benchmarking_demonstrations} discusses existing quantum optimization demonstrations that may build up to a set of benchmarks.

\subsection{Related Work}\label{sec:benchmarks_relatedwork}
Important collections of existing (classical) optimization benchmarks are publicly hosted benchmarking challenges. They have proven as valuable tools for fair and insightful comparisons between different classical solvers and/or platforms.
These challenges aim at understanding and improving the practical performance of algorithms for various optimization problems, particularly those that are theoretically hard. They aid in evaluating realistic algorithm performance conditioned on the resources available at the time of the competition.
Some of the most relevant, regularly held optimization challenges tackling problems with classical digital computers are the Discrete Mathematics and Theoretical Computer Science (DIMACS) implementation challenges \cite{dimacs23} and the Satisfiability (SAT) competition \cite{froleyks2021sat}.
While the problems targeted in the former vary from competition to competition, the latter always focuses on SAT problems which represent a well-defined class of problems that, beyond their theoretical intrigue, find practical applications across various fields \cite{yu2021progress, claessen2008sat}.
Another important resource for optimization benchmarks is the Mixed Integer Programming Library (MIPLIB) \cite{GleixnerHendelGamrathetal2021}. This library provides regularly updated pure and mixed integer programs that are chosen under the consideration of solvability and numerical stability.
Currently, it holds 240 benchmarking instances that are solvable with existing methods and an even larger collection of unsolved, and numerically difficult instances. The performance of many solvers regarding several classes of optimization problems is regularly compared in \cite{mittelmann2023bench}.

Optimization competitions and libraries of this form that are compatible with (near-term) quantum computers do not exist as of now. It remains an open task to set them up and, thereby, enable streamlined benchmarking of quantum optimization methods.
In fact, most existing related work focuses on application-centric quantum hardware benchmarking frameworks that employ optimization problems as exemplary application.
A benchmarking framework presented by Lubinski \emph{et al.}~\cite{lubinski2023applicationoriented} suggests the comparison of the output quality to circuit width and depth for implementations of Shor's factoring algorithm, Grover search, Hamiltonian simulation, etc., executed in numerical simulations or gate-based quantum hardware.
Another framework introduced by Tomesh \emph{et al.}~\cite{tomesh2022supermarq} measures the qubit connectivity, circuit depth, number of 2-qubit gates, ratio of gates which can be executed in parallel, number of times qubits are acted upon, and number of mid-circuit measurements for algorithms including QAOA, VQE, and Hamiltonian simulation algorithms executed on quantum hardware platforms from different providers.
Furthermore, Martiel \emph{et al.}~\cite{Martiel21BenchmarkingQuantumCoprocessors} present a metric that measures the maximum problem size for which a problem instance can be solved given a pre-defined accuracy level. Notably, the suggested metric does not incorporate time as an explicit factor. Instead, the metric is conditioned on algorithms to be run in polynomial time. This directly excludes the use of error mitigation and can easily fail to provide a fair and meaningful comparison between different platforms.
Further benchmarks that investigate different properties of quantum annealing methods for optimization problems -- some including comparisons to classical and quantum algorithms -- can be found in Refs.~\cite{Ronnow14Defininganddetectingquantumspeedup, Juenger2019,JuengerRendl,PerdomoQuantumOptIndust19, GrantBenchmarkingQAPortOpt21,tasseff2022emerging, lubinski2024optimization, Willsch_2020}. A comparison of Ising machines that checks success probabilities for finding the ground state and the corresponding time-to-solution is presented by Mohseni \emph{et al.}~\cite{mohseni2022ising}.

One particularly interesting work is introduced by Fin{\v{z}}gar \emph{et al.}~\cite{Finzgar22QUARK}. They compare various QUBO solvers with respect to the solution quality and time-to-solution and as such represents a good baseline for an optimization benchmark.
Although the presented publications do not offer holistic optimization benchmarking frameworks, they provide interesting insights that can help to define reproducible, representative, and fair benchmark libraries, which facilitate identifying advantages and disadvantages of various (quantum) solvers.

\subsection{Goals}\label{sec:benchmarking_goals}
Reasons for running a benchmark can be diverse. In the following section, we focus on benchmarking goals related to understanding the potential of quantum computers for solving optimization problems in terms of a fair comparison.

One of the main goals of benchmarking in the  present context is the comparison of classical and quantum algorithms --- aiming at demonstrating quantum advantage.
If classical platforms are compared amongst themselves, one often relies on benchmarking problems such as SPEC \cite{Muller2011Spec} and LINPACK \cite{dongarra1979linpack}, which are established in the community. However, these problems are not well suited to compare classical and quantum platforms. 
Instead, the community will have to agree on a set of problem-centric benchmarks that could enable the direct comparison of run times, solution qualities, etc., of classical and quantum solvers. Suggestions for suitable benchmarking problems are given in Sec.~\ref{sec:benchmarking_problems}.

Benchmarking may also be executed to track the progress in hardware developments (cf. Sec.~\ref{sec:benchmarks_relatedwork}) as well as algorithmic improvements. 
This is particularly important for monitoring the impact of the fast development in quantum hardware, supporting software, and algorithmic research.
For an example of tracking progress of classical hardware and algorithms, we refer the interested reader to Koch \emph{et al.}~\cite{KochBertholdPedersenetal2022}, which discusses the advancements in mathematical programming solvers over $20$ years.

\begin{figure}[ht!]
\includegraphics[width=0.5\textwidth]{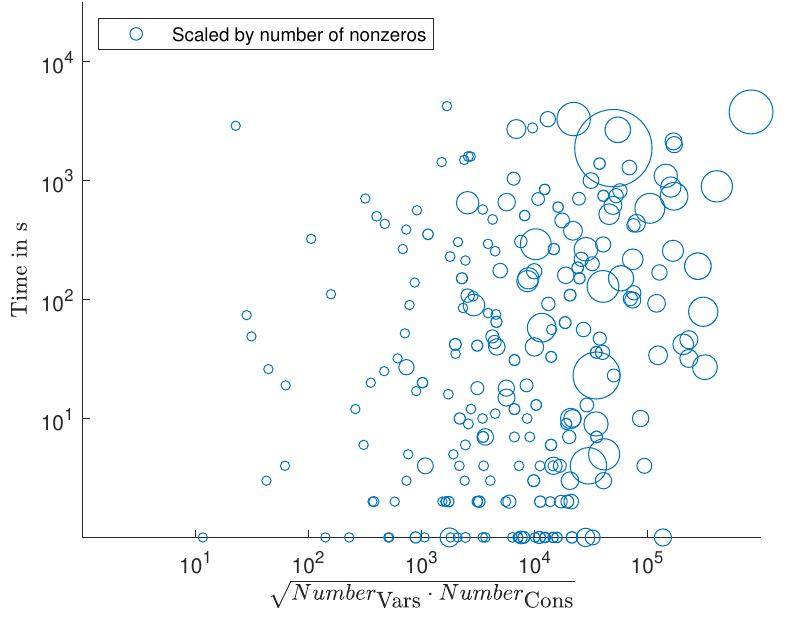}
\caption{The figure illustrates the time it takes the Gurobi solver \cite{gurobi10} to find the optimal solution for MIPLIB 2017 \cite{GleixnerHendelGamrathetal2021} instances of the form $\text{min}_x\:c^Tx$ subject to $Ax \leq b$, $x$ being partly integer. The size of the circles depicts the number of non-zero coefficients in the constraint matrix $A$ and the instances are ordered (on the x-axis) according to the square root of the number of variables times the number of constraints.} \label{fig:gurobi_scaling}
\end{figure}

Furthermore, benchmarking can give us insights into an algorithm's scaling behavior: Is there a problem scale at which algorithms fail to provide good solutions --- possibly due to hardware limitations? How do the resources needed to find solutions to a problem of a certain quality scale with the system size? An example of MIPLIB problem instances is illustrated in Fig.~\ref{fig:gurobi_scaling}. In this case, many -- but not all -- larger instances require longer solver run times.

\subsection{Design}\label{sec:settings}
To ensure reproducibility as well as replicability of a benchmark, it is important to clearly define the specifics in terms of model selection, pre-processing, computational platform, target metrics, and selected algorithm (provably exact, heuristic, etc.) including suitable hyperparameters.
A discussion of relevant algorithmic paradigms and quantum algorithms for optimization problem classes can be found in Sec.~\ref{sec:paradigms} and Sec.~\ref{sec:problem_classes}, respectively.
The remaining levels of design choices are discussed in the following section. Starting from the decision whether the benchmark is defined with respect to a particular model, and, if so, which one. Secondly, it usually is beneficial to apply pre-processing to reduce the problem to its challenging core. A possibly applied pre-processing procedure may already consider the computational platform that is employed afterwards to execute the optimization algorithm. Finally, the algorithm execution and resulting outcomes can be captured in a comparable fashion by choosing appropriate benchmarking metrics.

\subsubsection{Model-(In)Dependence}
At the core of a practically motivated optimization task is the solution of a concrete decision problem such as the selection of an optimal (or at least a good) tour for TSP. 
In order to treat this problem in a rigorous way a mathematical model must be developed. 
However, the model choice is usually not unique and may lead to differently behaving algorithms. In particular, certain models may be significantly better or worse suited for a particular optimization algorithm or may even contain characteristics that cannot be handled by a particular algorithm, such as non-linearities or constraints.
For benchmarks, it is therefore crucial to decide on the level a problem is defined: either \emph{model-independent}, i.e., on the level of the original optimization problem, which allows different mathematical modeling methods, or \emph{model-dependent}, i.e., on the level of a concrete mathematical model.

Model-independent benchmarks that place few limitations upon admissible solution strategies offer an interesting  method to identify quantum advantage for optimization. More specifically, quantum advantage may stem from novel formulations of problems that could not be captured on a model-dependent level.
Furthermore, quantum hardware is quickly evolving. To enable comparability and compatibility for different generations of hardware, model-independent benchmarks are better suited, since they may allow one to leverage new features that might not have been available in earlier generations, such as dynamic quantum circuits \cite{baeumer2023efficient}.
We expect these benchmarks to eventually show how quantum optimization algorithms are approaching, meeting, and exceeding the capabilities of classical techniques for solving particular problem instances.

Nevertheless, model-dependent benchmarks are  also a useful tool. In particular, they can help to study the relative performance of different algorithms and platforms for fixed model formulations.
At this point, it should be noted that model-specific solvers are often significantly faster for the target problem than general solvers.
For specific metrics to compare different types of solvers, we refer the interested reader to Berthold and Csizmadia~\cite{BertholdPrimalIntegral21}. 
Some model-dependent benchmarking test instances that are openly available, established, and unsolved are given in MIPLIB \cite{GleixnerHendelGamrathetal2021}, the Quadratic Programming Library (QPLIB) \cite{FuriniEtAl2018onlineQPLIB}, and the Satisfiability (SAT) Problem Library \cite{Hoos2000Satlib}.
As an example, we can consider (random) SAT instances which exhibit phase transitions. It may now be of particular interest to investigate whether a quantum algorithm is better suited to handle the difficulty of a phase transition.
In fact, it has been studied by Yu \emph{et al.}~\cite{yu2023solution} whether an adaptive bias version of QAOA might help to avoid transition phenomena. 
While the presented study focuses on small problem instances, where classical algorithms can provide solutions, the findings hint at the potential for advantages in the realm of large problem densities and, hence, motivate further investigation.

\subsubsection{Pre-processing}\label{sec:pre-processing}
The term \emph{pre-processing} or \emph{pre-solving} refers to a set of methods that are applied to a problem instance or data before the actual optimization algorithm is executed. 
Standard pre-processing approaches require classical computational effort. Therefore, the pre-processing should also be taken into account in a benchmark's performance evaluation.

The goal of pre-processing is to modify the problem in such a way that it allows the optimization algorithm to work more efficiently. How to implement pre-processing in practice depends on the mathematical formulation of the optimization problem, on the data, as well as on the considered optimization algorithm and computational platform. 
In fact, it is not uncommon for many problems such as Euclidean TSP or Steiner tree problems in graphs that pre-processing is able to remove 95\% of the variables and, hence, reduce the problem size as well as resource requirements significantly~\cite{ABCC2006,rehfeldt2023}. 
How this can be achieved often depends on the context and characteristics of the problem and data. 
A central aspect of pre-processing is the reduction of the search space, for example by decreasing the number of decision variables, or reducing the number of alternative optimal solutions, or by decomposing the problem into smaller subproblems. 

Considering the example of QUBO, various reduction techniques to eliminate optimization variables have been suggested in the literature. A recent review has been provided by Rehfeldt \emph{et al.}~\cite{rehfeldt2023}. For example, a set of analytic reduction rules is presented by Glover \emph{et al.}~\cite{glover2018} that allows one to conditionally fix certain optimization variables, while guaranteeing that at least one optimal solution will remain in the search space.
In addition to the number of optimization variables, their degree of correlation, which is determined by the sparsity of the QUBO matrix, is also of interest. In the case that certain groups of decision variables are not correlated, QUBOs can be decomposed trivially into smaller sub-QUBOs to be solved independently.

\subsubsection{Platforms}\label{sec:platforms}
Different platforms offer different strengths and suffer from different bottlenecks. Hence, some devices could be better suited for certain optimization problems than others.
It may not be directly evident which hardware is best suited for a given optimization problem. 
Furthermore, depending on the platform, the optimization models and applicable algorithms do vary, resulting in better or worse performance for a given problem.

\begin{figure}
\begin{center}
\includegraphics{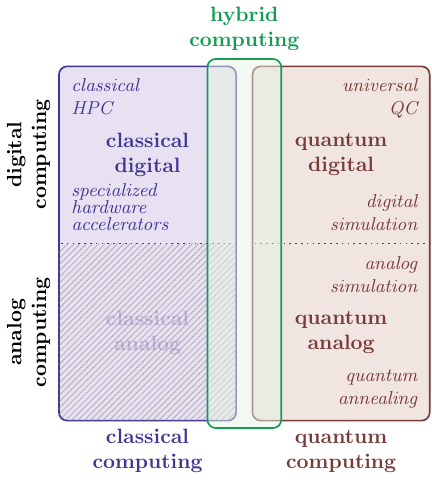}
\end{center}
\caption{Sketch of the optimization platform ecosystem. We focus on digital classical computing as well as digital and analog quantum computing.} \label{fig:platforms}
\end{figure}

Tailoring the choice of platform to the characteristics of the problem at hand -- if possible -- can lead to better outcomes. This approach aligns with the idea of \emph{platform-aware} optimization, ensuring that the selected platform is optimally matched to the unique requirements of the problem being addressed.
To understand the advantages and disadvantages of different platforms for different optimization problems, performing cross-platform benchmarking is crucial. At the same time, such benchmarks pose a particular challenge, as they must allow a fair comparison between systems that can function very differently.
In the following, we are going to elaborate on various hardware types that are relevant for optimization problems, i.e, classical vs.~quantum, and analog vs.~digital. 
A sketch of the platform ecosystem is shown in Fig.~\ref{fig:platforms}.

Classical digital systems are the workhorse of today's computer technology. The basic units of information are the binary values 0 and 1. The information in classical digital computations is processed by compositions of logical operations such as AND and NOT gates which can be expressed in a circuit model. Digital classical computational models often rely on a combination of Central Processing Units (CPUs), which excel at sequential and general data processing tasks, and Graphics Processing Units (GPUs), which specialize in parallel processing tasks.
There are several other types of processors and specialized hardware accelerators designed for specific tasks and applications, for example Tensor Processing Units (TPUs) for deep learning or Field-Programmable Gate Arrays (FPGAs) for signal processing and embedded systems, just to name two. The combination of CPU, GPU and special-purpose processors forms the foundation for classical high-performance computation in a wide range of  applications. Due to their widespread use, classical digital hardware devices span a wide spectrum of computing platforms, including desktop computers, computing clusters, cloud services, mobile phones, and many more. All of these platforms can be used to solve optimization problems.

In additional, special purpose systems for optimization are developed.
More specifically, these systems simulate quantum phenomena such as quantum annealing, see Sec.~\ref{sec:algorithms_qaa}, hence, aiming to gradually decrease an energy function. Well-known examples for this type of technology are the Fujitsu Digital Annealer \cite{TsukamotoDigAnnealer17, MatsubaraDigitalAnnealer20} that is based on a hardware-accelerated Markov Chain Monte Carlo approach, the Hitachi CMOS Annealing Machine \cite{YoshimuraCMOSAnnealing20} which corresponds to an in-memory computing architecture that represents an Ising model with local interactions, and the Toshiba Simulated Bifurcation Machine \cite{HayatoBifurcation16, WangBifurcation23}. It should be noted that the latter does not approximate an annealing process but instead implements a highly parallelizable approximation of Ising dynamics with classical hardware (GPU, FPGA). This allows one to compute heuristic solutions to problems with up to millions of variables in a time frame of seconds.

In digital quantum computing \cite{Krantz2019} the basic units of information are \textit{qubits} instead of bits. The information processing consists of consecutive quantum gate operations such as CNOT gates and single-qubit rotation gates which, together, form a quantum circuit, similar to a classical logical circuit. 
Important bottlenecks of existing digital quantum computers are limited coherence times, which result in limited circuit depths, limited qubit numbers, and limited qubit connectivity.
There are a variety of ways to quantify the performance of a digital quantum computer. For near-term hardware where noise is a dominant factor, one might be tempted to consider metrics such as the coherence times of qubits or the fidelity with which various quantum operations can be performed. Such metrics are valuable, however, they can fail to give an accurate description of a device's potential for useful computation. This has motivated proposals for more holistic error metrics, as discussed in Sec.~\ref{sec:hardware_benchmarks}.

Quantum analog hardware is realized by quantum annealing \cite{kadowaki1998quantum, farhi2000quantumAdiabatic, AlbashAdiabaticQC18, Ronnow14Defininganddetectingquantumspeedup, McGeoch2014AdiabaticQC} and quantum analog simulation \cite{hangleiter2022analogue, BlochQuantumSim12, TrotzkyRelaxationTowardsEquilibrium12, ChoiManyBodyLocalization16, MaxwellSite-ResolvedSpinCorr16, BernienProbingManyBody17, BlattSimulationTrappedIons12, ZhangManyBodyDynPhases17, ArrazolaDigital-AnalogTrappedIons16}.
In quantum annealing, the energy function or Hamiltonian of a system is gradually (but not necessarily perfectly adiabatically) deformed from an initial ground state to a final state that can be measured to find the solution to an optimization problem. These systems are designed for QUBOs. 
Important bottlenecks of these platforms include the coherence time, restrictions on the driver Hamiltonian, and challenges in embedding practical problems into the device connectivity.
The coherence time of current publicly available annealing machines is shorter than the annealing time \cite{king2022coherent}. 
This means that these platforms do not achieve closed system dynamics but exhibit strong coupling with the environment. Further hardware improvements are therefore required~\cite{King23QuantumCritDyn}. The primary computational resource is believed to be incoherent quantum tunneling events \cite{boixo2014evidence,denchev2016computational}.
Therefore, recent efforts are directed towards engineering quantum drivers that cannot be simulated classically \cite{nishimori2017exponential, King23QuantumCritDyn}.
Quantum analog simulation describes a setup consisting of physical systems with sufficiently controllable parameters such that one can manipulate and probe the dynamics underlying a family of Hamiltonians. More specifically, analog quantum simulators enable the emulation of continuous quantum dynamics of a target system with a simulator that is easily accessible in the lab.
This setting has proven itself valuable to simulate physical dynamics including quench problems, and ground state problems \cite{hangleiter2022analogue} but has not been studied in much detail as a platform for optimization applications, yet.

\subsubsection{Metrics}\label{sec:benchmarks_metrics}
Due to the technological differences between classical and quantum systems, the selection of a set of metrics that enable a fair comparison is crucial.
The four main metric categories that suffice said criteria are \emph{resource cost, run time, quality, and problem complexity}.

The resource cost includes any quantum or classical computational resources employed to solve a given optimization problem including any potential pre- and/or post-processing steps and resources that are being used in parallel.
The respective costs may be quantified in various ways.
Firstly, a fair metric would be the amount of energy that is used to solve a problem. However, in most cases, the user will not have access to this information, particularly, when specialized computational resources such as CPUs, GPUs, or quantum computers are accessed via the cloud.
Similarly, the monetary cost of running an algorithm -- including the pricing of specialized computational resources -- could provide an interesting metric but will be tough to evaluate on a fair basis as these depend on hardware providers and service contracts. 
Classical computing resources are typically measured with respect to factors such as the used memory, required memory bandwidth, and central processing unit (CPU) specifications, including clock speed and the number of cores.
Quantum computational resources can be measured in terms of the number of circuits that are run, the number of executed gates per circuit, the number of qubits and shots per circuit execution, and the quality of the qubits. Of course, the resource count should also consider resources that are required for problem-dependent calibrations and error mitigation or -- if applicable -- correction. The number of qubits in a circuit strongly depends on the mapping of problem variables to qubits. Hence, algorithmic improvements in mapping schemes can lead to a reduction in qubit requirements.

The run time provides an objectively fair metric to compare different computational platforms. It is, hence, of utmost importance to clearly define what determines a \emph{faster} algorithm.
The effective run time of an algorithm consists of time used for pre-processing, transpilation, embedding, compilation, execution, and post-processing. 
Now, it is not a priori clear, which of these time factors should go into a time benchmarking metric:
(i) only the time from start to end of the solver (ii) including the generation of the model for the specific solver, (iii) including pre-processing from the raw data. 
Arguably, a fair comparison considers the total time. Nevertheless, it can also be interesting to analyze the composition of the total time as this can help to identify the most costly steps and potential bottlenecks.
Given the current status of software for quantum computing transpilation, embedding, and compilation times will in fact play a non-negligible factor. Hence, understanding the relation of time-to-solution and QPU time gives us insights into the (progress in) efficiency of these steps.
However, these are also areas of active research and development and we expect them to be significantly reduced in the near future.
Another factor to take into account when measuring the time-to-solution is the potential use of parallelization, which effectively corresponds to a trade-off between computational resources, i.e., cost, and run time \cite{Edson_TTSvsETS2012}.

It should also be noted that the nature of a run time benchmark strongly depends on the benchmarking goal, e.g., aiming to reach a feasible, proven optimal, or $\epsilon-$close to optimal solution. If the goal is finding a proven optimal solution, then it has to be taken into account that certifying optimality typically requires a significant amount of time compared to merely finding the optimal solution. 
In the case of feasibility problems, we must consider what to do if one method finds a solution and one does not. A good approach is to have feasibility problems as a separate class and count how many problems can be solved within a certain time limit per instance --- even if sometimes none is found. Setting the right time limit is crucial. Short times $<1$s are often difficult to measure together with the setup time. Common choices for classical heuristics are therefore $<1$s, 10s, 60s, 600s, 3{\small,}600s, and 10{\small,}000s. 

Different forms of optimization methods, for example deterministic vs.~heuristic, also impact the specifics of a run time benchmark. Considering, e.g., non-deterministic methods that produce various solutions, one has to fix whether the run time is calculated as the total time, the average time to find a solution, the minimum time to find a solution, etc. One proposal to address this, is to explore the whole boundary of time-to-target and optimality-gap-at-time values~\cite{lubinski2023optimization, dunning2018works}. Another metric that aims at giving a holistic point of view by considering a combined measure for solution quality and total heuristic run time is suggested by Berthold~\cite{BERTHOLD2013611MIP}.

Quality metrics strongly rely on the nature of a problem.
Firstly, one requires a solution to be compatible with the problem constraints, i.e., to be feasible, which can be easily checked.
Ultimately, one wants to find the optimal solution or at least a solution that is $\epsilon-$close to the optimal solution. 
It should be noted that the evaluation of optimality may be resource-intensive or intractable. 
Given a feasible solution, the evaluation of the optimality gap may be identified via bounds given from continuous relaxations \cite{NoceWrig06} or dual representations \cite{bonnans2006numerical}. 
Another quality metric for optimization algorithms is solution diversity. Given multiple solutions with equal cost values, one would expect a fair sampling from these solutions. Interestingly, this property is not necessarily given in quantum annealing algorithms~\cite{YamamotoFairSamplingAnnealing20, KonzFairSamling19,nelson2022high}, and also not easy to realize on current gate-based devices~\cite{golden2022fairsampling,pelofske2021sampling}. 
Similarly, one can measure the success rate of a heuristic algorithm to find solutions that are feasible or suffice a certain optimality distance in a fixed number of algorithm execution runs.

The practical difficulty of a problem is often related to the problem size, density or number of variable connections, and the nature of the constraints. The difficulty may be changed by modifying one or more parameters that characterize the system as a whole. 
These changes can even lead to  computational phase transitions, which are marked by abrupt and significant shifts in the time to find a solution of certain quality, or even the feasibility of finding a solution. 
Understanding the range of problems and characteristics where phase transition behavior is observed presents a challenging task. 

In order to provide a holistic quantum optimization benchmark, metrics that describe algorithm and solution properties according to these four classes should be measured. Depending on the problem, model, algorithm, and execution platform the respective metrics should be wisely chosen and faithfully reported on.

\subsection{Problems}\label{sec:benchmarking_problems}
Next, we present a set of model-independent problems that show promise for establishing standardized quantum optimization benchmarks.
The problems should be chosen sufficiently hard for state-of-the-art classical solvers to leave a margin for quantum methods but within an intermediate size range where current and/or near-term quantum technologies can be effectively deployed. 
Ideal problem instances are those for which existing methods are unable to provide provably optimal or feasible solutions. 
Furthermore, it is crucial that the problem can be formulated as a quantum-compatible model, i.e., with sufficiently few variables/qubits, limited connection density, relatively small coefficients, etc. 
In other words, we are looking for a sweet spot for current quantum systems that enables the investigation of quantum advantage compared to adequately chosen digital algorithms.

It can also be valuable for tracking the progress in algorithmic and hardware development to identify a family of problems that are similar in their nature but differ in their difficulty.
Notably, complexity theory typically makes statements about the hardest instances in a problem class and, therefore, seldom provides descriptive arguments about the hardness of individual instances --- which very much depends on the size, structure, and precise coefficients of the instances.
Furthermore, the particularly chosen instances may correspond to crafted, random, or real-world problems:

\textbf{Crafted instances} are most likely the best way to find problem instances that are hard to solve for classical methods. However, there is a risk, that there is (or can be developed) also a classical method that takes advantage of the special problem structure, or that the instances are of no practical value.

\textbf{Random instances} are different since they lack structure. This can make them, depending on the problem, particularly easy or very difficult to solve. It certainly sets them apart, but how meaningful these instances are can be debated.

\textbf{Real-world instances} correspond in many practically relevant cases to MIPs that can be solved up to sufficient accuracy with existing solvers. Instances that are relevant and difficult to solve with existing methods are hard to find in a setting that fits the dimensionality and density limitations of current quantum hardware. In fact, these problems are generally hard to find because open problems of industrial relevance are rarely published or may be ill-defined. This highlights a common \emph{selection bias}: We rarely model practically relevant problems in a way we know we cannot solve.

In the following, we list a set of binary problems that are promising candidates for quantum optimization benchmarking problems. 
These problems are random or crafted instances and relate only to few practical applications.
Nevertheless, they offer a good test bed for algorithm development and progress tracking.
Classical methods that can be used to tackle these problems include Branch-and-Cut based Integer (linear, non-linear, semi-definite) programming (ILP) \cite{Wolsey2020}, Pseudo-Boolean (PBO) optimization \cite{DGDNS21}, SAT-Solving (SAT) \cite{froleyks2021sat, prasad2005survey}, Constraint Programming (CP) \cite{minizinc}, and several other general techniques \cite{localsolver, KuroiwaBeck2023} as well as a multitude of specific heuristic approaches.

\textbf{Maximum Independent Set (MIS)}. 
Given a weighted graph $G=(V,E,c)$, we are looking to find
\begin{mini}|s|[0]
    {x \in \{0,1\}^{|V|}}{\sum_{v\in V} c_v x_v\hspace{2.5cm}}{}{}
    \addConstraint{}{x_u+x_v\leq 1, \,\forall (u,v)\in E.}
\end{mini}
MIS can be formulated as a QUBO for a suitably large penalty factor $P$:
\begin{mini}|s|[0]
    {x \in \{0,1\}^{|V|}}{\sum_{v\in V} c_v x_v^2 - P \sum_{(u,v)\in E} x_u x_v.}{}{}
\end{mini}
This is a classic $\NPO$-hard problem even in case $c_v=1$ and there are known instances that are hard to solve to proven optimality with existing methods starting at problem sizes of several hundred variables \cite{IndependentSetsSloane2000}. 
The allure of this problem class for quantum optimization benchmarking is that it is well suited for translation into relatively sparse QUBOs.
    
A subclass are MIS with unit disc graphs $G$ (UD-MIS) which are defined by vertices $V$ on a two-dimensional plane with edges $E$ connecting all pairs of vertices within a unit distance of each other \cite{Clark1990, Andrist_2023}. These problems can be naturally mapped onto the structure of Rydberg quantum computers \cite{pichler2018quantum, Wu2021} via the Rydberg blockade \cite{EbadiEtal2022}. The optimization itself can then be carried out either in an analogue or digital fashion. It is also worth mentioning here that any QUBO can be mapped constructively to a UD-MIS with weighted vertices (UD-MWIS) \cite{Nguyen2023}.

Optimization problems that may be modeled as a MIS are, for example, \textit{Sudoku} and the \textit{multidimensional n-Queens problem}.
Interestingly, 3D $n$-Queens problem instances on a $14^3$ board with 2,744 binary variables, and most larger board sizes, have no known proven optimal solution. As there is by definition only one problem instance per board size, we can nicely track how far we get.
Furthermore, it should be noted that the \emph{maximum clique} problem can be solved as a MIS on the complementary graph.

\textbf{(Multi-dimensional) Knapsack / Market Share.}
Given a set $I = \{1, \ldots, n\}$ of items, weights $w_i$, $i\in I$, and a capacity $C$, the basic \emph{knapsack} constraint regarding a set of binary variables $x_i$, is 
$\sum_{i\in I} w_i x_i \leq C$.
Knapsack problems typically have an objective function, where a profit $p_i$ is assigned to each item, i.e.,
\begin{maxi}|s|[0]
    {x \in \{0, 1\}^n}{\sum_{i\in I} p_i x_i\hspace{2.5cm}}{}{}
    \addConstraint{}{\sum_{i\in I} w_i x_i \leq C.}
\end{maxi}
A quadratic objective $\sum_{i,j\in I^2} p_{ij} x_i x_j$ corresponds to the \textit{binary quadratic knapsack} problem~\cite{PISINGER2007623QuadraticKnapsack}.
In a \emph{multi-dimensional knapsack} problem, we have $J$ dimensions such that
\begin{mini}|s|[0]
    {x \in \{0, 1\}^n, s \in \R^J}{\sum_{j\in J} |s_j|\hspace{2.9cm}}{}{}
    \addConstraint{}{\sum_{i\in I} w_{ij}x_i+s_j = C_j,\, \forall j\in J.}
\end{mini}
If designed with carefully selected weights, we get so-called \emph{market share} problems~\cite{CornuejolsDawande1999} that are explicitly designed to be difficult to be tackled by classical branch-and-bound based methods. These problem instances are dense multi-dimensional knapsack problems \cite[e.g.~Chapter 31]{gonzalez2018handbook}. 
Still, there exist instances with less than $n=100$ variables and $m=8$ constraints which cannot be solved to proven optimality. However, for this problem class the difficulty depends strongly on the particular coefficients, the ratio $c_i/w_i$, and $C$. 
Problems with comparatively small $C$ can be solved quickly by DPs and if the $w_i$ are small compared to $C$, heuristics will give good results. For carefully chosen coefficients, however, these problems can be difficult to solve with classical methods to proven optimality.
We argue that this problem class provides a good quantum optimization benchmarking candidate because there is clear evidence that it is challenging for classical digital systems to find (approximate) solutions already --- even for small system sizes in the multidimensional case.

\textbf{Low Autocorrelation Binary Sequences (LABS).} 
This is a difficult binary non-linear optimization problem~\cite{Packebusch_2016}. 
Given a sequence $S=\left(s_1, \ldots, s_k\right)$ of length $k$ with binary $s_i\in\set{-1, +1}$, the autocorrelations of the sequence correspond to
\begin{align}
    \mathcal{A}_j\left(S\right) = \sum\limits_{i=1}^{k-j}s_is_{i+j}
\end{align}
for $j\in\{0, \ldots, k-1\}$.
Furthermore, the sequence \emph{energy} is defined as
\begin{align}
    E\left(S\right)=\sum\limits_{j=1}^{k-1} \mathcal{A}^2_j\left(S\right).
\end{align}
The goal of the optimization problem is to find a sequence $S$ that minimizes the energy function given above.
There is only one particular instance per size $N$. Hence, the frontier of what is possible is always well defined. So far problems up to size $N=66$ and $N=127$ (skew-symmetric) can be solved to optimality.  
Given the spin-like binary variables, one might be able to find a model formulation that could provide a natural fit for a quantum platform. However, the quartic objective function also imposes some challenges to do so \cite{shaydulin2023evidence}.

\textbf{Quadratic Assignment Problem (QAP).}
The QAP \cite{Burkard1997QAPLIB} has remained one of the great challenges in combinatorial optimization. 
For $I=\{1,\ldots,n\}$
\begin{mini}|s|[0]
    {x \in \{0, 1\}^{n \times n}}{\sum_{i,j,k,l \in I} a_{ij} b_{kl} x_{ik} x_{jl}\hspace{0cm}}{}{}
    \addConstraint{}{\sum_{j\in I} x_{ij} = 1,\, \forall i\in I}
    \addConstraint{}{\sum_{i\in I} x_{ij} = 1,\, \forall j\in I,}
\end{mini}
where $a_{ij}$ and $b_{kl}$ describe the problem instance.
It is still considered a computationally nontrivial task to solve to optimality even for modestly sized problems, i.e., $n\approx 30$. 
If modeled as a binary problem, the number of variables is squared. 
Modeling this problem as a QUBO results in a dense formulation. Nevertheless, the long and unsuccessful search for improvements in classical methods indicates that there might be room for quantum advantage.

\textbf{Sports Timetabling Problems.}
There are regular competitions held to find the best methodology for tackling timetable scheduling \cite{VANBULCK20231249ITC}.
A mathematical formulation that describes problems of this form has been presented, for example by de Werra~\cite{DEWERRA1981SportsScheduling}.
These problems describe scheduling problems of sports leagues or tournaments and are generated in a way that ensures at least one feasible solution exists per instance. 
These problems have a strong combinatorial structure meaning that in different feasible solutions many variables need to have different values. 
 
Interestingly, for several medium sized instances no known (and investigated) method was able to find even a single feasible solution \cite{VANBULCK20231249ITC, vanbulck2023algorithm}. There are different kinds of requirements, i.e., problems can be generated with increasing difficulty. From pure binary problems, to problems that rely on the counting of integer variables or binary representations thereof. 
What is particularly interesting about these problems is that one can generate instances at variable size and difficulty to track progress while it is known that sufficiently large problems are difficult for state-of-the-art methods.

\textbf{Spin glasses.} 
The study of optimization problems with respect to physics Hamiltonians, such as spin glass models~\cite{Ronnow14Defininganddetectingquantumspeedup,albash2018demonstration}, is of great importance and has inspired the development of classical solvers, including simulated annealing~\cite{kirkpatrick:1983} and simulated quantum annealing~\cite{santoro2002theory}.
Spin glasses defined over different graphs have already been used to compare the performance of classical optimization with quantum annealing methods~\cite{Ronnow14Defininganddetectingquantumspeedup,albash2018demonstration}. This is enabled by their natural representation as quantum Hamiltonian.
More specifically, the mapping of a spin glass problem onto a QUBO is trivial and leads to relatively sparse models. 
Although classical solvers perform well on spin glasses defined on regular lattices compared to quantum annealing, the possibility of digital quantum optimizers having better performance still exists --- possibly on non-regular lattice structures \cite{slongo2023quantum}. Moreover, spin glass problems have properties that make them provably average-case hard for various families of classical algorithms~\cite{basso2022performance}. It remains open as to whether quantum algorithms can do better in these average-case settings, and thus, superpolynomial speedups for spin glass problems are still possible. 
Considering all of the above, we argue that they offer an interesting candidate for quantum benchmarks, especially in a cross-platform setting.

\subsection{Demonstrations}\label{sec:benchmarking_demonstrations}
Experimental realizations of large scale quantum optimization algorithms have to address several challenges. Not only is it difficult to cope with hardware induced limitations -- see Sec.~\ref{sec:scaling} for further details -- but it is already a non-trivial task to choose an appropriate benchmarking problem -- see Sec.~\ref{sec:benchmarking_problems}.
In fact, most existing hardware demonstrations of quantum algorithms for optimization are considering QUBO-type problems with a grid structure.
Several papers argue in the following way: First, it is stated that a problem is difficult according to complexity theory, e.g., it is $\NPO$-hard. Then, a feasible -- but not provably optimal -- solution is found by applying a quantum algorithm.
However, finding a feasible but not necessarily optimal solution is not necessarily difficult for state-of-the-art classical methods~\cite{rehfeldt2023,Juenger2019,Punnen2022}.
Thus, there is an inconsistency in the argument, since classical algorithms are ruled out with a reference to complexity theory while allowing to use quantum heuristics.
These results are valuable capability demonstrations, i.e., they can be seen as hardware benchmarks rather than performance benchmarks.
Nevertheless, there is a potential for better performing quantum heuristics, and to show this, we require careful and fair systematic benchmarking.
In this section, we present an overview of selected state-of-the-art experimental quantum optimization results run on gate-based quantum computers with more than 16 qubits that represent excellent starting points for comprehensive quantum optimization benchmarks.

\begin{table*}[ht!]
\caption{An overview of state-of-the-art experimental realizations of optimization algorithms on gate-based quantum computers with more than 15 variables.
In cases where data was not made available in the corresponding publication or the accompanying data repository, we denote this in the respective field with \emph{N/A}. 
\emph{AR} denotes the approximation ratio, given based on the \emph{mean} and the \emph{best} sample value of the experiment, \emph{n.n.~grid} stands for nearest neighbor grid.
Furthermore, JSP, FVQE, QAMPA, GQAOA, and NDAR abbreviate job shop scheduling problem, filtering variational quantum eigensolver, 
quantum
alternate mixer-phaser ansatz,  
greedy QAOA, and noise-directed adaptive remapping, respectively.}
\label{tab:quant_opt_experiments}
\begin{tabularx}{\textwidth}{@{\extracolsep{\fill}}l|llllllll}
\hline
   \ & & & & \multicolumn{2}{c}{AR} \\
   \cline{5-6}
   Problem & Algorithm & Qubits & Density & mean & best & Depth & Year & Ref. \\ \hline
   Sherrington–Kirkpatrick & QAOA  & 17 & 100$\%$ & 0.61 & N/A  & $1\leq$ p $\leq 3$ & 2021 & \cite{Harrigan2021} \\
   MAXCUT (R3R) & QAOA  & 20 & 16$\%$ & 0.64 & 1 & p $=2$ & 2023 & \cite{Sack2023} \\
   MAXCUT (R3R) & QAOA  & 20 & 16$\%$ & 0.94 & 1  & p $\leq 10$  & 2023 & \cite{shaydulin2023qaoa} \\
   MAXCUT (R3R) & QAOA & 22 & 14$\%$ & 0.67 & N/A  & $1\leq$ p $\leq 3$ & 2021 & \cite{Harrigan2021} \\
   MAXCUT (n.n.~grid) & QAOA  & 23 & 13$\%$ & 0.72 & N/A & $1\leq$ p $\leq 5$  & 2021 & \cite{Harrigan2021} \\
   QUBO (JSP) & FVQE & 23 & N/A & 0.88 & N/A &  $1\leq$ p $\leq2 $ & 2022  &\cite{amaro2022case} \\
   MAXCUT (heavy-hex.) & QAOA  & 27 & 8$\%$ & N/A & 1 &  p $=2$ & 2022 & \cite{weidenfeller2022scaling} \\
   MAXCUT (R3R) & QAOA  & 30 & 10$\%$ & 0.59 & 0.83 & p $=2$ & 2023 & \cite{Sack2023} \\
   MAXCUT (R3R) & QAOA  & 32 & 10$\%$ & 0.88 & 1  & p $\leq 10$ & 2023 & \cite{shaydulin2023qaoa} \\
   MAXCUT (R3R) & QAOA  & 32 & 10$\%$ & N/A & 1 & p $=2$ & 2023 & \cite{moses2023race} \\
   MAXCUT (R3R) & QAOA  & 40 & 8$\%$ & 0.58 & 0.78 & p $=2$ & 2023 & \cite{Sack2023} \\
   Sherrington-Kirkpatrick & QAMPA &  50 & 100$\%$ &  0.55 & 0.83   & p $=2$  & 2023 &\cite{maciejewski2023design} \\ 
   Sherrington-Kirkpatrick & QAOA &  50 & 100$\%$ & 0.54 & 0.84  & p $=2$  & 2023 &\cite{maciejewski2023design} \\
   Sherrington-Kirkpatrick & GQAOA &  72 & 100$\%$ & N/A & 0.92  & p $=1$  & 2023 &\cite{Dupont_2023QuantumCombOpt} \\
   Sherrington-Kirkpatrick & NDAR-QAOA &  82 & 100$\%$ & 0.87 & 0.97  & p $=1$  & 2024 &\cite{maciejewski2024improving} \\
   QUBO (heavy-hex.) & QAOA &  127 & 2$\%$ & 0.67 & 0.85 & $1\leq$ p $\leq2$ & 2023 &\cite{Pelofske23QAOA} \\
   PUBO (heavy-hex.) & QAOA & 127 & 2$\%$ & 0.65 & 0.84 & $1\leq$ p $\leq2$ & 2023 & \cite{Pelofske23QAOA} \\
   PUBO (heavy-hex.) & QAOA &  127 & 2$\%$ & 0.73 & 0.89  & $1\leq$ p $\leq5$ & 2023 & \cite{pelofske2023scaling} \\
   QUBO (heavy-hex.) & QAOA &  414 & 0.6$\%$ & 0.57 & 0.69 & p $=1$ & 2023 &\cite{pelofske2023scaling} \\
   PUBO (heavy-hex.) & QAOA &  414 & 0.6$\%$ & 0.56 & 0.68 & p $=1$ & 2023 &\cite{pelofske2023scaling}
\end{tabularx}
\end{table*}

Tab.~\ref{tab:quant_opt_experiments} summarizes existing quantum optimization hardware implementations run with (variants of) QAOA or VQE and problems corresponding either to (weighted) MAXCUT, Sherrington-Kirkpatrick, or other instances of QUBO/PUBO that use more than 15 qubits.
The most commonly employed graph structures are random 3-regular (R3R) graphs or graphs that are  particularly fitted to hardware layouts such as heavy-hexagon or nearest neighbor grid layout \cite{ibm_quantum_platform, Harrigan2021}.
Besides the target problem and employed algorithm, the table also lists the problem size in terms of number variables (which equals the number of qubits in the considered cases), the density (which quantifies the number connections between variables), the best and mean approximation ratio, the depth of the ansatz in terms of the number QAOA operator repetitions (except for results from Amaro \emph{et al.}~\cite{amaro2022case}, which corresponds to the repetitions of a ``RealAmplitudes'' ansatz with pairwise entanglement \cite{Qiskit}), and the year of publication.
All presented instances with fewer than 50 qubits employ a standard QAOA or VQE schemes, i.e., they are optimizing the variational parameters using the quantum computer. Since the references report on shot numbers respectively number of parameter update iterations in various ways (or not at all), the respective numbers are difficult to compare and have, therefore, not been added to the table. However, for future benchmarks, providing such numbers in a comparable format would be important. The results of Maciejewski~\emph{et al.}~\cite{maciejewski2023design} with 50 qubits consider both QAOA, and QAMPA~\cite{larose2022mixer}, a hardware-efficient ansatz derived from QAOA and using the same number of entangling gates.
The 72 qubit experiment by Dupont \emph{et al.}~\cite{Dupont_2023QuantumCombOpt} on the other hand, employs a greedy procedure. In each iteration cycle, several variables are set to values that are evaluated with a classical procedure, hence, producing a smaller optimization problem to be solved on the quantum computer. Further, the demonstrations with more than 100 qubits/variables~\cite{Pelofske23QAOA,pelofske2023scaling} employ either a gridsearch over $\sim$7000 parameter configurations or \emph{concentration of parameters}~\cite{brandao2018fixed, Akshay_2021_concentration, galda2021transferability}, also cf.~Sec.~\ref{sec:unconstrained_discrete_optimization}, where QAOA on large instances can be run on quantum hardware using parameters found for classically trained smaller (e.g., 16-qubit) instances~\cite{pelofske2023scaling}. 
Finally, the recently proposed Noise-Directed Adaptive Remapping meta-heuristic improves 
algorithm performance on noisy hardware by adaptively gauge transforming the cost Hamiltonian, with its effectiveness demonstrated for QAOA with $82$ qubits~\cite{maciejewski2024improving}. 

The table illustrates that most experiments with higher qubit numbers have lower problem densities. Increasing the density for a larger number of qubits would require additional SWAP gates, cf.~Sec.~\ref{sec:scaling}, hence, resulting in increased hardware noise. This phenomenon manifests itself also in the resulting approximation ratios. Considering the approximation ratios, it should be noted that for common graph problems such as MAXCUT the objective values usually lie between 0 and a maximum value $C_{\max}$, while for general problems such as Ising models usually the objective values are inside an arbitrary interval $[C_{\min}, C_{\max}]$. Therefore, a fair comparison of approximation ratios necessitates that the data of each reference is expressed consistently. Thus, we first convert a problem into a minimization problem, and then, we normalize by the range $|C_{\max} - C_{\min}|$, i.e., we define the approximation ratio as $(C_{\max}-\braket{C})/(C_{\max}-C_{\min})$, where $\braket{C}$ denotes the achieved objective value. In fact, some published values do not correspond to range-normalized data, such as the ones from Harrigan \emph{et al.}~\cite{Harrigan2021}. Therefore, the values listed in the table are computed from the corresponding data repositories.
Moreover, several references give the approximation ratio based on the best sample (which may improve with a higher number of shots) and others on the mean sample value $\braket{C}$. Hence, these values are listed separately in Tab.~\ref{tab:quant_opt_experiments}.

In addition, we would like to highlight a work by Fuller \emph{et al.}~\cite{fuller2021approximate} and Moses \emph{et al.}~\cite{moses2023race}, where strategies have been investigated to address problems with more variables than qubits.
In the former,  MAXCUT problems on a planar graph with up to 40 variables are solved with QRAO using up to 15 qubits, cf.~ also Sec.~\ref{sec:unconstrained_discrete_optimization}. In fact, the approach achieves an approximation ratio of 0.905 for the 40 variable problem executed on quantum hardware.
In the latter, a variant of QAOA with mid-circuit measurements and qubit re-use has been tested which could be scaled to 130 variables on 32 qubits, achieving an approximation ratio around 0.8.
In general, the intrinsic differences between the method employed in this work and standard QAOA / VQE approaches make it difficult to compare these results with the ones presented in the rest of this section. Nevertheless, we believe that increasing the variable-to-qubit-ratio beyond one is a very interesting direction to scale towards relevant problems. 

To conclude, these demonstrations build a foundation for future quantum optimization benchmarks. They also highlight the difficulty as well as importance to agree on common metrics to achieve comparable and fair benchmarks. This section is a first attempt for a coordinated effort to achieve this goal.

\section{Illustrative Applications}
\label{sec:applications}

Identifying optimization problems that are promising candidates for a practically relevant quantum advantage is challenging. This is due to the many open questions about future performance of quantum optimization algorithms, as well as the heuristic nature of many of them. Further, benchmarking problem instances known to be hard for today's solvers are often (hand-)crafted. Thus, even if a quantum advantage could be shown for one, it would not necessarily generalize to other problem instances. While these benchmarks are still crucial to identify promising problem structures and track algorithmic progress, we also need to look into real-world problem instances. Further, there might be a \emph{selection bias}, as already discussed in Sec.~\ref{sec:benchmarking_problems}. We are often focusing on models where we have an idea about how to solve them. New formulations of a problem, for which no solver may exist yet, are often unconsciously discarded.

Thus, within this section, we discuss two exemplary industries, i.e., finance and sustainable energy --- two fields that have many outstanding (optimization) challenges. For general discussions on the potential applicability of quantum algorithms for finance and sustainability we refer the interested reader to the following resources \cite{egger20, Orus2019, HermanQuantComp23, berger21, zhou22, paudel22}. Here, we discuss families of optimization problems arising in the two industries with increasing model complexity. Importantly, the goal of the following discussion is not to claim that the selected use cases are the most likely candidates for near-term quantum advantage. Instead, they should serve as a playground of (possibly simplified) real-world problems that can help to set the course for research and link to problem classes discussed in Sec.~\ref{sec:problem_classes}. Further, the discussion highlights that being practically relevant requires taking into account many details and complications that most published results in the related quantum optimization literature are lacking. Thus, defining a family of problems with increasing complexity helps to determine progress on the quest towards quantum advantage in optimization.

\subsection{Financial Asset Allocation}\label{sec:applications_finance}
Optimization problems in finance span a wide range of applications, from the quantification and management of risks, to asset allocation, option pricing, macroeconomic modeling, algorithmic trading, lending and more~\cite{Zenios2007}. 
The goal of this section is to present a variety of challenging optimization problems from the financial realm that may provide interesting use cases for quantum algorithmic development, as the technology advances.

Firstly, we give a short overview of the field and a discussion of advancements in quantitative financial theory pertaining to optimization. We provide a few example problems and a discussion of their practical limitations in Sec.~\ref{sec:applications_finance_overview}. Then, we present the current state of quantum optimization and early investigations of use cases in finance in Sec.~\ref{sec:applications_finance_refs}. Next, we provide a deep-dive into a particularly hard problem with many unresolved questions and evident practical relevance: the \emph{asset allocation problem} and different formulations thereof, with increasing levels of complexity in Sec.~\ref{sec:applications_finance_deepdive}. Finally, in Sec.~\ref{sec:applications_finance_outlook}, we conclude and outline further research directions that could act as an exploration ground for quantum optimization applications in finance.

\subsubsection{Overview}\label{sec:applications_finance_overview}
The global financial system is the backbone of economics, embedding into every part of society through either direct digital channels, such as monetary transactions and contracts, or in the form of indirect physical carriers, such as transferable goods and services. It represents the world's largest human-made regulated stochastic network, hosting trillions of interactions between market makers, traders, consumers, regulatory control and monitoring entities. The system is powered by a variety of distributed computing technologies and operates on the shoulders of over 120~years of financial theory, heuristic analysis, and a broad range of mathematical tools and probabilistic models informing decision making --- within regulatory and business constraints. This enables one to manage risk, counter financial crime, derive investment and business decisions, and service consumer needs with personalized experiences~\cite{ibm2023embeddedfinance}.

Despite a century of progress, many models in regulated financial institutions suffer from different degrees of approximations and model trade-offs due to irregular and/or scarce information and the need to balance model complexity, computational feasibility, and regulatory, as well as business constraints. 
This is generally driven by challenges originating from fundamental unknowns underlying the interaction dynamics between market participants~\cite{bachelier1900theory, hayek1945economics}, data quality anomalies in observables and the overall fast-paced change of the world. 
The latter may gradually become a notable challenge for many modeling trade-offs relied on today, and thus, could influence the financial system's systemic resilience~\cite{imf2020systemicrisk, ecb2023fsr, imf2023stability}. For example, more complex probabilistic drivers are emerging from new consumer-banking interactions~\cite{euaiact2023}, climate change and energy transmission risk guidelines~\cite{bis2021climate}, to the economic impact of changing geopolitical dynamics~\cite{blackrock2023geopolitics}. This demands the development and analysis of new modeling techniques and optimization algorithms to better inform decision making under uncertainty. In many finance problems, even \emph{small improvements} can have a \emph{significant impact}. Although, it is still unclear if and how a potential quantum advantage could be realized for financial optimization problems -- even as the underlying technology advances -- a thorough understanding of the broad spectrum of classical models and their limitations in the financial industry could help guide the development of quantum optimization approaches for use cases that deal with uncertainty in objective functions, decision variables or constraints.

While not aiming for rigor or completeness, the following provides a bigger picture and short recap of advances and challenges in financial optimization problem domains for their respective input dependencies. This links to the roots of quantitative financial theory and the understanding that we cannot explain with confidence \emph{why} the prices of financial assets move, but only attempt to model \emph{how} they move. As a result, the modeling of price dynamics and volatility has become a core focal point in finance with many models aiming to capture its essence: from Brownian motion~\cite{bachelier1900theory} to perfect delta-hedging --- assuming no price jumps and financial crashes~\cite{black1973pricing}; capturing volatility clustering phenomena~\cite{Cont2007} with GARCH~\cite{Bollerslev1986} and Heston~\cite{heston1993} models --- assuming volatility fluctuations decay over one timescale; Multifractal Random Walks~\cite{stochvol2000} and Rough (Heston) Volatility models~\cite{volrough2018, Bouchaud2017, Dandapani2021} capturing most \emph{stylized facts}~\cite{Cont2001} from empirical observations of financial time series, such as heavy-tailed probability distributions~\cite{nair_wierman_zwart_2022}. This list summarizes just a few milestones over a century of financial modeling, effectively demonstrating that capturing market dynamics with such complex input patterns is difficult.

There are also many optimization problems in finance that do not need to account for input uncertainties to prove valuable for business, such as linear programming problems (see Sec.~\ref{sec:continuous_optimization}), quadratic programming problems (see Sec.~\ref{sec:discrete_optimization}), and (mixed) integer programming problems (see Sec.~\ref{sec:milp-miqp}). However, for most instances corresponding to these classes, there are established, well-performing classical solvers, such as interior-point and simplex methods for linear and quadratic programs, or branch-and-bound and combinations with cutting-plane methods for integer programs. Nevertheless, these problems form an excellent ground to benchmark and develop quantum optimization algorithms and potentially detect value in specific financial applications.

Financial optimization problems of substantial complexity that face limitations when being approached with state-of-the-art solvers may, hence, be interesting subjects for the exploration of quantum algorithms and are summarized here with a few examples:
\begin{itemize}
    \item Dynamic programming problems (see Sec.~\ref{sec:dynamic_programming}) for pricing and hedging of derivatives on binomial lattices, or structuring of collateralized mortgage obligations with maximized profits from issuance;

    \item Stochastic programming problems (see Sec.~\ref{sec:optimal_control}) for minimizing bond portfolio credit risk using a Conditional Value-at-Risk measure, or asset-liability management maximizing wealth, or retiring outstanding debt at minimal cost, or creating a synthetic option strategy to reach desired payoff at the end of a planning horizon;

    \item Robust programming problems (see Sec.~\ref{sec:robust_optimization}) for optimizing portfolios while taking estimation risk of input parameters into account, or determining lower and upper bounds on the price of a security;

    \item Multi-objective programming problems (see Sec.~\ref{sec:multi_objective_optimization}) for dynamic margin-volume balancing to price mortgages combining optimal revenue, balance sheet and business objectives, or optimizing an investment portfolio of stocks combining return, risk diversification, incentivizing decarbonization and Environmental, Social, and Corporate Governance (ESG)~\cite{oecdESG} objectives.
\end{itemize}

\subsubsection{Related Work}\label{sec:applications_finance_refs}
The majority of existing quantum optimization research in finance focuses on investigating quantum variants of linear and quadratic optimization problems and simplified (mixed) integer programs. Most publications employ a QUBO formulation, combined with a ground state solver such as QAOA or VQE (see Sec.~\ref{sec:discrete_optimization}) that is executed on numerical simulators, gate-based hardware, or annealers, typically for systems with less than $50$ variables. In the following, we present a selection of the corresponding papers.

Brandhofer \emph{et al.}~\cite{Brandhofer2023} study portfolio optimization using QAOA for quadratic binary optimization constrained by the number of assets. That is, the presence or absence of a particular stock in a portfolio has a value of 1 or 0 respectively, and the sum of the assets is the value of the investment, which is a (simplified) budget constraint. Risk is minimized by having the smallest combined covariance between all pairs of stock values as represented by a covariance matrix; this minimum ensures portfolio diversification.

In a numerical study, Hodson \emph{et al.}~\cite{Hodson2019} compare a Quantum Alternating Operator Ansatz to QAOA for models that incorporate trading and investment constraints. Baker and Radha~\cite{Baker2022} evaluated the solution quality as functions of qubit number and circuit depth. Mugel \emph{et al.}~\cite{mugel22} compare methods for dynamic portfolio optimization on quantum processors using 8 years of data for 52 assets with quantum annealing, VQE, and a quantum-inspired optimizer based on Tensor Networks using pre-processing to reduce the problem dimension. Venturelli et al.~\cite{Venturelli_2019} compare the performance of reverse quantum annealing for portfolio optimization problems formulated as QUBOs with classically executed genetic algorithms. Constrained QAOA solutions to portfolio optimization are also presented in Herman \emph{et al.}~\cite{herman23}, who use Zeno dynamics with rapid sampling to restrict the solutions to a constrained subspace. Giron \emph{et al.}~\cite{10250961} use QAOA to study the problem of collateral optimization --- the optimal allocation of financial assets to satisfy obligations at minimum cost. In a broader context, a quantum optimized portfolio allocation could feed into a risk model evaluated on a quantum computer~\cite{Woerner2019, Egger2019, Stamatopoulos2020}.

Other finance optimization applications that have been proposed for quantum algorithms include transaction settlements, where the goal is to find the maximum number of cash transactions between multiple parties given their liquidity~\cite{braine2021quantum}, and Anti Money Laundering, which matches network motifs on snapshots of transaction graphs~\cite{calude17}.

\subsubsection{Deep Dive: Optimal Asset Allocation}\label{sec:applications_finance_deepdive}
Among the hardest problems in the financial industry is the market adaptive search for an optimal portfolio composition with an investment allocation in financial assets that maximizes the return of investment within a given risk appetite and time frame, while complying to possibly other objectives and constraints. Reasons for its complexity relate to the stochastic nature of the financial system and various statistical and computational challenges as discussed in Sec.~\ref{sec:applications_finance_overview}. It is also a problem where marginal improvements in practical solutions can have substantial impact. In this context, the term \emph{solution} should be discussed. It is difficult to find a provable or universal solution, instead one accepts an outcome with different forms of trade-offs to encompass particular use case requirements, market environments, asset class restrictions or investment scenarios. While over the past $70$ years, both academic researchers and industry practitioners have created a large repertoire of targeted heuristic approaches and rigorous mathematical tools, it remains one of the most hunted challenges in finance~\cite{Kolm2014}. This partially explains why the quantum computing community is exploring this use case in various quantum optimization for finance publications~\cite{Brandhofer2023, herman2023, dalzell2023quantum, Hodson2019}. However, we stress that it remains unclear how to systematically advance towards the complexity and value frontier at which a real-world practical impact can be realized.

\begin{figure}
    \includegraphics[width=0.95\columnwidth]{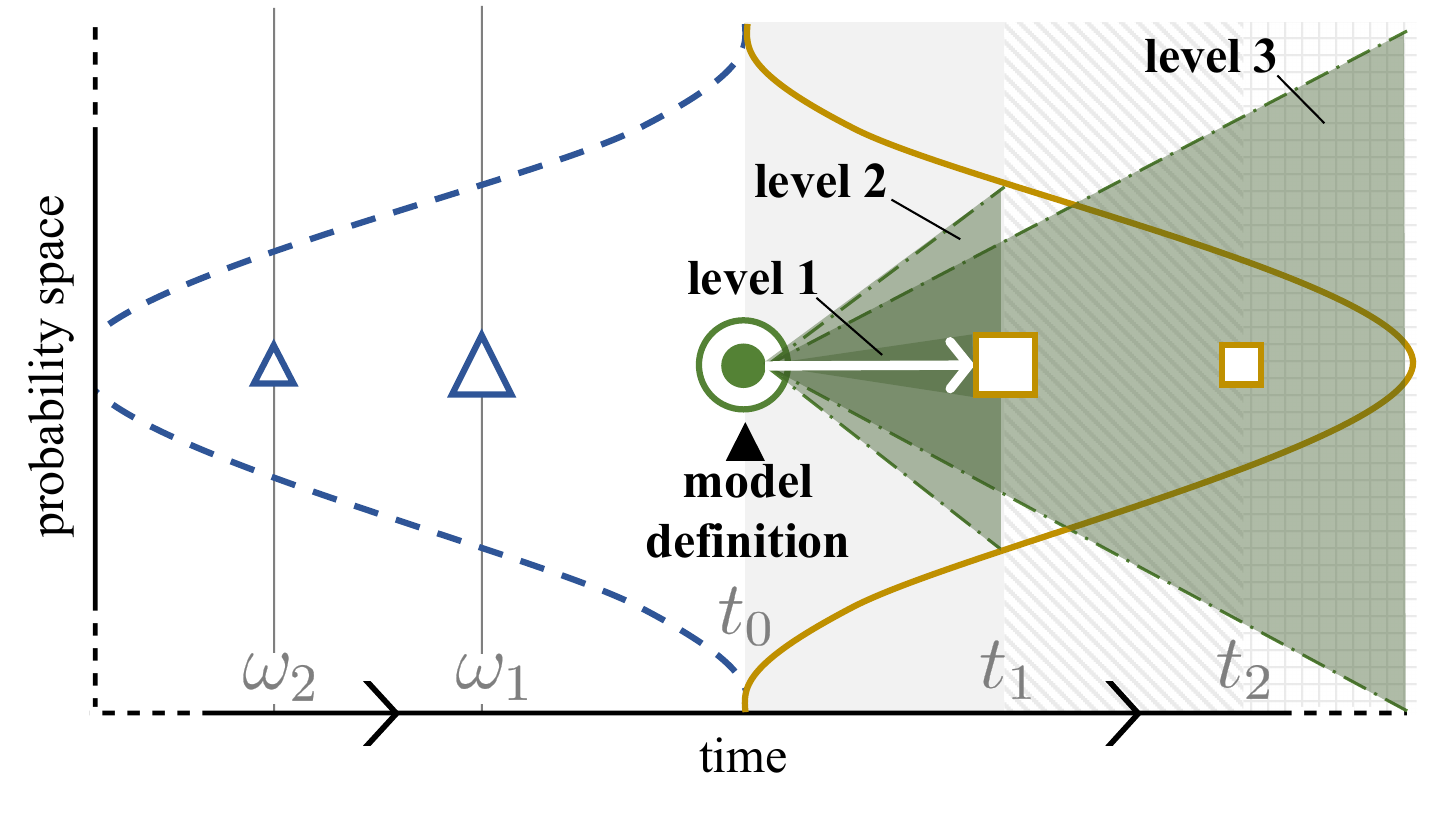}
    \caption{An illustration of the optimal asset allocation at time $t_0$ (model definition), considering perfect predictors of asset valuation at the end of investment horizon $t_1$ given prior statistics $\omega_1, \omega_2, \ldots$ (level 1), then taking uncertainties of these predictors into account but still with a limited view until $t_1$ (level 2), before anticipating a multi-period allocation strategy and respective uncertainties beyond $t_1$ (level 3).}
    \label{fig:finance_assetalloc_illustration}
\end{figure}

In this section, we introduce the concept of value-guided \emph{levels} with examples representing incremental increase in use case complexity by respective problem formulations --- for reasons of simplification in the mean-variance framework, as illustrated in Fig.~\ref{fig:finance_assetalloc_illustration}. The aim is to systematically explore different challenges of this use case and possibly inspire the exploration and research of quantum optimization algorithms in this domain.

\textbf{Model Definition.} From the ground up, it needs to be well understood within which environment we are attempting to construct an asset allocation problem. For instance, we obviously cannot comprehend all the markets with interactions between millions of traders with each a different investment style, behavior, risk limits or trading frequency. But rather than accepting the entire universe of possibly feasible portfolios $\mathcal{S}$, a preliminary \emph{steering of the downstream complexity} for all other levels can help build up and realistically frame the use case. The field of investing is found with very different business domains, expectations and requirements that yield various degrees of complexity to start with. For example, a wealth manager for retail, affluent or high-net-worth individuals may have different requirements for composing or rebalancing investment portfolios than say, a large asset manager. While this seems rather obvious, it is key to formulate the model purpose, and thus, its expected performance. In particular, an informed pre-selection of diversified asset classes, industry exposures, risk profiles or expected dynamics for a given investment time horizon helps build a more robust starting point, $t_0$, for the optimization problem.

\textbf{Level 1.} At this stage we begin with the simplest problem formulation based on Harry Markowitz's introduction of modern portfolio theory in $1952$~\cite{Markowitz1952}, by formalizing portfolio diversification and optimal return-risk trade-offs with a mean-variance framework. It assumes a risk-averse myopic investor in a frictionless, one-period financial market with exact knowledge of parameters capturing asset price dynamics. The aim is to find an optimal portfolio $x = (x_1,...,x_N) \in \mathbbm{R}^N$ from permissible portfolios $\mathcal{S}$ with fractions $x_i$ of a fixed budget to be fully invested into $N$ pre-selected, non-redundant assets at time $t_0$, based on estimates of expected returns $\mu \in \R^N$ and their positive semi-definite covariance matrix $\Sigma \in \R^{N \times N}$ at the end of one investment period $t_1$. This can be formulated in three equivalent variants as \emph{quadratic programming problems} with linear constraints and without uncertainty in the parameters:
\begin{mini}|s|[0]
    {x \in \R^N}{ \frac{1}{2} x^\top \Sigma x }{}{}\label{eq:finance_level1}
    \addConstraint{}{\mu^\top x \geq r_{\text{min}}}
    \addConstraint{}{\mathbbm{1}^\top x = 1}
    \addConstraint{}{x \in \mathcal{S}},
\end{mini}
where the convex quadratic objective minimizes the portfolio risk $x^T \Sigma x$ with an expected minimal portfolio return $r_{\text{min}}$. Instead, one can also maximize the expected return $\mu^T x$ while bounding the risk $x^T \Sigma x \leq \gamma^2$ as a quadratic constraint, which can be transformed into conic form~\cite{BenTalNemirovski2001}. Another alternative is to maximize a concave quadratic utility function $\mu^T x - \lambda/2~x^T \Sigma x$ with the risk-aversion factor $\lambda$ for steering the return-risk trade-off and variable focus on tail risk. However, all of these formulations are relatively trivial to solve, especially in the continuous setting, where a closed-form solution may be derived using Lagrangian multipliers~\cite{Heston1972} considering equality constraints~\cite{Gould2001} and with Karush-Kuhn-Tucker conditions considering also inequality constraints~\cite{Karush1939}. Notably, realistic market conditions are not captured by this model formulation. Nevertheless, this simple model offers excellent benchmarking opportunities to evaluate advances in quantum technology. In fact, the model complexity can also be increased in a controlled manner by adding additional constraints that might match a pre-defined investment strategy and, as the term $x \in \mathcal{S}$ is meant to reflect, to further constrain the universe of portfolio configurations and test with an increasing number of additional linear constraints that match a pre-defined investment strategy. In practice, for a sufficiently large number of constraints one often has to rely on numerical approaches. As discussed in Sec.~\ref{sec:applications_finance_refs}, so far, most quantum approaches to this use case employ QUBO formulations with binary or integer encodings, which result in an \textbf{NPO}-hard combinatorial problem and constraints that are converted into penalty terms~\cite{Glover2019} of the objective function. Since these penalty terms only facilitate approximate encoding of the constraints, additional substantial post-processing would be required to achieve reliable results.

\textbf{Level 2.} A severe limitation of Level~1 is the ignorance of uncertainty in the inputs and instability of the optimization~\cite{Michaud2015}. Thus, it is of limited practical value to derive actual investment decisions. In fact, the optimization is highly sensitive to \emph{small changes} in the estimated expected returns $\mu$ and covariance matrix $\Sigma$, leading to \emph{large differences} in the resulting optimal portfolio $x$. At this stage, the problem formulation is extended to take the stochastic nature of underlying observables into account. The root challenge comes from the portfolio-specific operating regime of the observation ratio $q = N/T$ given the prior number of empirical (sample) realizations $T$ for the $N$ considered assets at $t_0$. In the large limit with $q \rightarrow 0$, the sample estimators $\mu$ and $\Sigma$ would converge to the \emph{true} realized (population) $\mu_\text{true}$ and $\Sigma_\text{true}$, respectively. But in reality most asset allocation problems suffer from a lack of data (except maybe in high-frequency trading) with $q > 0$. For example, a realistic scenario for stocks could be $N = 600$ with $T = 2500$ of about 10~years of daily returns and, thus, $q = 0.24$. In case of a few thousand stocks, the issue of stock lifetime and structural evolution of markets over time becomes relevant, shifting the problem into the large dimensional limit regime with $q \sim 1$ and fast diverging estimators with $q > 1$. If one would aim to increase the $T$ statistics with intra-day observations, one may not capture the dynamics for lower frequency trading strategies anymore and possibly introduce biases in the resulting optimal $x$. This makes the problem very complex, even for a small number of assets, and usually forces the industry to simplify $\mathcal{S}$ at the level of model definition. 
In the following, we are going to present several approaches to address this challenge, ordered by increasing levels of complexity: first we de-bias and reduce the estimation error of $\mu$ and $\Sigma$ prior to the optimization, then adding regularizing techniques to the optimization itself, before combing both. 
Among various methods, the Black-Litterman model~\cite{BlackLitterman1992} could be leveraged to create an expected return prediction by combining actual market returns with independent expert beliefs about their future, taking into account, for instance, recent news, economic forecasts, analyst ratings, or financial statements~\cite{AvramovZhou2010, CorPeTue2018}. 
Alternatively, a robust optimization problem can be formulated to achieve a stabilizing effect on $\mu$ during the optimization itself. For this purpose, we define the ellipsoidal uncertainty set $\mathcal{U}_\mu = \{ \mu~|~(\mu - \mu_0)^\top Q_\mu^{-1} (\mu - \mu_0) \leq \kappa^2 \}$, where $Q_\mu$ is the covariance matrix of estimation errors in $\mu$, and $\kappa$ is the uncertainty aversion defining the width of the uncertainty. This induces a new optimization problem by replacing $\mu$ with:
\begin{eqnarray}
    \tilde\mu(x) &\coloneqq& \argmin_{\mu \in \mathcal{U}_\mu} \mu_0^\top x - \kappa \sqrt{ x^\top Q x }.
\end{eqnarray}
Even if the resulting improved estimate of $\mu$, denoted by $\tilde{\mu}$, still substantially deviates from $\mu_\text{true}$, finding the minimal out-of-sample risk portfolio for it may still be valuable in practice. 
The sample covariance matrix $\Sigma$ for $q \in (0,1]$ can suffer from small eigenvalues near zero, leading to a potential underestimation of the realized (out-of-sample) risk exposure at $t_1$. To address this issue, Random Matrix Theory~\cite{Mehta2004, Wigner1955, NeumannGoldstine1947} provides efficient tools to de-bias the eigenvalues of $\Sigma$ such that the resulting $\widetilde{\Sigma}$ is closer to $\Sigma_\text{true}$~\cite{Ledoit2011, RIE2016, BenaychGeorges2021, BunBouchaudPotters2017, LedoitWolf2017}. 
A factor modeling method to represent the portfolio returns with a small number $M$ of common factors~\cite{ChamberlainRothschild1983} can additionally help to circumvent the curse of dimensionality if $M \ll N, T$. 
Furthermore, one may want to further regularize $\Sigma$ by adding $L_1$- or $L_2$-norm constraints to the optimization. 
Arguably one of the most important ones is the consideration of a transaction cost function $\mathcal{C}(x \mid x_0)$ that estimates the cost for rebalancing the portfolio $x_0$ at $t_0$ towards the optimal portfolio $x$ at $t_1$, taking for instance broker fees and implementation shortfall into account. This results in the following problem formulation:
\begin{maxi}|s|[0]
    {x \in \R^N}{ \tilde{\mu}^\top x - \mathcal{C}(x \mid x_0) - \frac{\lambda}{2} x^\top \widetilde{\Sigma} x }{}{}\label{eq:finance_level2_trx}
    \addConstraint{}{\mathbbm{1}^\top x + \mathcal{C}(x \mid x_0) = 1}
    \addConstraint{}{x \in \mathcal{S}},
\end{maxi}
where $\mathcal{C}$ determines the complexity of the problem and depends on how realistic transaction costs should be represented. 
For example, a piecewise linear non-convex function capturing transaction volume and fixed costs for buying and selling an asset can be formulated, or 
a more realistic variant with stronger penalization on turnover considers costs proportional to the $L_1$-norm of rebalancing with $\mathcal{C}(x \mid x_0) = \beta \| x - x_0 \|_1$, where $\beta > 0$ denotes a cost parameter.  
While formulations of this problem that are based on $\mathcal{C}$ employ strong regularization on $\Sigma$, depending on the respective $q$ regime set out in the model definition, the resulting $x$ may still be biased and sensitive to estimation errors. 
This can be further addressed with two methods. The first one uses a non-parametric bootstrap resampling approach~\cite{MichaudR2008} 
and the second one uses a nested clustering approach~\cite{Prado2019} that partitions $\widetilde{\Sigma}$ by forming clusters of highly-correlated assets. 
The latter approach can help to prevent error propagation of intra-cluster noise across clusters. Due to a reduction of dimensionality and better control of noise isolation with smaller sample sizes, the exploration of near-term variational quantum algorithms with realistic stock portfolio sizes may become possible in this setting.

\textbf{Level 3.} Until this point we have restricted ourselves to a static one-period market environment. At this level, we introduce the next complexity layer by considering market dynamics and temporal dependency. For example, similar to volatility, correlations fluctuate, evolve with time and jump or even flip signs~\cite{REIGNERON20113026, CCorr2021}. However, our understanding of correlation dynamics is nowhere near our understanding of volatility, which creates an interesting domain to explore with quantum computing. Similarly, this is the case for multi-period optimization, which could help capture inter-temporal effects, such as accounting for time-varying constraints, return forecasts, cost-favorable positions for trading in following periods, or foreseeable events impacting risk, trading volume or liquidity. Although there is a lot of research interest, given the computational hardness of the multi-period domain, most practitioners, as well as classical enterprise solver suites, cannot effectively tackle these use cases~\cite{Kolm2014}. Essentially, the $h$-period optimization problem can be \emph{generally} formulated as follows:
\begin{maxi}|s|[0]
    {x_{t + 1},\ldots, x_{t + h}}{ \mathbbm{E} \left[ \mathcal{X}(x_{t+1},\ldots,x_{t+h}) \mid \mathcal{F}_t \right] }{}{}\label{eq:finance_level3_general}
    \addConstraint{}{x \in \Theta},
\end{maxi}
where $\mathcal{X}$ is the inter-temporal utility function given the filtration $\mathcal{F}$ associated to the probability space at the $t^{\text{th}}$ period with a set $\Theta$ of constraints. If we can assume that $\mathcal{X}$ is separable in time, then the objective function can be written as $\text{min}_x \{ a(x) + b(x) \}$ with a static (forward-looking) part $a(x) = \sum_\tau a_\tau(x_\tau)$ and a dynamic (coupling) part $b(x) = \sum_\tau b_\tau(x_{\tau-1}, x_\tau)$ with $\tau = t+1,\ldots,t+h$, and their respective separation of constraints becomes $x \in \Theta^{(a)} \cap \Theta^{(b)}$. Here, $a_\tau$ represents a one-period optimization program such as the mean-variance terms discussed in Levels~1 and 2, while $b_\tau$ serves as an intra-period regularization penalty such as the $\mathcal{C}(x_\tau | x_{\tau - 1})$ transaction cost functions discussed in Level~2. 
Problem formulations of this form and other multi-objective settings are gaining substantial interest in the financial industry, and will play a key role in the coming years. Thus, it is interesting to investigate new approaches for this use case with quantum technology and inspire the development of new quantum algorithms. 

This concludes our short deep-dive illustration with increasing levels of complexity and selected examples within the mean-variance framework. But notably, this can be extended as our exploration of quantum solutions evolve. In particular, the development of new quantum optimization solutions should transition from current Level~1 investigations to Level~2 and 3 challenges, and with that drive research towards handling optimization problems under uncertainty in order to aim for valuable contributions to the financial industry.

\subsubsection{Outlook}\label{sec:applications_finance_outlook}
The financial industry offers a broad spectrum of hard optimization problems and demands the exploration of new solution approaches to tackle many of today's trade-offs. This short condensed review provides a satellite view of open optimization challenges relevant to the financial industry. A selection of use cases are discussed along with their classical challenges and how optimization problems can be constructed with increasing complexity levels. The main challenge remains to effectively combine both the modeling and optimization of problems to achieve practically relevant results for industry use cases. However, there are many more directions to investigate, such as handling skewed, non-Gaussian heavy tailed probability distributions, Bayesian approaches and industry sector or subgroup specific priors, or non-Markovian decision processes. All of these aim to inspire and guide the exploration of research, development and benchmarking of potential industry-specific quantum optimization applications.

\subsection{Sustainable Energy Transition}
\label{sec:applications_sustainability}

Sustainability describes the goal of a responsible use of resources to guarantee a long-term existence of humanity on earth in a healthy environment. While sustainability has multiple dimensions, a common focus is put on environmental problems, such as countering climate change. For this paper, we focus on a subsection of sustainability, that of sustainable transition of the energy sector. It is a particularly relevant area as it is rich with optimization problems with societal benefit.
We discuss multiple problems in this domain that can serve as illustrative test cases with different challenges to benchmark optimization algorithms. The use cases described here serve as an interesting playground to generate examples derived from real-world data to study corresponding quantum optimization algorithms. While they might not necessarily be candidates for a practical quantum advantage in near future, they can help to improve our understanding about problem characteristics where quantum optimization may have some potential over classical algorithms and where not.

\subsubsection{Power Grid Overview}
\label{sec:future_energy_systems}

For decades, society has relied on carbon-intense and centralized energy sources \cite{Smi19}. Few powerful generation sites were responsible for meeting energy demands, and adjusting energy generation was relatively simple with fossil fuels. However, the energy sector is changing rapidly. Technological advancements and climate-change awareness have led to an increase in renewable energy sources. Power grid digitization  \cite{en13020494, BVS+18} has resulted in greater connectivity and interactions between customers, producers, utilities, and grid operators. This has revolutionized how we manage and optimize energy resources by enabling real-time monitoring, data-driven decision-making, and enhanced control over energy consumption and distribution. 

However, the transition to renewable energy sources and the digitization of the power grid are not without challenges. 
Renewable energy sources such as solar and wind are intermittent and probabilistic in nature, and, they require the coordination of a large number of assets; for example, to match energy supply with demand. Energy storage solutions are essential for managing the variability associated with renewable energy production, but they present their own set of challenges, including cost and scalability. Additionally, power grid digitization introduces multiple challenges including cybersecurity and data privacy. Adapting grid systems to this new paradigm will necessitate substantial investment and innovative solutions, potentially surpassing the capabilities of classical computer systems.

In this section, we discuss three types of optimization problems in power grids. We start with a brief overview of the area in Sec.~\ref{sec:future_energy_systems} and related quantum optimization work, in Sec.~\ref{sec:sustainability_related_work}. In Sec.~\ref{sec:e_mobility}, we introduce \emph{electric mobility} (e-Mobility) and present three families of e-Mobility use cases and how they connect to the problem classes and algorithms in Sec.~\ref{sec:problem_classes}. 
Finally, Sec.~\ref{sec:sustainability_outlook} concludes and outlines potential research directions.

\subsubsection{Related Work} \label{sec:sustainability_related_work}
Research in quantum optimization for power grids is currently in its early stages, with a predominant focus on the unit commitment problem (UCP) \cite{1295033,van2018large}.
Koretsky \emph{et al.}~\cite{koretsky21} consider the UCP for power networks by combining quantum and classical methods: QAOA handles the binary on-off variables for each unit and a classical optimizer handles the continuous variables for how much power each unit should provide. Takahashi \emph{et al.}~\cite{takahashi23} writes the UCP as a QUBO problem for network switches equal to 0 or 1 if open or closed, subject to connection, voltage, and maximum current constraints.
They model the constraints as penalties in the objective function and solve the problem with annealing. Similarly, Halffmann \emph{et al.}~\cite{halffmann22} and Braun \emph{et al.}~\cite{braun23b} formulate the UCP as a QUBO with penalty terms. Mahroo and Kargarian \cite{mahroo22} decompose the UCP into a sequence of quadratic continuous and binary (unconstrained) subproblems, similarly to ADMM~\cite{gambella2020mixed, gambella2020multiblock}, cf.~Sec.~\ref{sec:milp-miqp}. 
Other non-UCP quantum optimization works in power grid include \cite{10247202, 9863874}, which formulate a power flow problem as a QUBO with constraints encoded as penalties in the objective, and \cite{SCHWORM202347}, which addresses an energy supply scheduling problem with storage, where the optimization is formulated as a constrained quadratic program and solved with annealing. 

All of the works mentioned above have in common that they focus on small, illustrative problems and that they explore simple mathematical models (mostly QUBOs) that do not capture the complexity of real-world use cases. Although these works may not have a direct practical impact yet, they could help to pave the way for advancing the corresponding theory, problem formulations, algorithms, and to understand practical requirements on optimization solutions to have practical impact.

\subsubsection{Leveraging the Flexibility of e-Mobility}\label{sec:e_mobility}
As outlined in Sec.~\ref{sec:future_energy_systems}, the world is progressively adopting carbon-neutral energy generation methods \cite{EU-report23-05,france-nuclear}. While renewable sources like wind and solar power play a vital role, they often produce energy when demand is low. To address this disparity, it is crucial to invest in energy storage solutions, such as batteries, and the optimized coordination thereof. 

Electric vehicles (EVs) are an energy storage solution of increasing importance. In 2022, the number of EVs on the road exceeded 26 million \cite[pp.\ 14]{ev-outlook-2023}, and the global fleet of EVs is expected to grow to about 240 million in 2030 --- 10\% of the road vehicle fleet \cite[pp.\ 109]{ev-outlook-2023}.
To put this opportunity in perspective, EV batteries can typically store  20 to 100 kWh and can have an output power of 10 kWh when connected to a Type 2 charger. Thus, 400 EVs, in principle, can deliver 4 MWh of energy to the grid, which is approximately the same output power of an energy storage substation \cite{substation-3.8MWh}. 

To leverage e-Mobility's potential to stabilize the energy network and balance supply and demand, it is crucial to address three challenges: The first challenge is to increase the EVs' penetration, for example, by facilitating access to charging infrastructure \cite{en12142739}. This is particularly important in densely populated areas where overnight charging may only be available for some EVs. The second challenge is to effectively control how EVs store energy \cite{LUND20083578}. Unlike conventional energy storage systems, EVs consume energy from the grid, and their batteries must be sufficiently charged by strict deadlines. The final challenge is to integrate EVs into the power grid infrastructure \cite{LUND20083578}, enabling energy retailers to store energy in EVs that they can then inject back into the grid, i.e., an EV acts as a ``virtual power plant.''

Next, we discuss three optimization-related e-Mobility use cases, each addressing one of the challenges mentioned earlier. 
We start with a ``core'' mathematical problem and then extend this to a family of problems with increasing practical relevance, linking it to the problem classes discussed in Sec.~\ref{sec:problem_classes}.

\begin{figure}
\includegraphics[width=0.85\columnwidth]{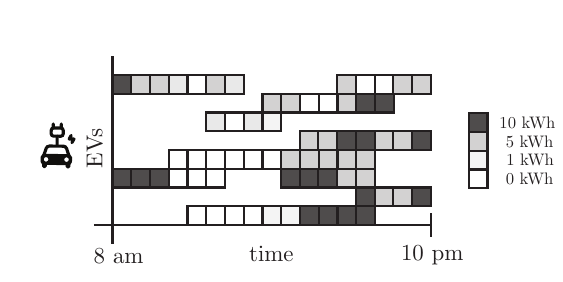}
\caption{Illustrating a possible instance for the problem of scheduling the charging of EVs in a parking site (use case 1).}
\label{fig:EV_scheduling}
\end{figure}

\textbf{Use case 1: Scheduling EVs in a parking site.} 
This use case consists of allocating EVs to parking spaces in a parking site \cite{SZL+21, SYH+22}. In short, consider a parking site that has two types of parking spaces: regular parking spaces and parking spaces where EVs can be plugged in. EVs arrive at the parking site indicating the kWh they want to replenish at each time, e.g., in each hour. With that information, the parking site administrator decides where to allocate the EVs to maximize the charging volumes and thereby profit. To simplify the problem formulation even further, let us assume that the EVs arrival and departure times are known in advance. Figure \ref{fig:EV_scheduling} shows the problem schematically, where the $y$-axis in the figure represents the EVs. The goal is to select EVs such that the number of EVs at a given time does not exceed the number of chargers and the input power of the site.

We can formalize the problem as follows. Let $N$ be the number of EVs and $K$ the total number of intervals (e.g., 24 intervals, where each interval corresponds to an hour). Next, let $u_n \in \{0,1\}^K$ be a vector that indicates the (known) presence of EV $n \in \{1,\dots,N\}$ at time $k \in \{0,\dots,K-1\}$, and $v_n \in \R^K$ be the value that the parking site will obtain by admitting EV $n$ (e.g., dollars). Thus, the inner product $v_n^T u_n$ denotes the total value for the parking site if admitting EV $n$. 
Also, let $d_n \in \R^K$, where the $k$-th entry indicates the energy requested by EV $n$ at time $k$. The parking site has a maximum input power of $E$ kWh and has a maximum capacity of $M$ EVs. 
Then, the optimization problem is the following:
\begin{maxi}|s|[0]
{x\in \{0, 1\}^N}{ \sum_{n=1}^N x_n ( v_n^T  u_n)\hspace{1.75cm}}
{}{} \label{eq:mult-dimensional-knapsack}
\addConstraint{\sum_{n=1}^N x_n u_{n,k} \le M, \, \forall k} 
\addConstraint{ \sum_{n=1}^N x_n d_{n,k} u_{n,k} \le E, \, \forall k,}
\end{maxi}
where $x_n$ denotes the decision variable of whether to accept EV $n$ to a parking space where it can be plugged in or not. 
That is, Eq.~\eqref{eq:mult-dimensional-knapsack} is a binary optimization with $2K$ knapsack constraints; $K$ knapsacks for the parking spaces, and $K$ knapsacks for the energy available in the parking site.

Some instances of the multi-dimensional knapsack problems can be challenging to solve, even when the problem instances are relatively small, cf.~Sec.~\ref{sec:benchmarking_problems}.
Thus, studying problems of the form in Eq.~\eqref{eq:mult-dimensional-knapsack} can be of interest for near-term quantum optimization algorithms with a potential practical application. Of course, we need to benchmark specific practically relevant problem instances to assess whether they are difficult classically. If the instances prove difficult, we can proceed to explore whether there is an opportunity for a practically relevant quantum advantage. Since reality and real world scenarios are often more delicate, there will likely be additional constraints and cost function adjustments to Eq.~\eqref{eq:mult-dimensional-knapsack} thereby increasing its general difficulty for classical solvers. 
Some interesting problem extensions for Eq.~\eqref{eq:mult-dimensional-knapsack} towards practically relevant applications are the following:
\begin{enumerate}
    \item The power available in a parking site may vary over time; for example, when the parking site shares its connection to the grid with another infrastructure or a local renewable energy source is considered, such as a solar power plant.
    \item The decision of admitting/rejecting EVs is carried out via a pricing policy that affects the willingness of the drivers to use the parking site. 
    \item Increase the EV charging flexibility. EVs can have a desired state of charge at one or multiple points in time (e.g., at least 60 kWh by 6 am).
    \item The EV arrivals and departure times may not be known in advance, as described at the beginning of the section. Additionally, the forecast of renewable energies includes uncertainties if considered in (1).     
\end{enumerate}

The original problem in Eq.~\eqref{eq:mult-dimensional-knapsack} can be modeled as a QUBO. The original objective function is linear, i.e., the corresponding cost matrix will be a diagonal matrix, cf.~Sec.~\ref{sec:unconstrained_discrete_optimization}, and the constraints can, for instance, be encoded as penalties into the objective, cf.~Sec.~\ref{sec:unconstrained_discrete_optimization}, which, however, will immediately imply a dense cost matrix. 
The extensions (1)-(4) affect the difficulty of Eq.~\eqref{eq:mult-dimensional-knapsack} in different ways: 
(1) imposes additional constraints that may result in an increasing problems size, which may result in more qubits or having to encode more binary variables per qubit. 
(2) will imply adding continuous variables and results in a MIP. 
(3) adds flexibility to how EVs charge and discharge, which may further increase the number of decision variables.
Finally, (4) introduces uncertainty to the optimization problem, potentially necessitating a shift in the formulation from offline optimization to online or stochastic optimization such as MDP, cf.~Sec.~\ref{sec:optimal_control}.

\textbf{Use case 2: Charging and discharging EVs.} 
Consider the scenario where a fleet of EVs are plugged in from 6pm to 6am the next day, e.g., delivery vehicles. The fleet owner allows the energy supplier to use the EVs as energy storage subject to the constraint that the EVs are fully charged at departure time (i.e., by 6am the next day). The goal of the energy supplier is to provide electricity to the end customers at minimum cost. In particular, the energy supplier would like to store energy in the EVs when this is cheap, and use the energy in the EVs to power the grid when energy production is low or expensive to buy from the energy markets. The problem is challenging because the energy demand and energy prices may vary over time.

We can formulate the problem above as a \emph{deterministic} discrete-time control problem, cf.~Sec.~\ref{sec:optimal_control}. In short, we divide time in slots $k \in \{0,\dots,K-1\}$ of equal duration, where $K$ defines the total considered time horizon. The duration of a slot can be, for example, 1 hour, and the horizon 12 hours. 
Next, let $N$ be the number of EVs and $x_k \in \R^N_{\ge 0}$ represent the state of charge of the EVs at time $k$. In each time slot $k$, the controller selects an action from the action set $A(x_k) \subset \R^N$ containing the possible input/output powers of each EV, e.g., $\pm10$ kWh. The action set at time $k$ depends on the state $x_k$ since, for example, it is not possible to discharge an EV if its battery is empty. The choice of action $a_k \in A(x_k)$ affects the EVs' state of charge in the next time slot $k+1$. In particular, $x_{k+1} = F(x_k, a_k)$ where $F$ is a function that deterministically maps a state $x_k$ given an action $a_k$ to the next state $x_{k+1}$. 
Next, let  $T_k(x_k, a_k)$ be the cost obtained by selecting action $a_k$ in state $x_k$ at time $k$. 
The resulting optimization problem is given as following:
\begin{mini}
{a_k \in A(x_k)}{\sum_{k=0}^{K-1} T_k(x_k,a_k)\hspace{0.5cm}}
{}{} \label{eq:EV_control}
\addConstraint{x_K = x'}
\addConstraint{x_{k+1} = F(x_k, a_k),}
\end{mini}
where $x_0 \in \R^N_{\ge 0}$ denotes the EVs' initial state of charge and $x'$ the state that all EVs are fully charged. 
Thus, the goal is to select actions $a_k \in A(x_k)$ to minimize the sum of the costs given that $x_{k+1} = F(x_k, a_k)$ and the final state, $x_K$, is equal to $x'$. 
We have modeled the problem as a discrete optimal control problem, cf.~Sec.~\ref{sec:optimal_control}, but other formulations are possible as well. For instance, we can formulate it as a MIP with $K+1$ constraints for the dynamics, plus the constraints for enforcing feasibility of actions.
However, unlike the previous use case, Eq.~\eqref{eq:EV_control} is a less promising problem candidate for demonstrating industrial quantum advantage in the short term. 
The problem formulation can be mapped to a UCP (via DP) \cite{1295033}, which commercial solvers can solve efficiently with hundreds of units \cite[Sec.~6]{doi:10.1287/ijoc.2019.0944}. Nonetheless, we can add features to Eq.~\eqref{eq:EV_control} to make it more difficult to solve classically, and at the same time, closer to real-world scenarios:
\begin{enumerate}
    \item Energy constraints that affect how an EV or group of EVs can charge or discharge at a given time. 
    \item EVs arriving and departing at different (stochastic) times instead of being parked overnight.    
    \item Stochastic cost function that captures, for example, the uncertainty in the electricity demand. 
\end{enumerate}

We can approach the problem in Eq.~\eqref{eq:EV_control}, for instance, with DP.
As mentioned in Sec.~\ref{sec:dynamic_programming}, DP requires searching over all possible states and actions, which is often exponential in size. One approach to speedup DP is using Grover’s search algorithm as a subroutine. Similar to \cite{ABIKPV19} for the TSP, the quantum algorithm could first use DP to compute multiple paths from state $x_0$ to $x_K=s'$ and then use Grover's search to find a combination of those paths that yields lower cost. Nonetheless, such an approach is unlikely to yield practical speedups in the short term since Grover's search requires FTQC and a quadratic speedup might not be sufficient for the considered problems and scales. 
An alternative is to model the problem as a MIP, and try algorithms discussed in Sec.~\ref{sec:milp-miqp}.

The extensions (1)-(3) add different complications to Eq.~\eqref{eq:EV_control}: 
(1-2) add constraints to enforce certain target states at given times or feasible actions, e.g., minimum constant charge or discharge times, which affects both, DP and MIP formulations as shown for uni-directional charging by Federer \emph{et al.}~\cite{Federer2022}.
In case of stochastic arrival/departure times, (2) will require more complex approaches, e.g. MDP, to handle the uncertainty.
Similar to the stochastic case in (2), (3) will require approaches like MDP to compute a policy that minimizes the costs under uncertainty.
MDPs quickly become very difficult to solve due to the curse of dimensionality, cf.~Sec.~\ref{sec:optimal_control}.
Thus, this might be an interesting domain for first small test cases to analyze the potential of corresponding quantum approaches.
Since the result of MDP is a policy, the optimization and inference are split, which may allow one to relax requirements on the time-to-solution.

\textbf{Use case 3: Energy retailing with storage.} 
Consider an energy retailer that delivers energy to end customers. The energy supplier buys energy from energy producers in advance, e.g., on the day-ahead energy trading market, in one-hour time intervals, where the electricity price depends on several factors such as energy demand, renewable energy generation, etc. Additionally, the energy supplier has access to an energy storage system, e.g., a fleet of EVs that act as a \emph{virtual power plant}, that can accumulate energy when this is cheap, to later use it when energy demand is high. The energy supplier's goal is to deliver energy to its customers at the lowest possible cost. Thus, it is incentivized to leverage the flexibility provided by its energy storage assets. The problem is challenging because the electricity demand and the energy available in storage are not exactly known in advance. Furthermore, if the energy demand is not met, the energy supplier has to buy energy from other retailers that often sell energy at a higher price.

We can formulate the problem mathematically as follows. Let $e_k \in \R_{\ge 0}$ be the decision variable indicating the MWh the energy supplier buys in each hour $k \in \{0,1, \dots, K-1\}$ in advance, usually with $K=24$, and $p^{\text{low}}_k \in \R$ the price of the \emph{cheap} energy (which sometimes might even be negative!). Similarly, let $D_k$, $B_k$, $P_k^\text{high}$ be random variables that capture, respectively, the electricity demand at time $k$ in MWh, the energy available in storage at time $k$ in MWh, and the price of the \emph{expensive} energy at time $k$. The optimization problem is the following: 
\begin{mini}
{e_k \ge 0}{\sum_{k=0}^{K-1}  \left(p^{\text{low}}_k e_k   +  \mathbf E \left[ P^{\text{high}}_k [D_k - e_k - B _k ]^+  \right]   \right)}
{}{} \label{eq:energy_sell}
\end{mini}
where $[\cdot]^+:= \max\{0,\cdot\}$ and  the expectation is with respect to $D_k$, $B_k$, and $P_k^\text{high}$. That is, we have a stochastic unconstrained optimization where the objective function at time $k$ consists of two terms: the cost of buying cheap energy, and the expected cost of buying expensive energy. Note that $[D_k - e_k - B_k]^+$ captures the energy demand that cannot be met with the energy bought (i.e., $e_k$) and the energy in storage (i.e., $B_k$). 

The optimization in Eq.~\eqref{eq:energy_sell} is convex since the objective function is piece-wise linear and convex.
Thus, if $D_k$, $B_k$, and $P_k^\text{high}$ are random variables \emph{independent} of the decision variables, we can efficiently solve the problem with conventional convex optimization tools, in fact, by solving $K$ separate problems --- one for every term in the sum in Eq.~\eqref{eq:energy_sell}. Hence, Eq.~\eqref{eq:energy_sell}, in its current form, is \emph{not} a promising problem candidate for demonstrating quantum advantage. Yet, we can make it more challenging and closer to real-world problems with the following three extensions:

\begin{enumerate}
    \item The energy available in storage at time $k$ depends on the previous decisions taken; for example, on how EVs charge and discharge, as in the previous use case 2. 
    \item Convex or concave cost functions, i.e., the cost to buy energy might depend on the quantity ordered.
    \item The energy supply has to meet the energy demand.
\end{enumerate}

The extensions add different layers of complexity. In particular, (1) \emph{couples} the random variables with the decision variables and ``links'' the $K$ time steps, i.e., terms in the objective function Eq.~\eqref{eq:energy_sell}.
This is similar to the system dynamics function in optimal control, cf.~Sec.~\ref{sec:optimal_control}. 
This extension is important since assuming that the energy in storage (i.e., $B_k$) is independent of the decision variables does not allow us to capture, for example, that the energy retailer buys energy in advance to keep in storage for later use, i.e., the decisions in the past affect the events in the future.  We can recast Eq.~\eqref{eq:energy_sell} as an optimal control problem where the system dynamics are given by the algorithm in the second use case, i.e., the energy available by how EVs charge and discharge. That is, we are coupling this use case with the previous one, and, indirectly, inheriting the quantum challenges already discussed. In other words, we are adding the complications of use case 2 on top of use case 3. 
(2) adds a constraint of the form $g(D_k - e_k - B_k) \le 0$, where $g$ is a function such as CVaR that measures risk. The problem falls within the realm of robust optimization, see Sec. \ref{sec:robust_optimization}, or chance constraints \cite{nemirovski_2007_chance_constraints}.
In case of (3), depending on the exact properties of cost functions, the problem may remain convex or not. In case of a non-convex problem, this can quickly become very challenging to be solved.

\subsubsection{Outlook}\label{sec:sustainability_outlook}
Quantum computing has the potential to significantly advance optimization problems related to the sustainable energy transition.
However, identifying the most promising applications is an ongoing challenge. 
Here, we have introduced illustrative examples of optimization problems that may serve as test cases derived from real-world data. 
Under certain assumptions on the problem data, some instances may prove to be classically challenging even with few variables.
We have shown how the problem classes introduced in Sec.~\ref{sec:problem_classes} relate to these families of use cases with increasing complexity and of increasing relevance to real-world applications.
The discussion underlines the need for additional research on quantum optimization algorithms to approach the complications required for practical impact. It further shows that it is crucial to benchmark the problem instances to identify which ones are truly difficult classically and where there is room for a potential quantum advantage.

\section{Conclusion \& Outlook}\label{sec:conclusion}

In this paper, we provided a comprehensive overview of the potential, challenges, and emerging research areas in quantum optimization. We observed that, while complexity theory is useful to guide towards provable performance guarantees, it may not always be useful for finding practical quantum advantage. This underscores the need to develop and analyze quantum optimization heuristics to better understand their effectiveness. We highlighted key problem classes in optimization, reviewed existing algorithms and suggested new research directions. Further, we discussed the challenges of executing and scaling these algorithms on noisy hardware and the importance of benchmarking for identifying quantum advantages. Lastly, we explored two illustrative application domains, emphasizing the need to expand our focus beyond problem classes like QUBO to truly impact these fields.

As quantum hardware evolves with more qubits, reduced errors, and faster circuit execution, a new era is emerging where progress in quantum algorithm research comes from both theoretical analysis and empirical methods. While there remains a strong need to discover new algorithms with provable performance guarantees, 
we must also prioritize the development of quantum algorithms with no such guarantees, and benchmark their performance on real quantum computers. 
The improving abilities to test ideas in practice unlocks unprecedented opportunities for the advancement of quantum optimization. Even if it does not immediately imply a practical quantum advantage, the intuition gained from running experiments on real hardware empowers one to quickly validate proposals.   

In addition to technical advances in quantum optimization, there is also the question of responsible research and use of this new technology. There are a variety of causes, arising from different applications. Thus, further research should be done with the awareness that optimization use cases occur in a multitude of contexts --- from those with positive societal impact, to those with nefarious intent. We hope that open scientific discussions encourage advancement in quantum optimization and advocate responsible use of the insights presented in this work. While we showcase where benefits for optimization algorithms may lie, it is important not to overstate or misinterpret the results of this paper and its \textit{potential} applications. Such considerations motivate the need to establish clear benchmarks, cf.~Sec.~\ref{sec:benchmarks}, which enable a reliable interpretation of the scientific insights by a broader audience. Moreover, where use case applications of quantum optimization algorithms are chosen and funded, we encourage the prioritization of those with positive social impact, for example, sustainable energy transition as discussed in Sec.~\ref{sec:applications_sustainability}.

In conclusion, quantum optimization holds vast potential for various applications, but significant challenges remain in demonstrating a practical quantum advantage. Once a tangible quantum advantage is demonstrated, we expect that quantum optimization would rapidly influence many domains given the widespread relevance of optimization. This paper has provided a blueprint to define and measure progress in quantum optimization, which unquestionably has exciting prospects.

\vspace{3mm}


\noindent\textbf{Acknowledgments.}
The authors thank Jens Eisert and Mark Wilde for their valuable feedback and suggestions to further improve this paper.

Andris Ambainis acknowledges the support of the Latvian Quantum Initiative under European Union Recovery and Resilience Facility Project No.~2.3.1.1.i.0/1/22/I/CFLA/001 and the QuantERA II ERA-NET Cofund projects QOPT and HQCC.
Brandon Augustino and Swati Gupta acknowledge support from the US Defense Advanced Research Projects Agency (DARPA) Contract No.~HR001120C0046.
Vedran Dunjko acknowledges the support of the Dutch Research Council (NWO/OCW), as part of the Quantum Software Consortium programme (Project No.~024.003.037), and of the Dutch National Growth Fund (NGF), as part of the Quantum Delta NL programme.
Stuart Hadfield was supported by NASA Academic Mission Services Contract No.~NNA16BD14C, and by DARPA under interagency agreement IAA 8839, Annex 114. 
Thorsten Koch acknowledges support by the BMBF Research Campus MODAL (05M14ZAM, 05M20ZBM).
Steve Lenk was funded by Free State of Thuringia (Th\"uringer Aufbaubank) through project Quantum Hub Th\"uringen (2021 FGI 0047), by
Bundesministerium für Wirtschaft und Energie, Germany through the project “EnerQuant” (Project No.~03EI1025C), and  by the European Union under Horizon Europe Programme. Views and opinions expressed are however those of the author(s) only and do not necessarily reflect those of the European Union or European Climate, Infrastructure and Environment Executive Agency (CINEA). Neither the European Union nor the granting authority can be held responsible for them. Grant Agreement 101080086 - NeQST.
Jakub Marecek acknowledges the support of the Czech Science Foundation (23-07947S). 
Stefano Mensa, Emre Sahin and Benjamin Symons work was supported by the Hartree National Centre for Digital Innovation, a UK Government-funded collaboration between STFC and IBM. IBM, the IBM logo, and ibm.com are trademarks of International Business Machines Corp., registered in many jurisdictions worldwide. Other product and service names might be trademarks of IBM or other companies. The current list of IBM trademarks is available at https://www.ibm.com/legal/copytrade.
Giacomo Nannicini acknowledges support by ONR (N000142312585).
Patrick Rebentrost acknowledges support by the National Research Foundation, Singapore.
The work of Jon Yard was supported in part by the NSERC Discovery under Grant No.~RGPIN-2018-04742, the NSERC project FoQaCiA under Grant No.~ALLRP-569582-21 and the Perimeter Institute for Theoretical Physics. Research at Perimeter Institute is supported by the Government of Canada through Innovation, Science and Economic Development Canada and by the Province of Ontario through the Ministry of Research, Innovation and Science.
Los Alamos unlimited release LA-UR-23-33327.

\bibliography{references}
\end{document}